\documentclass[11pt,a4paper]{article}

\usepackage{iftex}
\ifPDFTeX
  \usepackage[utf8]{inputenc}
  \usepackage[T1]{fontenc}
  \usepackage{lmodern}
\else
  \ifXeTeX\fi
  \usepackage{fontspec}
\fi

\usepackage{amsmath,amssymb,amsthm,mathtools}
\usepackage{bm}

\usepackage{graphicx}
\graphicspath{{figures/}}
\usepackage{float}
\usepackage[section]{placeins}  % FloatBarrier at section boundaries

\usepackage{caption}
\usepackage{subcaption}

\usepackage{booktabs}
\usepackage{multirow}
\usepackage{array}
\usepackage{tabularx}
\usepackage{longtable}
\usepackage{threeparttable}
\usepackage{calc}
\usepackage[most]{tcolorbox}

\usepackage[a4paper,margin=1in]{geometry}
\usepackage{setspace}
\newcommand{\notepar}[1]{\par\begingroup\fontsize{9}{11}\selectfont\justifying\noindent #1\par\endgroup}

\usepackage{microtype}
\usepackage{ragged2e}  % \justifying for Notes
\usepackage{fancyvrb}
\usepackage{fvextra}
\fvset{breaklines=true,breakanywhere=true,fontsize=\small}
\DefineVerbatimEnvironment{verbatim}{Verbatim}{breaklines=true,breakanywhere=true,fontsize=\small}

\providecommand{\real}[1]{#1}

\usepackage[round,authoryear]{natbib}
\setcitestyle{aysep={,}}

\usepackage[hyphens]{url}
\usepackage[colorlinks=true,linkcolor=blue,citecolor=blue,urlcolor=blue,linktocpage=true,bookmarks=true,bookmarksopen=false,breaklinks=true]{hyperref}

\title{\textbf{Simulating Macroeconomic Expectations in Survey Experiments with LLM-based Economic Agents}%
\thanks{We thank Yuriy Gorodnichenko, Kai Li, Tracy Xiao Liu, Xiaobin Liu, Bin Miao, Carlo Pizzinelli, Thomas J. Sargent, Yan Shen, Mingzhu Tai, Lin Wang, Shan Wang, Johannes Wohlfart, Liyan Yang, Yang Yang, Ji Zhang and seminar and conference participants at The University of Hong Kong, Tsinghua University, Shanghai Jiao Tong University, Nanjing University, Xiamen University, Sun Yat-sen University, 2025 Global AI Finance Research Conference, The 25th China Economic Annual Conference, The 22nd Chinese Finance Annual Meeting, 2025 Future Scholars in Finance Forum and 2025 Symposium on Recent Developments in Time Series Econometrics and Macroeconomics for many helpful comments and suggestions. We gratefully acknowledge the support of the National Natural Science Foundation of China (Grant No. 71991474, 72073148, 72273156, 72303258) and National Social Science Foundation of China (Grant No. 22AZD121, 24ZDA042).}}
\author{\textbf{Jianhao Lin}\thanks{Lingnan College, Sun Yat-sen University, China, 510275. Email: linjh3@mail.sysu.edu.cn.}\qquad
\textbf{Lexuan Sun}\thanks{Corresponding author: Lingnan College, Sun Yat-sen University, China, 510275. Email: sunlx7@mail2.sysu.edu.cn; Tel: (+86) 18392757553.}\qquad
\textbf{Yixin Yan}\thanks{Lingnan College, Sun Yat-sen University, China, 510275. Email: yanyx33@mail2.sysu.edu.cn.}}
\date{First Version: May 2025\\
This Version: June 2026}

\begin{document}

\maketitle

  \noindent
  \normalsize
\begin{abstract}
  \noindent
  \normalsize
We introduce a framework for simulating macroeconomic expectations in survey experiments using LLM-based economic agents (LLM Agents). We construct LLM Agents equipped with several functional modules that retrieve personal characteristics, prior expectations, and dynamic external information. We validate our framework by recapitulating three representative survey designs covering various expectations across different types of respondents. Our results show that LLM Agents generate expectation distributions highly similar to human data and capture human-aligned qualitative patterns in open-ended responses. Evaluation reveals that priors are crucial for matching distributions, whereas personal and external information drive human-like thought processes. Our findings offer guidance for narrowing the belief gap between generative AI and humans at the aggregate level while delineating the boundaries of the framework.

\end{abstract}


\bigskip

\addtocontents{toc}{\protect\setcounter{tocdepth}{-1}}
\clearpage
\thispagestyle{plain}
\section{Introduction}\label{introduction}\label{sec:1}

How agents form expectations is a crucial question for analyzing macroeconomic dynamics \citep{coibion2015information,coibion2018formation}. Survey experiments are widely used to study this question, with randomized controlled trials (RCTs), hypothetical vignettes, and randomized information experiments being the three most common designs \citep{haaland2023designing, binder2026information, coibion2026expectations}. However, collecting survey responses presents several challenges, including high costs, slow data collection, and low response rates \citep{meyer2015household, cavallo2016billion, jarmin2019evolving}. Recent research attempts to use \textit{in silico} samples to address these constraints \citep{zarifhonarvar2026generating,wu2025llm}. These studies show that Large Language Models (LLMs, also referred to as \emph{foundation models}\footnote{{} In this paper, we use \emph{foundation models} interchangeably with general-purpose LLMs pre-trained on massive datasets and applicable to a broad range of downstream tasks. Examples include the GPT series by OpenAI and DeepSeek-R1 by DeepSeek.}) can simulate survey experiments on household inflation expectations. Despite the rapid growth of this literature, two gaps remain. First, existing methods are effective primarily for RCT designs and focus on household inflation expectations, but they overlook other types of designs, respondents, and macroeconomic expectations. Second, although these studies show that LLMs can recover human-comparable treatment effects, \citet{zarifhonarvar2026generating} notes that expectations generated by LLMs with personas are often highly concentrated around the mean. Consequently, they may lack the key heterogeneity that characterizes real-world expectations data.

To fill these gaps, we propose a novel framework for constructing LLM-based economic agents (LLM Agents) to simulate macroeconomic expectations held by different types of respondents across several representative survey designs. The concept of LLM Agents in this paper is similar to that of \emph{Homo silicus} proposed by \citet{horton2023large}. Specifically, they are LLM-based computational models whose design is informed by economic theory and tailored to model the expectations of specific agent types. Before constructing LLM Agents, we first analyze why the outputs of foundation models with personas are overly homogeneous, and identify three possible reasons \citep{zarifhonarvar2026generating, wang2025large, xie2026evaluating}. First, foundation models trained on static data cannot retrieve dynamic information driving expectations. Second, personas provide insufficient context for models to generate heterogeneous outputs. Third, alignment via human feedback embeds annotators' beliefs, anchoring outputs of models that are not calibrated to real data within a narrow range. Inspired by methods developed in recent work in AI simulation \citep{park2023generative, li2024econagent, piao2025agentsociety, mou2026individual}, we incorporate agentic module design into our framework. This new method uses the foundation model as the ``brain'' and customizes task-specific modules as the ``hands'' and ``feet'' based on the characteristics of agents. This design enables LLM Agents to pull in external dynamic information, condition on richer context, and use real expectations data for calibration, thereby allowing them to generate more heterogeneous expectations and role-consistent open-ended responses. We further validate our framework across two other types of representative survey designs and a large-scale expectation survey to demonstrate its extensibility.

This paper introduces and validates our framework through four parts. First, we describe the construction of LLM Agents. We focus on two representative agent types commonly encountered in expectation surveys---households and experts. Accordingly, we develop LLM Agents to simulate household expectations (referred to as \emph{Household Agents}) and expert expectations (referred to as \emph{Expert Agents}). For households, personal characteristics, prior expectations, and social media information play important roles in shaping expectations. Thus, Household Agents are equipped with three modules: a Personal Characteristics Module (PCM), a Prior Expectations \& Perceptions Module (PEPM), and a Social Media Information Module (SMIM), which together draw on household surveys and social platform data. In contrast, expert expectations are primarily shaped by professional background and domain knowledge. In addition to the PEPM, Expert Agents add two further modules: a Professional Background Module (PBM), populated from institutional websites and LinkedIn, and a Knowledge Acquisition Module (KAM)\footnote{{} This paper uses a number of abbreviations. For the reader's convenience, Supplementary Appendix Table~\ref{tab:abbrev} provides a complete list of all abbreviations used in this paper, together with their full terms.}, which retrieves domain knowledge from web search. Then, we initialize each LLM Agent with a minimal role specification and a task description (termed a \emph{naive persona}) in a prompt (i.e., initialization prompt). Meanwhile, we specify a module-invocation rule in this prompt so that agents generate their expectations by weighing priors against new signals according to Bayesian updating, with the weight governed by their assigned levels of confidence.

Second, we introduce the experimental designs. We draw on three representative survey experiments that cover common types of macroeconomic expectations. We select these experiments because their designs are transparent, replicable, and widely recognized. They also report well-identified qualitative findings and provide public microdata, enabling us to compare the simulation results with human data. Specifically, the first is a hypothetical vignette experiment designed by \citet{andre2022subjective} on inflation and unemployment expectations of households and experts. The second is a randomized information experiment introduced by \citet{chopra2025home} on home price expectations of homeowners and renters. Moreover, to assess the extensibility of our framework to a more general survey design and simulation performance of LLM Agents beyond the knowledge cutoff, our third experiment builds on the Michigan Surveys of Consumers (MSC)\footnote{{} The Michigan Surveys of Consumers is one of the longest-running household surveys in the world. It is conducted by the University of Michigan to assess U.S. consumer attitudes and expectations regarding personal finances, business conditions, and economic outlook. Established in 1946, the survey collects data from approximately 600 respondents each month and is widely used in many studies \citep{curtin1982indicators,dacunto2023data}.}. We use this third design to simulate long- and short-term inflation and home-price expectations in the 2025 MSC and compare the results with the real data to examine their post-knowledge-cutoff simulation performance. Then, we use LLM Agents to recapitulate these three experiments. Following \citet{horton2023large}, we use the term ``recapitulate'' rather than ``replicate'' because we do not aim to exactly reproduce the expectations of individual subjects in the original experiments. Instead, we examine whether, at the population level, LLM Agents can capture the key heterogeneity in expectations within and across groups.

Third, we analyze the simulation results. By comparing the simulation results with the original human data across three experiments, we further summarize the similarities and differences between the two and illustrate the capabilities and boundaries of this method. The results indicate that LLM Agents produce expectation distributions highly similar to those observed in human respondents. Although these distributions are slightly more homogeneous than those of humans, they still capture key heterogeneity within and across different types of agents. However, matching the distributions is necessary but not sufficient to reflect their simulation capability. We therefore examine whether they capture the key patterns in the thought processes underlying expectation formation. Using multidimensional text analysis of open-ended survey responses generated by human respondents and LLM Agents, we find that LLM Agents exhibit a thinking pattern similar to that of humans, namely selective recall, though the channels and content they recall are more limited. In addition, by extracting Directed Acyclic Graphs (DAGs) from these responses (i.e., our operationalization of \emph{mental models}), we find that LLM Agents trace causal paths similar to those of humans, though both the breadth of nodes and the diversity of paths are somewhat lower. These findings explain why LLM Agents generate distributions similar in shape to human data but slightly more homogeneous.

Fourth, we evaluate the contribution of each component in LLM Agents to the simulation. Specifically, by ablating one component at a time and holding the rest fixed, we investigate the source of LLM Agents' ability to simulate expectation distributions and capture key patterns in the thought processes. The results indicate that removing any single component degrades performance relative to the full agent, with the magnitude of degradation varying by module. This suggests that all components contribute to simulation capabilities across different dimensions. Across all modules, prior expectations from PEPM contribute the most to matching the distributions, whereas personal information from PCM and PBM, together with text data from human society extracted by SMIM and KAM, is essential for recapitulating human-like selective recall and mental models. Moreover, both LLM Agents with only initialization prompts (equivalent to foundation models with only naive personas\footnote{{} When all components of LLM Agents are removed except for initialization prompts, the module-invocation rules within these prompts cease to function. Consequently, the agents are equivalent to foundation models with only naive personas.}) and LLM Agents without initialization prompts fail to capture key heterogeneity and patterns. These findings suggest that: (i) foundation models alone fail to capture the latent mapping between simulated agents and their expectations, and thus cannot directly serve as \emph{Homo silicus}; (ii) without explicit guidance from roles, task objectives, and module-invocation rules, LLM Agents cannot effectively utilize the rich information extracted by their modules. Consequently, effective simulation depends not only on the volume of information acquired, but also on initialization grounded in clear objectives and economic theory. These findings offer insights into what information to incorporate and how to do so in order to calibrate foundation models as tools capable of effectively simulating expectations.

This paper makes two contributions to the literature. First, we contribute to the rapidly growing literature on how generative AI (GenAI) simulates economic agents. These studies attempt to use LLMs to simulate human beliefs \citep{bybee2023ghost,zarifhonarvar2026generating,wu2025llm}, behaviors \citep{horton2023large,tranchero2024theorizing,kazinnik2026bank}, and decisions \citep{li2024econagent,hansen2025simulating}. To the best of our knowledge, this is the first study to simulate the macroeconomic expectations among different types of respondents in various representative survey experiments by constructing LLM Agents. The two most related studies are \citet{zarifhonarvar2026generating} and \citet{wu2025llm}, but we are fundamentally different from both of them. Both studies examine whether LLMs can recover information-treatment effects on household inflation expectations within RCT designs. By contrast, we (a) construct LLM Agents that capture both distributional and qualitative heterogeneity across and within agent types, (b) cover macroeconomic expectations beyond inflation, and (c) accommodate experimental designs beyond RCTs. In addition, \citet{zarifhonarvar2026generating} notes that foundation models with personas generate highly homogeneous expectations due to static training data and a lack of real-life experience. By using multiple modules to incorporate information beyond what persona-only prompts can provide, our framework offers a composite paradigm grounded in GenAI and also calibrated with expectations survey data and text data from human society, which are rich in dynamic information and personal experience. Our goal is not to replace human samples in survey experiments; rather, our framework aims to establish a \emph{complementary} relationship with them. Specifically, while relying on these survey data for calibration, the framework holds future potential to provide low-cost pre-experimental simulations or impute a small number of missing observations for these surveys. Beforehand, however, we should first validate the simulation capabilities of this framework through several representative experiments, which constitutes one of the main contributions of this paper.

Second, we contribute to the emerging literature on AI behavioral science \citep{meng2024ai}, which primarily examines the similarities and differences between GenAI and humans in behavior and cognition, including rationality \citep{chen2023emergence,bini2025behavioral}, biases \citep{chen2024recovering,hagendorff2024deception}, and preferences \citep{goli2024frontiers,ouyang2024ethical}. However, this literature largely characterizes GenAI's behavior and cognition themselves, with little attention to the reasons GenAI offers for them or how these reasons compare with those of humans. Following recent research in behavioral economics that elicits underlying reasons through open-ended questions \citep{haaland2025understanding}, we collect LLM Agents' open-ended responses, which reflect the thoughts underlying their expectation formation, and compare them with those of human respondents in the original experiments. We do not aim to match thoughts at the individual level. Rather, we examine, at the population level, whether the generated thoughts reflect the key mental patterns of the corresponding human groups, specifically, selective recall and mental models. Through these comparisons, we identify both the capabilities and boundaries of our approach. While LLM Agents qualitatively recapitulate the mental patterns underlying expectation formation, they exhibit quantitative gaps relative to human data, including concentrated selective recall, simplified mental models, and reduced thought diversity. Therefore, these LLM Agents are abstractions of the key characteristics of real-world agents, akin to humans with smoothed-out idiosyncrasies.
These findings offer a feasible path for using agentic module design to better simulate the mental patterns underlying expectation formation, while delineating its boundaries also yields insights for distinguishing GenAI-generated from human-generated open-ended responses.

The rest of our paper is organized as follows. Section~\ref{sec:2} provides a framework. Section~\ref{sec:3} describes the construction of LLM Agents. Section~\ref{sec:4} introduces the experimental designs and prompts. Section~\ref{sec:5} presents the simulation results and analysis. Section~\ref{sec:6} evaluates the contributions of each component in LLM Agents. Section~\ref{sec:7} concludes.

\section{A Framework}\label{a-framework}\label{sec:2}

\begin{figure}[h]
\centering
\includegraphics[width=0.8\linewidth,keepaspectratio]{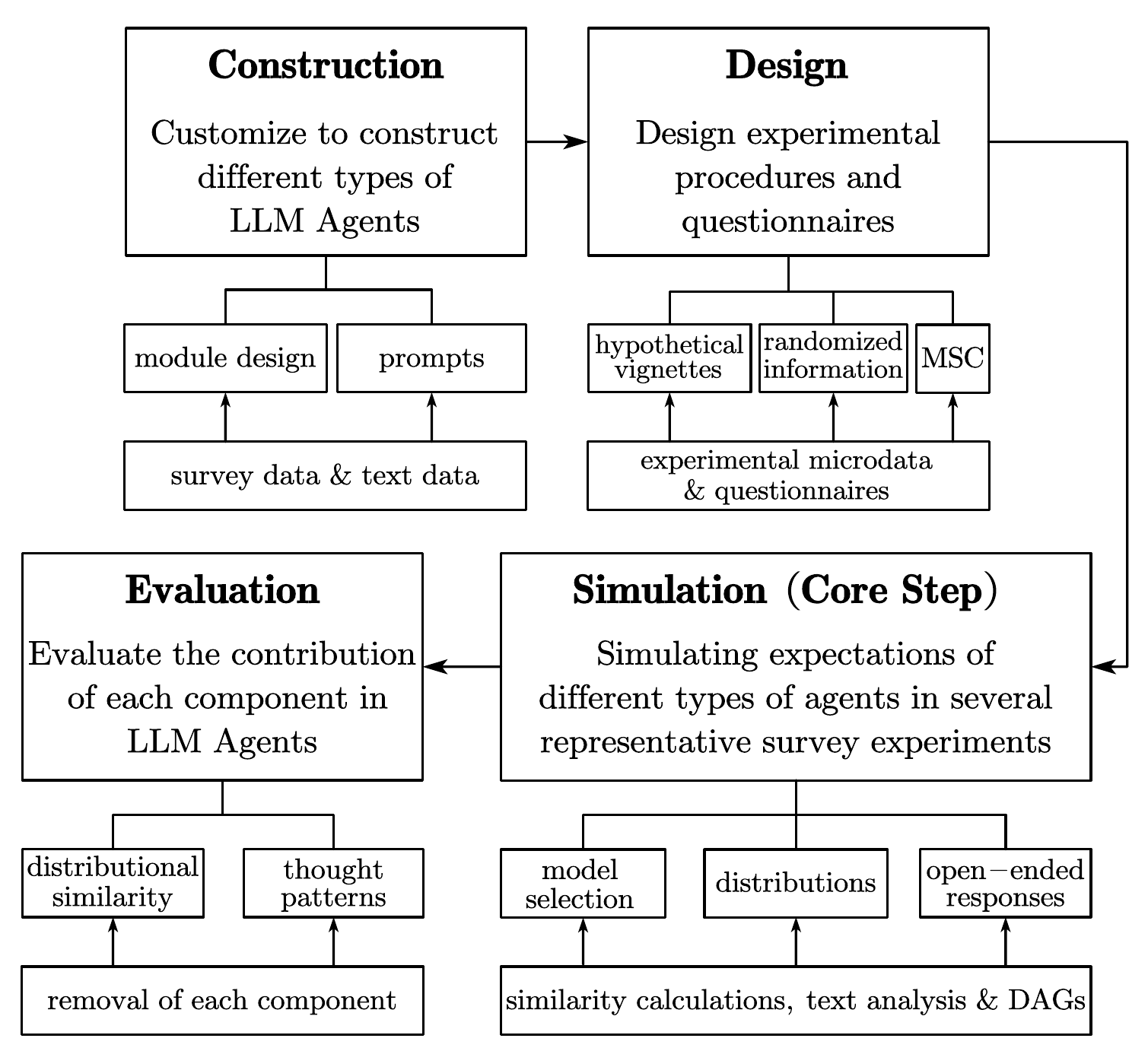}
\caption{The framework for simulating macroeconomic expectations}\label{fig:1}
\notepar{%
Notes: This figure illustrates a framework we propose for simulating macroeconomic expectations in survey experiments. The framework consists of four main steps in sequence: Construction → Design → Simulation → Evaluation. Each large box at the top provides a brief overview of the corresponding step, the smaller boxes in the middle summarize the key points of each step, and the rectangular strip at the bottom outlines the specific tasks, methods, or data used in each step.
}
\end{figure}

In this section, our goal is to propose a framework that enables economists to simulate macroeconomic expectations of different types of respondents in survey experiments using customized LLM Agents. As shown in Figure~\ref{fig:1}, this framework consists of four steps in sequence: Construction → Design → Simulation → Evaluation. It provides a practical methodology and operational procedures for simulating macroeconomic expectations in survey experiments.

The first step is to construct LLM Agents. In this step, we design two types of LLM Agents with distinct architectures to simulate the typical subjects in expectations surveys: households and experts. Each LLM Agent is equipped with modules that retrieve the key information required to simulate its target group. For instance, social media contains rich dynamic information and real-life experiences that can shape household expectations but are not readily accessible to foundation models. We therefore design the SMIM for Household Agents to scrape and clean the relevant data from social media platforms. In most cases, these modules extract information directly from actual survey data and text data on the internet. However, when the survey sample is too small to support the simulation, our LLM Agents generate synthetic data\footnote{{} The use of LLMs to generate synthetic data is becoming widespread in academic research, and its underlying rationale has gained increasing acceptance; it offers a viable alternative for constructing research datasets when empirical data are scarce \citep{yu2023large,halterman2025synthetically,ge2025scaling}.} from the available real samples and merge the two to construct a semi-synthetic dataset, thereby expanding the limited sample size. We then tailor the prompts in each module to fit the specific survey experiment, keeping their key wording consistent with the original questionnaire items to mitigate potential subjectivity. Finally, we use a standardized template of initialization prompts to initialize all agents\footnote{{} For the full prompts and related instructions for the LLM Agents across the three experiments, see: \href{https://drive.google.com/file/d/1PPiuPod1Ar38R8k43AUjul3rnGyOQw8j/view?usp=sharing}{{https://drive.google.com/file/d/1PPiuPod1Ar38R8k43AUjul3rnGyOQw8j/view?usp=sharing}}.}.

The second step is to design survey-experiment procedures and questionnaires. Before running simulations, we specify the experimental details and questionnaire designs according to research needs, and collect microdata from the corresponding survey experiments to calibrate LLM Agents or validate simulation results. Specifically, in this paper, we draw on three widely recognized designs to test the effectiveness of our framework: a hypothetical vignette experiment from \citet{andre2022subjective} that examines how four canonical macroeconomic shocks affect household and expert expectations of inflation and unemployment; a randomized information experiment from \citet{chopra2025home} on homeowners' and renters' home price expectations; and selected items from a large-scale expectations survey (MSC). All three survey experiments are highly representative of their respective design types and provide relatively complete microdata for comparison with our simulation results. Importantly, we do not aim to replicate every detail of these experiments one-to-one. Rather, we use these representative designs to assess whether our LLM Agents can recapitulate key patterns of expectations within and across certain real agents. Accordingly, we focus on qualitative comparison rather than strict statistical tests in the steps that follow \footnote{{} We focus on qualitative comparisons rather than statistical tests for three reasons. First, even when a fresh sample of human subjects is drawn from the same population to replicate the original experiment, the results may still fail a strict statistical test because of randomness and other uncontrollable factors \citep{camerer2016evaluating, camerer2018evaluating}, let alone the simulated results from LLM Agents. By contrast, the qualitative patterns that reflect core empirical regularities and their theoretical foundations are usually robust across replications, and these are what we seek to recapitulate. Second, by construction, LLM Agents cannot exhaust all unobserved heterogeneity among human respondents and are therefore likely to differ from human data in quantitative terms, often failing statistical tests. Yet whether a test is passed only indicates whether the two differ in a statistically significant sense; it does not speak to whether LLM Agents capture the features of respondents' expectation formation in an economic sense. Using statistical tests as the criterion may therefore obscure the simulation capability of LLM Agents. Third, much of the emerging literature on using LLMs to simulate human behavior likewise emphasizes the replication of qualitative findings rather than the passing of statistical tests \citep{horton2023large,argyle2023out,aher2023using}.}.

The third step constitutes the core of our framework. In this step, we use the constructed LLM Agents to simulate three experiments. First, we compare the responses of LLM Agents based on different foundation models with those of human data to guide model selection. Second, we assess how closely the simulated expectation distributions match the real ones by computing the shape similarity between them in the corresponding experiments. Section~\ref{sec:5-1} explains in detail why we adopt distributional shape similarity as the measure of closeness and how we compute it. Depending on research needs, we can not only simulate expectations for contemporaneous samples but also simulate expectations in future periods. For instance, in the first two experiments, we recapitulate results from contemporaneous samples to assess the extent to which responses generated by LLM Agents resemble those of human respondents. In the third experiment, we use sample data covering the full year before the foundation models' knowledge cutoff to simulate the expectation distributions over a subsequent period, thereby testing the post-knowledge-cutoff simulation performance of LLM Agents. Finally, by conducting textual analysis of open-ended responses and extracting DAGs from them, we compare the thought processes underlying expectation formation in LLM Agents with those in human respondents, examining whether the agents can recapitulate the key patterns of selective recall and mental models exhibited in the human samples.

The fourth step is to evaluate the contribution of each component of the LLM Agents. This step aims to assess how the modules and the initialization prompts added in Step 1 contribute to the simulation performance. By removing one component of the LLM Agents at a time while holding the others fixed, we identify the sources underlying different dimensions of simulation performance. Specifically, we compare the decline in distributional shape similarity and the distortion in the thought patterns underlying expectation formation after each component is removed, thereby assessing which components contribute more to capturing distributional heterogeneity and which are more important for recapitulating the qualitative patterns in human thoughts. This evaluation helps verify the soundness of the LLM Agents' architectures and provides insights for constructing them.

In summary, our framework does not aim to replace human samples in traditional surveys. Rather, it calibrates LLM Agents using real-world data, including data from these traditional surveys, with the aim of serving as a complementary tool to the three types of survey designs in the future. Before that, however, we first need to validate the framework's capabilities and define the boundaries of its applicability by recapitulating the core findings of several representative experiments, which is the focus of the subsequent sections of this paper. This helps us understand the similarities and differences between GenAI-generated expectations and human expectations, and offers a potential roadmap for more effectively reducing the systematic differences between the two in the future.

\section{Construction of LLM Agents}\label{construction-of-llm-agents}\label{sec:3}

In this section, we present the detailed procedures of Step 1 in our framework, explaining how we construct LLM Agents that represent different types of respondents. Specifically, we develop LLM Agents that simulate the expectations of households and experts, which serve as the subjects in the experiments described in the subsequent sections.

\subsection{LLM Agents for Simulating Household Expectations}\label{llm-agents-for-simulating-household-expectations}\label{sec:3-1}

Households are the most common subjects in expectation surveys, and they are included in all subsequent experiments. Before constructing the Household Agents, we first draw on existing economic theory and empirical findings to identify the main factors that shape household expectations.

First, a large literature suggests that economic expectations or perceptions are closely linked to various demographic characteristics. Studies have found significant differences in economic expectations across individuals of different ages, genders, political affiliations, education levels, and income groups \citep{souleles2004expectations,ehrmann2017consumers,bendavid2018expectations,coibion2022monetary,dacunto2024household}. Second, the prior expectations or perceptions of economic agents regarding economic variables serve as a crucial determinant of their future expectations, particularly their most recent perceptions of these variables \citep{jonung1981perceived,coibion2020inflation}. Because foundation models may be influenced by human feedback data used in post-training, the expectations they generate tend to be anchored within a narrow range. Therefore, feeding in expectation data from real surveys conducted during a given period helps correct this systematic bias. Third, media coverage exerts a significant influence on households' macroeconomic expectations \citep{carroll2003macroeconomic,lamla2012role}. In particular, with the rapid rise of social media, most households now get news primarily from platforms such as X (formerly Twitter) and increasingly consider these sources as more credible than traditional news media \citep{coibion2022monetary,ehrmann2022central,angelico2022can,gorodnichenko2024central}. Consequently, continuously updated social-media information has become an increasingly important factor shaping household expectations. Social media texts contain abundant dynamic information and real-world experiences. They can provide Household Agents with up-to-date information and rich contextual content, helping them generate associative responses that resemble those of real humans.

\begin{figure}[h]
\centering
\includegraphics[width=\linewidth,keepaspectratio]{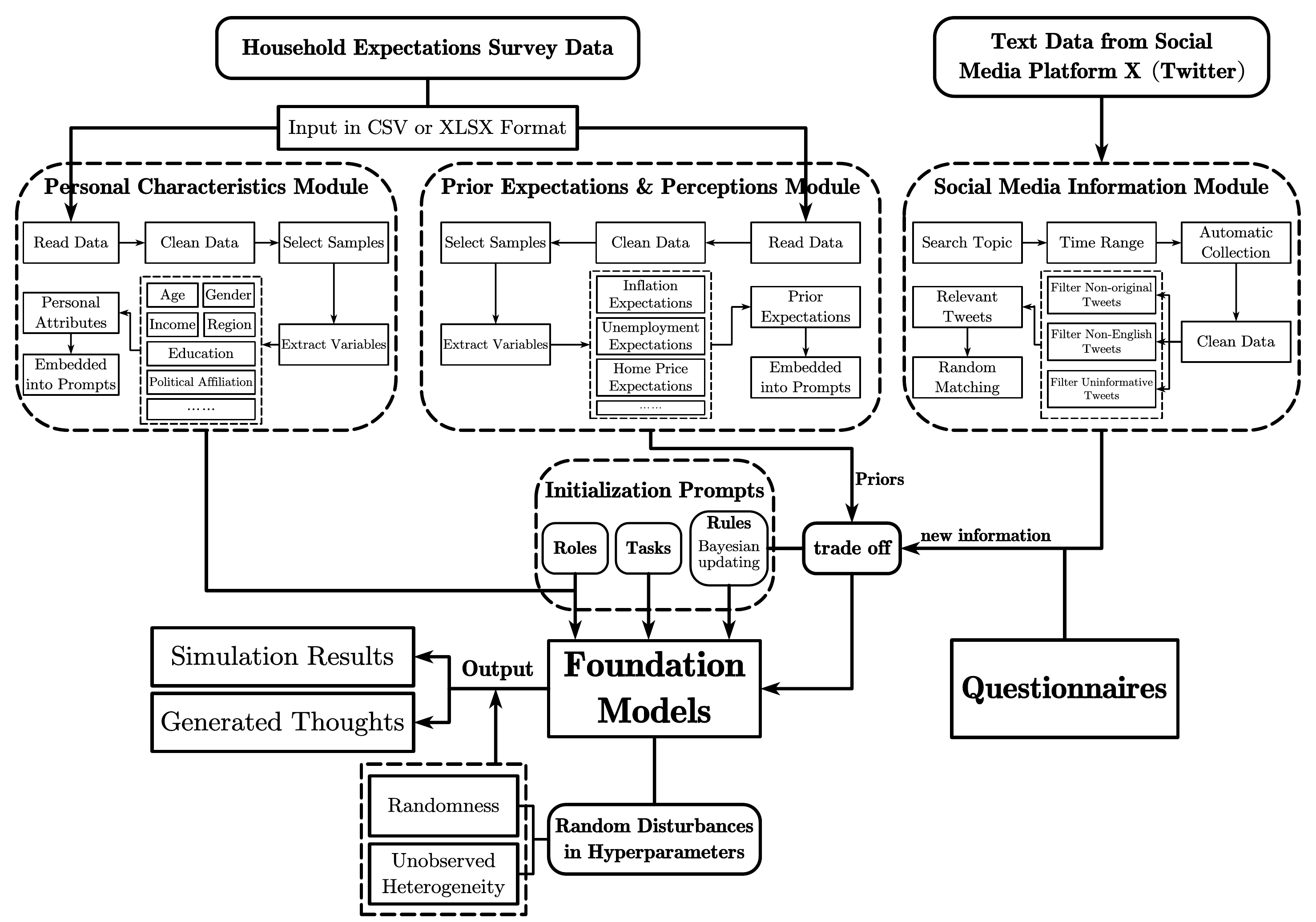}
\caption{LLM Agents for simulating household expectations}\label{fig:2}
\notepar{%
Notes: This figure presents the detailed architecture of the LLM Agents for simulating household expectations. The Household Agents consist of six main components: Personal Characteristics Module (PCM), Prior Expectations \& Perceptions Module (PEPM), Social Media Information Module (SMIM), Random Disturbances (RD), initialization prompts, and foundation models. Both the PCM and PEPM draw on data from household expectation surveys. SMIM collects tweet text data from the social media platform X. These modules automatically extract and process information, with their operational rules defined in the initialization prompts. For input questionnaires, Household Agents can engage in role-playing and perceive external environment through these components, ultimately outputting heterogeneous expectations along with the underlying thoughts.
}
\end{figure}

Based on this literature, we construct the PCM, the PEPM and the SMIM (see Figure~\ref{fig:2}) to incorporate information on households' personal characteristics, prior expectations and social media into LLM Agents. Specifically, the PCM includes key attributes such as age, gender, political affiliation, and education level of respondents. The PEPM captures their prior expectations about economic variables such as inflation, unemployment, and home prices. These data originate from household expectation surveys and are typically provided to the PCM and PEPM modules in CSV or XLSX format. Each module automatically reads the files, cleans the data, selects samples, extracts key variables, and embeds their numeric or textual values into prompts submitted to the Household Agents. The wording of the prompts varies with the designs of the experiments, but should remain largely consistent with the corresponding formulations in the original questionnaires. 

The SMIM automatically retrieves and processes text data from relevant posts on social platform X according to the experimental requirements. For instance, if the experiment focuses on U.S. inflation expectations, the user can set the search topic to ``US Inflation'' and specify a time window (i.e., the experimental period, typically aligned with the data range used in the PCM and PEPM). Empirically, it is difficult to identify the specific tweets read by individual respondents. Therefore, SMIM collects tweets related to a specified search topic that exhibit relatively high views, retweets, or replies\footnote{{} This design builds on the core premise of attention economics: in an information-rich world, human attention is the truly scarce resource \citep{Simon1971, loewenstein2025economics}. \citet{sims2003implications} formalizes this concept as rational inattention, where cognitively constrained economic agents allocate limited attention only to the most salient information sources and systematically filter out marginal information. Social media algorithms further amplify this scarcity. High-engagement tweets reflect user preferences and receive exponential exposure through algorithmic distribution, thereby driving the mainstream narrative \citep{loewenstein2025economics}. 
Consequently, using tweet popularity as a proxy for information exposure aligns with attention allocation theory and ensures empirical feasibility. It closely approximates the information set households likely observe and use to form their expectations during that period. Conversely, collecting low-engagement marginal tweets systematically selects information the public already deems low-value. This contradicts the theory of attention allocation and would introduce excessive noise into the simulation.}. Importantly, we do not manually filter these tweets. The X platform automatically aggregates them using its algorithm based on the entered search keywords. SMIM then performs data cleaning, including filtering out non-original tweets, non-English tweets, and uninformative tweets (e.g., very short or promotional tweets). Furthermore, SMIM randomly matches these tweets to each agent without replacement. This matching procedure is theoretically justified within our analytical perspective \footnote{{} There are two reasons. First, our research simulates expectations at the population level using a large sample, rather than perfectly replicating micro-level individuals. We aim to capture how social media information shapes the overall distribution of household expectations and do not require an individual agent's information-reception trajectory to precisely match its real-world counterpart. By the law of large numbers, randomly assigning a heterogeneous set of tweets to a heterogeneous set of agents in a sufficiently large sample generates an information exposure distribution that converges to the true population distribution, ensuring a valid aggregate fit. 
Second, from the perspective of information availability, we focus on mainstream public discourse as the primary information input for aggregate households, rather than individual private information sets. Random matching ensures SMIM supplies Household Agents with diverse, representative tweets, broadly exposing the simulated population to the dominant public opinions of the period.}. Finally, to verify that this randomization does not compromise empirical robustness, we report additional experimental results in Supplementary Appendix Section~\ref{sec:b}. 

In addition, to clarify the roles and tasks of Household Agents, we include a simple persona prompt (e.g., ``\textit{Suppose you are an ordinary individual (household)}") and a specific task description in the initialization prompts. Although simple, this prompting method is widely used in the relevant literature \citep{horton2023large,mei2024turing,cui2025large}, and we refer to this approach as the \textit{naive persona}. Further, we explicitly formulate module-invocation rules in the initialization prompts based on theories of economic expectations. This guides Household Agents to effectively use the information extracted by the modules, simulating the expectation formation of real households.

Specifically, the expectation formation of economic agents (including households) regarding a macroeconomic variable $\theta$ can be modeled as Bayesian learning under incomplete information \citep{baley2023bayesian,coibion2015information}. Assume their prior expectations before receiving external signals follow a normal distribution $\theta \sim N(\mu_0, \tau_0^{-1})$, where $\mu_0$ is the mean of the priors and $\tau_0 \equiv 1/\sigma_0^2$ is the precision of the priors. During the survey, agents receive new information, such as contemporaneous social media content and the information provided in the questionnaire. We represent this external information as noisy signals about $\theta$: $s = \theta + \varepsilon$, where the noise term $\varepsilon \sim N(0, \tau_s^{-1})$ and $\tau_s$ is the signal precision. Following Bayesian updating, agents combine their prior beliefs with the new signals to form the posterior expectation $\theta | s \sim N(\mu_1, \tau_1^{-1})$. 

Under the Normal-Normal conjugate model, the posterior mean is the precision-weighted average of the mean of the priors and the signals, and the posterior precision is the sum of prior and signal precisions:

\begin{equation}\label{eq1}
E\left[ \theta |s \right] \equiv \mu_1=\frac{\tau_0}{\tau_0+\tau_s} \mu_0+\frac{\tau_s}{\tau_0+\tau_s} s, \quad \tau_1=\tau_0+\tau_s .
\end{equation}

Defining the weight $\omega = \frac{\tau_s}{\tau_0 + \tau_s}$, we rewrite Equation~(\ref{eq1}) as:

\begin{equation}\label{eq2}
\mu_1 = (1 - \omega) \mu_0 + \omega s.
\end{equation}

Equation~(\ref{eq2}) captures how agents trade off between prior beliefs and new signals in reality, with the determination of the updating weight $\omega$ being central. Since prior beliefs are private information, and agents differ in their estimates of their private signal precision $\tau_0$, determining $\omega$ essentially requires determining $\tau_0$. The mainstream literature documents a mathematical and economic correspondence between agents' levels of confidence and their estimated precision of private information: overconfident agents tend to overestimate the precision of their private signals \citep{daniel1998investor,coibion2021you,broer2024forecaster}. Let $\gamma$ denote an agent's confidence level, which implies the relationship $\tau_0 \propto \gamma$. Therefore, when agents are highly confident in their prior expectations (i.e., $\gamma \to \infty$), it follows that $\tau_0 \to \infty$, then $\omega \to 0$. In this case, agents overweight their priors and underreact to new information (i.e., conservatism). Conversely, agents lacking confidence underweight their priors and over-rely on new signals, leading to over-updating of beliefs (i.e., base-rate neglect) \citep{chan2025prior,benjamin2019errors}.

To translate the above theoretical foundation into module-invocation rules that can be input into Household Agents, we assign them one of five confidence levels\footnote{{} When survey data on human respondents lack information regarding their confidence in (prior) expectations, we can employ random stratified sampling to divide the total sample into five subsamples that are approximately equivalent in both demographic structure and size. Each of these subsamples is then randomly assigned one of five distinct confidence levels.}, ranging from extremely weak to extremely strong, and instruct them to follow the directive below:

\begin{tcolorbox}[colback=white,colframe=black,boxrule=0.8pt,arc=0pt,breakable]
\begin{verbatim}
Suppose you are an ordinary individual (household) with {CONF} confidence + (task description) ......
IMPORTANT INSTRUCTIONS: Your responses should trade off among the various pieces of information mentioned above in accordance with your level of confidence: If you are confident, your answers will rely on Prior Expectations & Perceptions, and will not be influenced by other information, such as the Social Media Information. On the other hand, if you lack confidence, your answers are more likely to be influenced by other information.
\end{verbatim}
\end{tcolorbox}
\vspace{-2mm}
\noindent{\fontsize{9pt}{9pt}\selectfont Notes: CONF (i.e., confidence level $\gamma$) is a categorical variable comprising five categories: extremely weak, weak, moderate, strong, and extremely strong.}

Furthermore, given the close link between demographic information and the expectations generated by real agents, Household Agents should be required to fully take into account the information extracted by the PCM when role-playing to generate responses. Therefore, we instruct them to follow:

\begin{tcolorbox}[colback=white,colframe=black,boxrule=0.8pt,arc=0pt,breakable]
\begin{verbatim}
In addition, your responses should fully reflect the Personal Characteristics (such as age, gender, educational level, political affiliation, etc.) of the role you are portraying.
\end{verbatim}
\end{tcolorbox}

However, unobserved heterogeneity and various random factors remain, exerting unknown effects on respondents' answers. Although such factors are difficult for LLM Agents to capture and characterize, we can introduce some of them by adjusting the hyperparameters in foundation models that govern the diversity of generated text, such as \texttt{temperature} and \texttt{top-p}\footnote{{} Both \texttt{temperature} and \texttt{top-p} are hyperparameters that control the diversity of text generated by LLMs. The difference lies in that \texttt{temperature} modulates the shape of the probability distribution over the next token by rescaling logits before the softmax transformation, while \texttt{top-p} dynamically adjusts the size of the candidate token set based on cumulative probability mass. Together, they capture complementary dimensions of output stochasticity, and their joint variation induces cross-agent heterogeneity in generated text.}. Much of the existing literature overlooks these settings, simply applying default values or fixing them to integers like 0 or 1 for each LLM Agent \citep{horton2023large,cui2025large,zarifhonarvar2026generating}. This practice treats the text generation process as entirely identical across agents, thereby artificially increasing the homogeneity and rigidity of simulated outputs at the population level. To incorporate greater heterogeneity and randomness in population-level simulations, a growing body of computer science literature introduces or recommends diverse or even randomized hyperparameter settings across agents, rather than adopting a uniform fixed value \citep{anthis2025position,bui2025mixture,cecere2025monte}. 
Therefore, drawing on the concept of the disturbance terms in econometric regression models, we introduce normally distributed \footnote{{} The choice of the normal distribution rests on two considerations. First, the normal distribution is the canonical specification for unobserved disturbances in econometric models, being the maximum-entropy density given a finite mean and variance \citep{cover2006elements}; absent additional structural information on the latent factors at play, it imposes the fewest auxiliary restrictions on the shape of the disturbance and thus offers the most parsimonious description of unstructured between-agent variation \citep{greene2018econometric}. Second, the adoption of a normal disturbance is further supported on asymptotic grounds. The residual heterogeneity injected into each hyperparameter setting can be interpreted as the aggregate of a large number of small idiosyncratic factors operating within an individual agent; under standard regularity conditions, the central limit theorem implies that such aggregates are approximately Gaussian when the number of contributing factors is large \citep{hayashi2011econometrics}. The cross-sectional sample sizes used in our simulations, comparable to those of large-scale household and expert expectation surveys, are sufficient for the resulting population-level distribution of disturbances to be well described by this approximation, irrespective of the distributional shape of any individual contributing factor.} random disturbances to the two hyperparameters, \texttt{temperature} and \texttt{top-p} \footnote{{} We assign \texttt{temperature} a normal distribution with a mean of 1.0 and a standard deviation of 0.5, and \texttt{top-p} a normal distribution with a mean of 0.5 and a standard deviation of 0.25. Values falling outside the specified ranges ({[}0, 2{]} for \texttt{temperature} and (0, 1{]} for \texttt{top-p}) are winsorized to the corresponding endpoints, and both parameters are rounded to two decimal places.}. The results in Section~\ref{sec:6} demonstrate how these random disturbances contribute to characterizing heterogeneity. The results in Supplementary Appendix Section~\ref{sec:b} indicate that these disturbances do not compromise the robustness of the simulations.

\subsection{LLM Agents for Simulating Expert Expectations}\label{llm-agents-for-simulating-expert-expectations}\label{sec:3-2}

Some studies compare the heterogeneity in expectations between experts and households \citep{carroll2003macroeconomic,lamla2012role,andre2022subjective,andre2026narratives}, such as the hypothetical vignette experiment to be discussed in later sections. Therefore, it is necessary to construct LLM Agents for simulating expert expectations.

\begin{figure}[h]
\centering
\includegraphics[width=\linewidth,keepaspectratio]{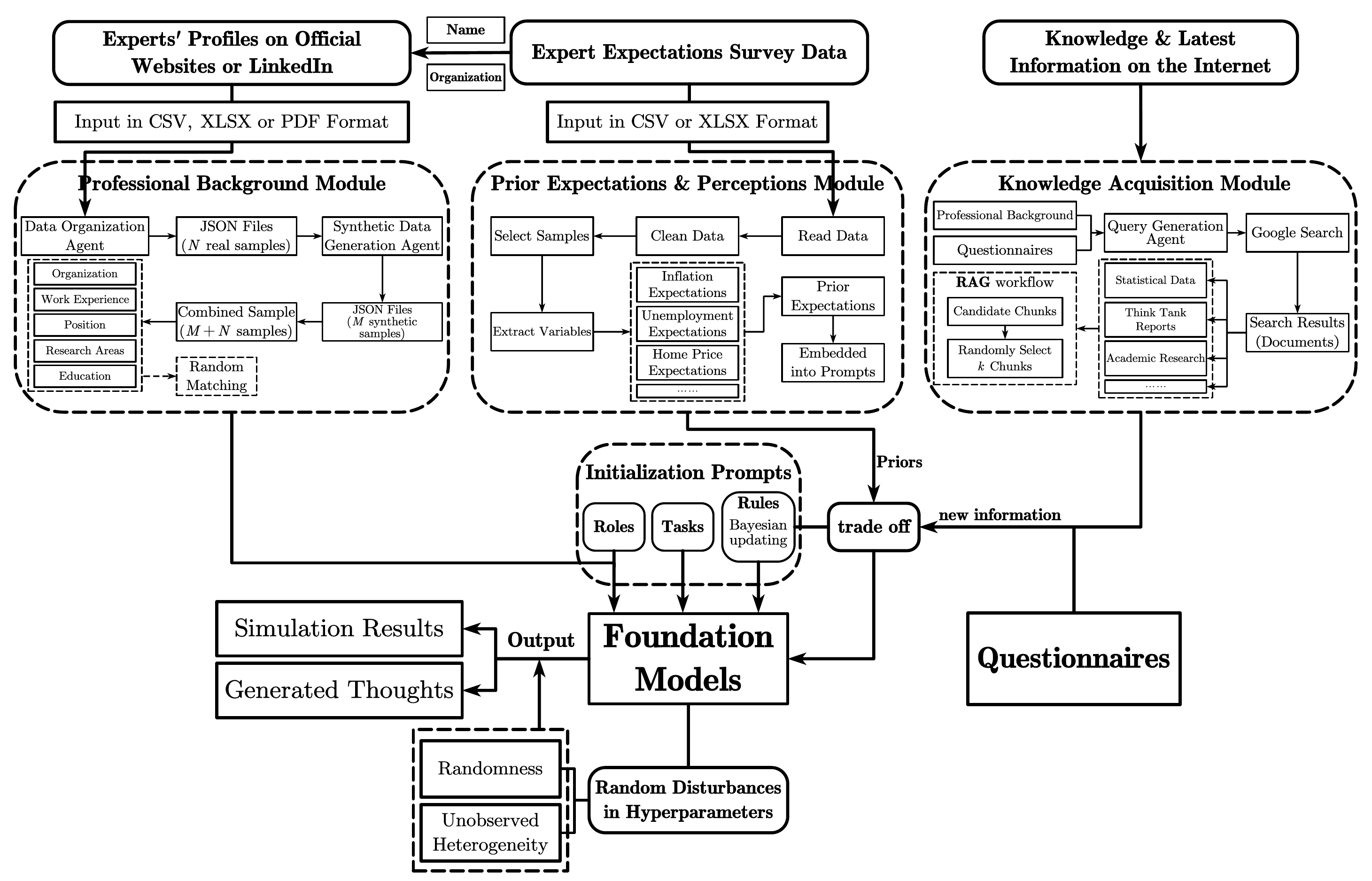}
\caption{LLM Agents for simulating expert expectations}\label{fig:3}
\notepar{%
Notes: This figure presents the detailed architecture of the LLM Agents for simulating expert expectations. The Expert Agents consist of six main components: Professional Background Module (PBM), Prior Expectations \& Perceptions Module (PEPM), Knowledge Acquisition Module (KAM), Random Disturbances (RD), initialization prompts, and foundation models. PBM utilizes actual experts' profiles from official websites or LinkedIn, and can generate synthetic data when the sample size is insufficient. PEPM derives data from expert expectation surveys. KAM retrieves and acquires relevant knowledge or the latest information from the internet on a personalized basis. These modules automatically extract and process information, with their operational rules defined in the initialization prompts. For input questionnaires, Expert Agents can engage in role-playing and perceive external environment through these components, ultimately outputting heterogeneous expectations along with the underlying thoughts.
}
\end{figure}

Research has shown that, compared to households, experts' beliefs or decisions are primarily influenced by their professional background (e.g., work experience, education, field of expertise), while the impact of demographic characteristics is relatively minor and unstable \citep{benchimol2022expert}. Furthermore, experts typically possess professional training, greater specialized knowledge, and stronger capabilities in retrieving professional information \citep{ericsson2018cambridge,gordon2013views}. Based on the above literature, we develop two new modules for the Expert Agents---PBM and KAM (see Figure~\ref{fig:3}), which correspond to the PCM and SMIM modules in the Household Agents, respectively. The design and functionality of the other components in the Expert Agents are analogous to those in the Household Agents.

The PBM utilizes collected information from experts' profiles on official websites or LinkedIn. Key information such as names and affiliated organizations is obtained from expert expectation surveys. Samples with missing or insufficient information are filtered out. The PBM then inputs the expert profile dataset into the Data Organization Agent, which processes each expert's profile into a coherent, uniformly formatted paragraph of approximately 500 words, outputting the results in JSON format. Given that survey-based expert samples are often limited, PBM employs the Synthetic Data Generation Agent to generate synthetic samples that closely resemble real expert profiles. These synthetic profiles exhibit high similarity to real ones in terms of writing style and structure, and can be merged with real samples to form a semi-synthetic dataset. This dataset includes essential expert information such as company/organization, work experience, position, research areas, and educational background. If the expert survey is anonymized, we cannot ascertain the priors corresponding to each expert. Therefore, the PBM randomly pairs expert profiles with priors to construct a semi-synthetic dataset.

The KAM automatically retrieves, crawls, and matches relevant knowledge and information from the internet. First, the Query Generation Agent generates five personalized queries for each expert based on their professional background and the target questionnaire. Subsequently, the Expert Agents collectively employ \emph{Google Search Engine} and the web search \& scraping tool \emph{Tavily}\footnote{{} See URL: \href{https://www.tavily.com/.}{{https://www.tavily.com/.}}} to extract and download the top 10 most relevant search results for each query within a specified time frame, saving the full text of webpage contents as documents. These documents comprise diverse data sources, such as statistical data, financial news, think tank reports, and academic research. Finally, to ensure that Expert Agents can retrieve key information from the extensive personalized knowledge base, we implement a workflow based on Retrieval-Augmented Generation (RAG)\footnote{{} Retrieval-Augmented Generation (RAG) is a technique that enhances the outputs of LLMs by integrating information retrieval models. It retrieves relevant information from external data sources and feeds it to the LLMs, which then generate more accurate and contextually relevant responses. This method combines the strengths of both retrieval and generation, allowing for dynamic and precise text generation tailored to specific queries \citep{gao2024retrieval}.} (see Supplementary Appendix Figure~\ref{fig:a-1}), enabling it to utilize \emph{k} filtered and randomly selected chunks of the most relevant and high-quality information.

\section{Experimental Design}\label{experimental-design}\label{sec:4}

In this section, we detail the design of three representative expectation survey experiments in Step 2 of our framework, and the data used in their corresponding simulations. Our experiment adopts the designs of these experiments to ensure comparability between our simulation results and those from human experiments\footnote{{} If you are interested in the detailed questionnaires for our three experiments, please refer to the link: \href{https://drive.google.com/file/d/1o6QtEfehZJnIbZ5gX2wOjyaMPQV4n6WZ/view?usp=sharing}{{https://drive.google.com/file/d/1o6QtEfehZJnIbZ5gX2wOjyaMPQV4n6WZ/view?usp=sharing}}.}.

\subsection{Hypothetical Vignette Experiment}\label{hypothetical-vignette-experiment}\label{sec:4-1}

The design of the first experiment (denoted as ``Experiment 1'') draws on the hypothetical vignette experiment\footnote{{} Hypothetical vignette experiments are commonly used to measure subjects' beliefs in hypothetical scenarios, such as those that could occur in the future but have not yet materialized. This method allows researchers to effectively control the specific information presented to respondents, thus facilitating the simulation and pre-assessment of the potential effects of proposed policies or anticipated shocks \citep{andre2022subjective,hainmueller2015validating}.} introduced by \citet{andre2022subjective}, an approach that has been widely adopted in many studies on macroeconomic expectations \citep{binder2026central,dibiasi2025uncertainty,bruschi2025subjective}. The experiment investigates how households and experts update their inflation and unemployment expectations in response to several common macroeconomic shocks (oil price shocks, government spending shocks, monetary policy shocks, and income tax shocks) through a series of sub-experiments, offering strong extensibility and generalizability.

\begin{figure}[!htbp]
\centering
\includegraphics[width=0.8\linewidth,keepaspectratio]{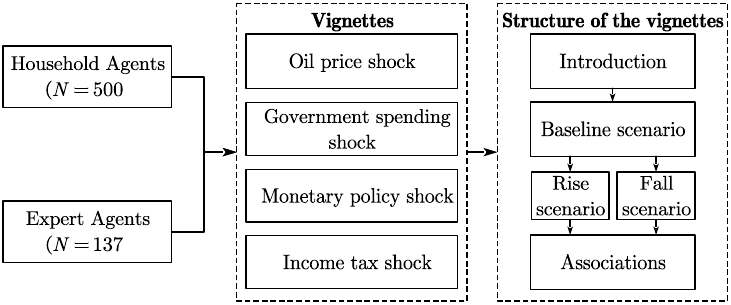}
\caption{Overview of the experimental procedure and structure of the hypothetical vignette experiment}\label{fig:4}
\notepar{%
Notes: This figure illustrates the framework of the experimental procedure and structure of the hypothetical vignette experiment. On the left panel, it presents the two types of agents participating in the experiment along with their respective sample sizes. The middle panel displays the four vignettes corresponding to different macroeconomic shocks. On the right panel, the figure outlines the specific structure of each vignette.
}
\end{figure}

We adopt and integrate the designs from Wave 1 through Wave 3 of the survey experiment by \citet{andre2022subjective}, which enables the simulation of all outcomes within a single wave. For each shock, we design corresponding hypothetical vignettes, with the core content of the questionnaire closely aligned with that of \citet{andre2022subjective}. The detailed experimental procedure and the survey structure are presented in Figure~\ref{fig:4}.

First, we describe the survey data used for simulations with Household Agents and Expert Agents, respectively. For Household Agents, the survey data inputs for PCM and PEPM are drawn from the 2019 MSC. After data cleaning and stratified sampling, a representative sample of 500 households is obtained\footnote{{} {The surveys of Wave 1 and Wave 2 in \citet{andre2022subjective} were both conducted in 2019, while Wave 3 was carried out during the COVID-19 pandemic (early 2021) and may have been subject to uncontrollable factors. Although \citet{andre2022subjective} consider this issue in their design and attempts to mitigate the impact of the pandemic, to avoid added complexity, we set the temporal context of this experiment in 2019. Therefore, all data for the modules used in this experimental simulation are sourced from 2019, contemporary with \citet{andre2022subjective}, to ensure that our developed LLM Agents accurately recapitulate the respondents' overall state during the original experiment---that is, their personal characteristics, priors, and the social media information they were exposed to at the time. Additionally, the purpose of the stratified sampling is to obtain a sample closely aligned with the demographic proportions of the 2019 American Community Survey (ACS), ensuring broad representativeness. The survey data from \citet{andre2022subjective} also maintains demographic alignment with the ACS.}}. The two variables input into PEPM are categorical measures (e.g., increase, decrease, or remain unchanged) related to inflation (price) expectations and unemployment expectations\footnote{{} Since the MSC data on unemployment expectations only provides categorical variables (direction of change) rather than continuous variables (point forecasts), all expectation variables in the PEPM for both Household and Expert Agents are standardized as categorical variables in this experiment. This ensures uniformity in input variable types and comparability of simulation results.}. For the Expert Agents, the input survey data for the PEPM are obtained from the 2019 Survey of Professional Forecasters (SPF). After data cleaning and sample selection for the specified year, 137 expert forecasts on the personal consumption expenditures price index and unemployment are retained. Although these forecasts are collected anonymously, the acknowledgments section of the quarterly SPF reports lists the names and affiliations of most participating experts. We therefore manually collect profiles of these experts from official websites or LinkedIn, compiling a dataset of 47 real samples. This dataset is input into the PBM to generate a semi-synthetic dataset (comprising 90 synthetic samples), which is randomly matched with the priors\footnote{{} We do not use the original data publicly released by \citet{andre2022subjective} in our simulations for two main reasons: (1) the published dataset lacks respondents' prior expectations and provides only limited personal characteristics; (2) the expert survey is fully anonymous and contains limited information, which prevents the construction of an expert profile dataset. Therefore, we employ the widely recognized and representative MSC and SPF datasets, which offer diverse informational dimensions and clear variable documentation, thereby facilitating data cleaning and analysis.}.

Then, we instruct the LLM Agents to respond to both the rise and fall scenarios within each hypothetical vignette\footnote{{} LLM Agents participate in and respond to each scenario, as opposed to being randomly assigned to different scenarios like human respondents in \citet{andre2022subjective}. This design is primarily motivated by two reasons: (1) Requiring human respondents to complete multiple scenarios at once may degrade response quality through fatigue and thus compromise experimental outcomes---an issue not present with LLM Agents. (2) Human participants retain memory of previous experiments, meaning that the order of scenarios and exposure to varying information across scenarios may introduce interference. In contrast, each API call to an LLM is independent, ensuring that the samples simulated by LLM Agents across scenarios strictly satisfy the assumption of independent and identically distributed (i.i.d.) data, free from interference caused by memory retention. These advantages of LLM Agents help control for the influence of extraneous factors, such as demographic characteristics, across different experimental scenarios.}. Following the approach of \citet{andre2022subjective}, each vignette adopts the same structure and begins with a brief introduction to familiarize respondents with the vignette's context. For example, in the oil price vignette, respondents are informed about the average price of crude oil per barrel in the past week. They then proceed to the baseline scenario, where the core variable (e.g., oil price) is assumed to remain unchanged. Under this scenario, we collect respondents' expectations regarding the unemployment rate in 12 months and the inflation rate over the next 12 months. Next, respondents are prompted to predict the unemployment rate and inflation rate under a scenario where an exogenous economic shock is introduced. Specifically, they are assigned to a rise scenario in which the shock variable increases (e.g., the average oil price rises by \$30) and a fall scenario in which the shock variable decreases (e.g., the average oil price falls by \$30). To simplify the analysis, \citet{andre2022subjective} reverse the sign of all predictions in the fall scenarios and merges them with the data from the rise scenarios. The main outcome variable is respondents' perception about the effect of a shock, measured as the difference between their predictions under the shock scenario and those under the baseline scenario.

Finally, we ask LLM Agents about their associations when making their predictions through structured and open-ended questions, thereby allowing us to directly measure their thought processes.

\subsection{Randomized Information Experiment}\label{randomized-information-experiment}\label{sec:4-2}

The design of the second experiment (denoted as ``Experiment 2'') draws on the randomized information experiment introduced by \citet{chopra2025home}. Unlike the first experiment, their approach directly presents subjects with information, thereby eliminating the need for constructing elaborate hypothetical scenarios. This type of experiment is commonly adopted in related studies and is also considered generalizable \citep{haaland2023designing,armona2019home,armantier2016price}. The experiment in this paper consists of two sub-experiments that investigate, respectively, how different types of home price forecasts influence the long-term home price expectations of homeowners and renters, and how an increase in expected home price growth affects their economic outlook. For our simulation, we directly use the survey data on homeowners and renters in 2024 provided by \citet{chopra2025home} for calibration, which includes detailed individual-level information such as respondents' priors (e.g., home price expectations and housing transactions intentions), confidence in those priors, and homeownership status. For the following two sub-experiments, we use the architecture of Household Agents to simulate homeowners (Homeowner Agents) and renters (Renter Agents), respectively.

In the first sub-experiment, a random half of respondents are assigned to the high-forecast group and receive a 10-year average annual home-price growth forecast of 6\%, while the remainder are assigned to the low-forecast group and receive a 2\% forecast. To quantify post-treatment differences in expectations across groups, we elicit each respondent's subjective probability distribution for the average annual growth rate of a representative U.S. home over the next ten years. Respondents assign probabilities to mutually exclusive and collectively exhaustive bins representing ranges of future home price growth. For each respondent, we then calculate the implied mean of its distribution using the bins' midpoints. This approach of eliciting agents' expectations through distribution forecasting serves as a complement to the point forecasting method used in the first experiment.

In the second sub-experiment, we focus on respondents' main considerations when confronted with changes in the long-term home price growth rate. To measure these considerations, respondents receive information prompting them to imagine that they revise upward their expectations on home price growth. They are then asked to indicate how this change in home price expectations would affect their own economic situation: improving, remaining unchanged, or worsening. Additionally, open-ended questions are used to collect explanations for their responses, allowing us to examine the mechanisms underlying expectation formation.

\subsection{Large-Scale Expectations Survey}\label{large-scale-expectations-survey}\label{sec:4-3}

In the first two experiments, we simulate expectation distributions for contemporaneous samples, rather than simulating macroeconomic expectations for future periods. To extend our study, we design the third experiment (denoted as ``Experiment 3'') to evaluate the post-knowledge-cutoff simulation capability of LLM Agents. Unlike the previous survey experiments, a large-scale household expectations survey (MSC) typically features broader temporal coverage, higher frequency, and more extensive scope, making it one of the most representative and comprehensive approaches for studying expectation dynamics. In this experiment, LLM Agents are employed to simulate MSC expectations data and the underlying thought processes for January 2025 and beyond\footnote{{} We select the period starting from January 2025 as the post-knowledge-cutoff test window because the knowledge cutoff dates of the advanced foundation models examined in this study mostly fall before January 2025 (see Supplementary Appendix Table~\ref{tab:a-1}). Therefore, using 2024 data to simulate the distributions of expectations in the MSC from January 2025 onward constitutes a rigorous test of post-knowledge-cutoff performance.}. This experimental design is difficult to achieve with traditional methods, whereas our framework accomplishes it efficiently.

Specifically, we focus on evaluating the ability of LLM Agents to simulate the distributions of households' short-term (one-year) and long-term (five-year) inflation and home price expectations. The input data used for calibration in the PCM and PEPM are drawn from a stratified sample of the 2024 MSC (sample size is 3,000, with demographic characteristics aligned with the full 2024 sample). Simultaneously, the SMIM automatically collects and processes hot-topic tweets related to ``US Inflation'' and ``US home price'' from platform X in 2024. The LLM Agents are tasked with responding to questions in the 2025 MSC survey regarding both short- and long-term inflation and home price expectations, providing explanations for their answers via open-ended questions. The simulated inflation and home price expectations will then be compared against human responses from the 2025 MSC (sample size is also 3,000, with demographic characteristics aligned with the full 2025 sample) to assess post-knowledge-cutoff simulation performance.

\section{Simulation Results and Analysis}\label{simulation-results-and-analysis}\label{sec:5}

In this section, we perform Step 3 of our framework. Specifically, we compare the similarity in shapes between distributions of expectations simulated by LLM Agents and those formed by human subjects, in order to evaluate simulation fidelity and post-knowledge-cutoff performance, respectively. Furthermore, we analyze the simulation results of LLM Agents regarding open-ended responses to examine whether they can recapitulate key patterns in human selective recall and mental models.

\subsection{Simulation Results for the Expectation Distributions}\label{simulation-results}\label{sec:5-1}

To compare the shape similarity between the distributions generated by LLM Agents and those produced by humans, we discretize the probability distributions of both sets of expectations data into probability vectors by constructing histograms\footnote{{} The number of bins for the histograms corresponding to the two datasets is determined according to the following rules: (1) In general, the Freedman--Diaconis rule is applied by default to automatically determine the bin count. (2) When the sample sizes of both datasets are large (substantially exceeding the bin count derived from the Freedman--Diaconis rule), the number of bins is set to approximately equal to or slightly exceed the sample size, so as to identify differences in the distribution shapes at a finer granularity. This approach facilitates automatic selection of an appropriate bin count across varying sample sizes, thereby mitigating subjectivity in bin number specification.}. These two vectors share the same dimensionality, with each element representing the distribution probability of the corresponding group's data within a specific numerical interval, thereby forming discrete approximations of the original continuous distributions. Subsequently, we compute both the Pearson correlation and cosine similarity between these two vectors as metrics to assess the shape similarity between the two distributions\footnote{{} We adopt the similarity of distributional shapes rather than standard statistical tests, such as the Kolmogorov-Smirnov (K-S) test or Wasserstein distance-based tests, for three reasons. First, our simulation aims to assess whether LLM Agents can recapitulate the heterogeneity in the human expectation distribution at the population level. Specifically, we evaluate how well the simulation results fit the geometric shape of the human distributions, rather than whether their generated samples are statistically indistinguishable from human data. Second, the $p$-values from K-S and Wasserstein distance-based tests are highly sensitive to sample size. In small samples, these tests often lack sufficient statistical power to detect economically meaningful distributional deviations; in large samples, they tend to over-reject null hypotheses due to minor, economically negligible differences. Because sample sizes vary substantially across our three experiments, these statistical metrics do not provide a reliable basis for cross-experiment comparisons. Finally, the similarity calculated from histogram-based probability vectors is bounded, dimensionless, and scale-invariant, which allows us to directly compare heterogeneity outcomes measured in different units.}.

As the ``brain" of LLM Agents, foundation models are central to simulation performance. Selecting appropriate foundation models for our framework is therefore essential before running simulations. Since foundation models are diverse and rapidly iterating, exhaustively testing all of them is infeasible. A common practice in the existing literature is thus to pre-test several mainstream models and select a suitable one for the primary analysis \citep{horton2023large,wu2025llm,zarifhonarvar2026generating,kazinnik2026bank}. Following this approach, we compare the simulation performance of LLM Agents across various mainstream foundation models. Based on this comparison, we propose the following model selection guidelines for our framework: (1) High distributional similarity is a necessary condition for good simulation performance; therefore, models yielding higher distributional similarity should receive priority. (2) Reasoning models achieve better distributional fit overall than non-reasoning models, and are thus more suitable for both types of LLM Agents in this paper. (3) When models perform comparably, open-source models are preferred to lower costs, mitigate the risk of hidden data leakage, and enhance reproducibility. (4) Given that experimental conditions vary, preliminary testing is recommended to compare model performance prior to final selection\footnote{{} No existing literature claims that its framework-based conclusions apply to all models, as doing so is unrealistic. Therefore, our objective is not to evaluate the universality of our framework across an exhaustive set of models. Instead, we identify the models suitable for our framework and synthesize this process into a preliminary guide for model selection. We first compare the simulation results of LLM Agents across multiple vignettes. These agents are built on six foundation models released by leading developers between March and July 2025. As shown in Figures~\ref{fig:a-2} and \ref{fig:a-3} of the Supplementary Appendix, reasoning models (Qwen3-235B-A22B-Thinking-2507, DeepSeek-R1-0528, GPT-o4-mini, and Gemini-2.5-Pro) outperform non-reasoning models (DeepSeek-V3-0324 and GPT-4.1-mini) on average in distribution fitting for both Household and Expert Agents.
Second, the pre-test results across Figures~\ref{fig:a-2} to \ref{fig:a-5} collectively indicate that the simulation performance varies minimally among these reasoning models in most cases. Consequently, we consider three additional criteria: (i) Data leakage risk: \citet{ludwig2026large} argue that open-source models mitigate the risk of hidden data leakage because researchers can verify that experimental results are not artifacts caused by prior exposure in proprietary training data. (ii) Reproducibility: open-source models offer stronger reproducibility, as their weights, tokenizers, and inference parameters are fully documented and accessible over the long term, whereas the outputs of proprietary models often drift silently as providers push updates \citep{spirling2023open,chen2024chatgpt}. (iii) Cost: open-source models generally have lower per-token prices, making them more cost-effective when simulation performance is comparable. Both Qwen3-Thinking-2507 and DeepSeek-R1-0528 are open-source reasoning models that satisfy these criteria. However, the former produces distributions that are on average closer to the human ones, while parameters such as \texttt{seed} are not supported in API calls to the latter (see the DeepSeek official documentation: \url{https://api-docs.deepseek.com/guides/reasoning\_model}), which makes the reproducibility of its simulation results difficult to control. We therefore select Qwen3-Thinking-2507 as the foundation model for our framework.}. Based on these guidelines and the model comparisons in Supplementary Appendix Figure~\ref{fig:a-2}--Figure~\ref{fig:a-5}, we select an open-source reasoning model from the Qwen series as the foundation model for our LLM Agents and report its simulation results\footnote{{} The key patterns in our results can also be captured by other models in this series. Due to space constraints, we only report the simulation results based on Qwen3-Thinking-2507; results from the remaining models are available upon request.}.

\begingroup
\captionsetup{type=figure}
\centering
\subcaptionbox{Simulation performance of Household Agents in Experiment 1\label{fig:sub-5-a}}{
  \includegraphics[width=0.9\linewidth,keepaspectratio]{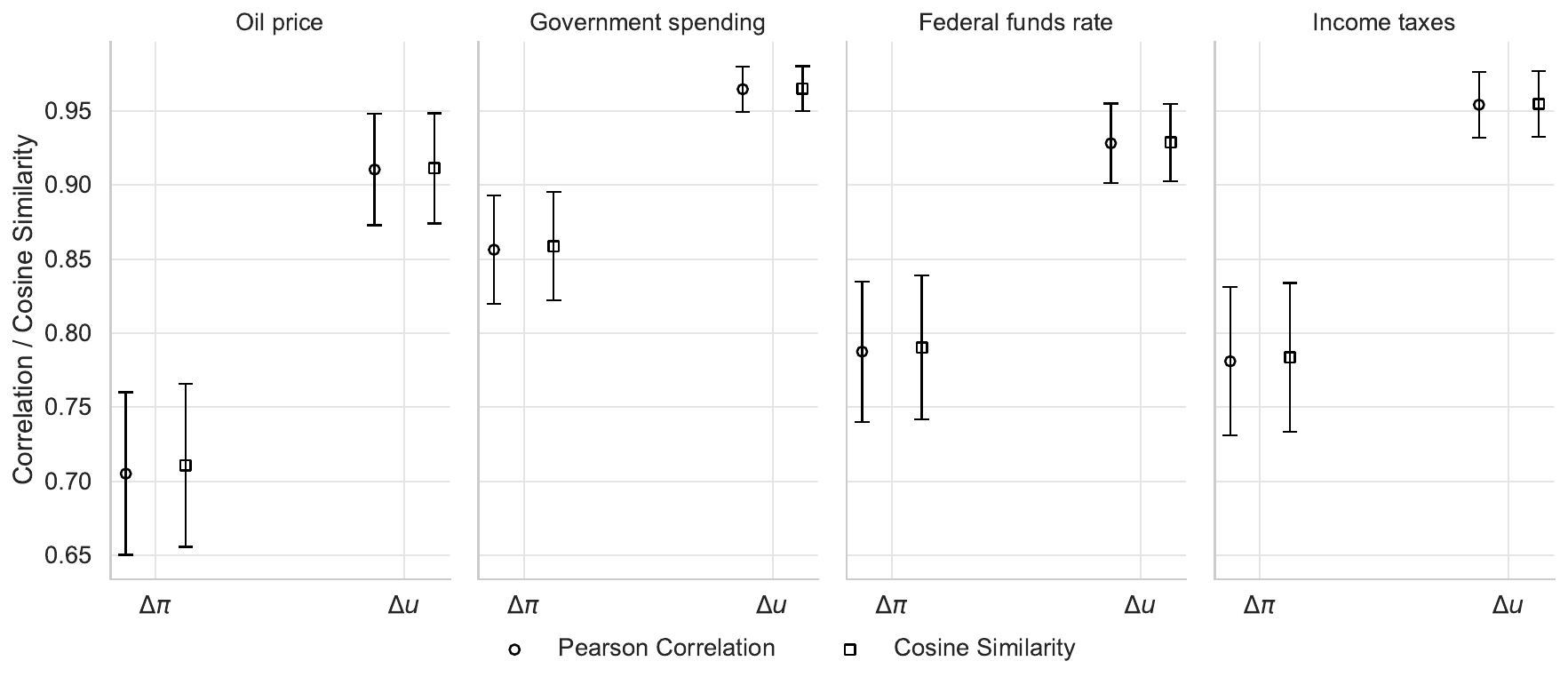}
}\par
\subcaptionbox{Simulation performance of Expert Agents in Experiment 1\label{fig:sub-5-b}}{
  \includegraphics[width=0.9\linewidth,keepaspectratio]{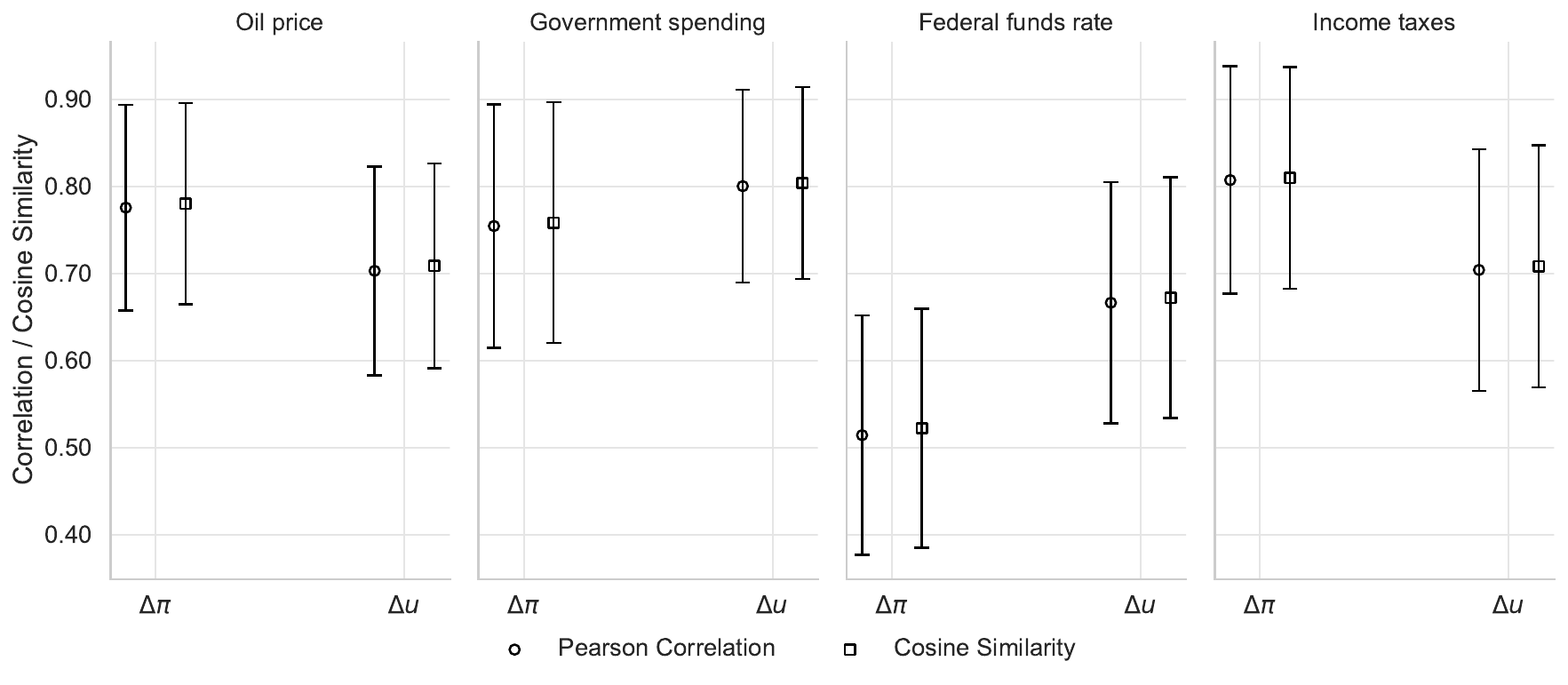}
}\par
\subcaptionbox{Simulation performance of LLM Agents in Sub-Experiment 1 of Experiment 2\label{fig:sub-5-c}}{
  \includegraphics[width=0.9\linewidth,keepaspectratio]{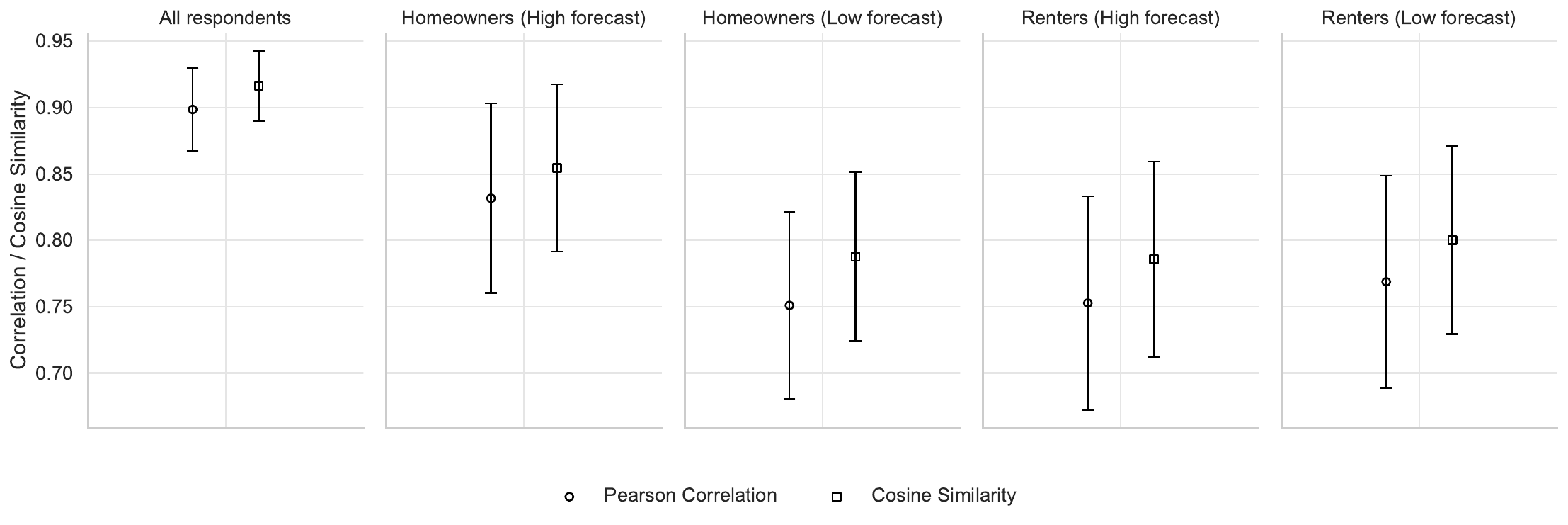}
}\par
\subcaptionbox{Post-knowledge-cutoff performance of Household Agents in Experiment 3\label{fig:sub-5-d}}{
  \includegraphics[width=0.9\linewidth,keepaspectratio]{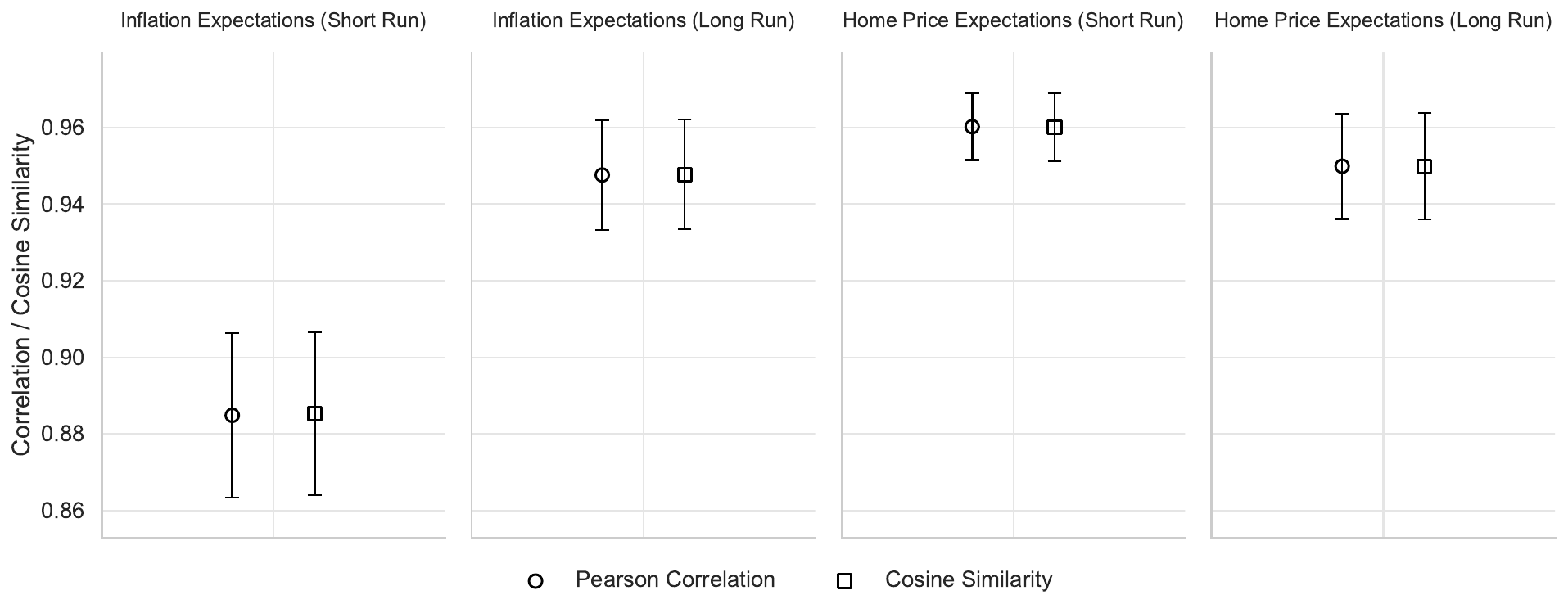}
}\par
\addtocounter{figure}{-1}%
\captionof{figure}{Shape similarity between the expectation distributions generated by LLM Agents and those generated by humans in three representative experiments}
\label{fig:5}
\notepar{%
\linespread{0.60}\selectfont
Notes: Panel (a) and Panel (b) display the distributional shape similarity, as measured by Pearson correlation and cosine similarity, between the changes in inflation expectations (\(\Delta\ \pi\)) and unemployment expectations (\(\Delta\ u\)) generated by Household Agents (Panel (a)) and Expert Agents (Panel (b)) and those of humans under four different vignettes. Panel (c) presents simulation performance from Sub-Experiment 1 of the randomized information experiment: the Homeowner and Renter Agents' simulated home price expectations for homeowners and renters in the high-forecast and low-forecast treatment groups, and the LLM Agents' simulated home price expectations for all respondents. Panel (d) displays post-knowledge-cutoff performance of Household Agents for long- and short-term inflation expectations and home price expectations of respondents in the 2025 Michigan Surveys of Consumers. Error bars present two-sided 95\% confidence intervals for the similarity metrics, obtained by bootstrap over histogram-based probability vectors.
}
\endgroup

The results in Figure~\ref{fig:5} demonstrate that, across three representative experiments, our LLM Agents consistently achieve strong performance in simulating the distributions of various macroeconomic expectations across different types of agents under varying scenarios\footnote{{} A potential concern is that the strong performance of the LLM Agents in this paper may simply result from the LLMs recalling or restating outcomes from existing survey experiments based on their extensive training data. However, this concern is unfounded for three reasons: (1) The survey data used in both the randomized information experiment and the large-scale expectations survey were officially released online only after January 2025---i.e., after the knowledge cutoff of all foundation models used in this paper---making it impossible for such data to have been included in their training. (2) Even if the data from the hypothetical vignette experiment were published before the models' knowledge cutoff, general-purpose foundation models are unlikely to have directly used individual-level survey data during training. This is due to the typical use of processed, unstructured text data in LLM training, as opposed to raw structured survey data, as well as privacy protection policies adopted by some developers \citep{zhao2023survey,yang2025qwen}. Unless specifically fine-tuned for such purposes, these models do not incorporate personally identifiable survey records. This also explains why many existing studies directly employ foundation models to replicate classic human experiments without considering this issue \citep{chen2023emergence,horton2023large,cui2025large}. (3) The results in Section~\ref{sec:6} indicate that even when incorporating all information extracted from modules, the LLM Agents still fail to generate effective simulation outcomes without explicit definitions of their roles, task objectives, and Bayesian updating rules for module invocation. This finding further demonstrates that their strong performance does not stem from the memorization of foundation models triggered by rich contextual inputs, but rather reflects deeper underlying mechanisms of comprehension and reasoning.}. The shape similarity (whether Pearson correlation or cosine similarity) between the simulated distributions and those generated by humans averages around 0.8 in most cases. 

The shape similarity between distributions generated by the naive persona approach and the true distributions is mostly below 0.4, as shown by the ``Naive Persona" results in Section~\ref{sec:6}. This indicates that our LLM Agents provide a better fit to the true distributions than this commonly used approach. Supplementary Appendix Figures~\ref{fig:a-7} to \ref{fig:a-9} plot the LLM-generated and human expectation distributions across all experiments, further supporting this finding. Although the simulated distributions are slightly more concentrated than the human distributions, they do not exhibit the severely distorted homogenization seen in distributions from the naive persona approach. Instead, they accurately capture the shape characteristics of expectation distributions within and across different agent groups; for instance, across all vignettes, the expectation distributions of households are more dispersed, while those of experts are more concentrated.

Moreover, despite some quantitative gaps from the true distributions, the expectation distributions generated by LLM Agents qualitatively reflect the key heterogeneity across agent types. Supplementary Appendix Figure~\ref{fig:a-6} shows that, similar to human households and experts, the directions of expectation changes produced by Expert Agents align more closely with textbook theoretical predictions, whereas Household Agents generate more counter-theoretical responses and exhibit more diverse directions of change. For example, most experts believe that an increase in government spending raises inflation and lowers unemployment, while only about half of households share this view. Similarly, Supplementary Appendix Figure~\ref{fig:6} demonstrates that LLM Agents capture the key qualitative differences between homeowners and renters: when anticipating future house price increases, most renters perceive that their household's future economic situation would worsen, whereas most homeowners believe it would improve or remain unchanged.

\subsection{Analysis of Open-Ended Responses}\label{analysis-of-open-ended-responses}\label{sec:5-2}

While achieving a distributional fit is a necessary condition for simulation, it is insufficient to demonstrate that LLM Agents capture the key heterogeneity in expectations across different agent types. Therefore, in this subsection, we analyze the open-ended explanations provided by LLM Agents for their expectations, to examine whether these agents recapitulate key patterns in the thoughts underlying human expectation formation.

\subsubsection{Selective Recall in LLM Agents vs. Humans}\label{selective-recall-of-llm-agents}\label{sec:5-2-1}

First, we evaluate the similarities and differences in selective recall between LLM Agents and their human counterparts. Some research indicates that selective recall plays a crucial role in shaping human cognition and behavior \citep{tversky1973availability,bordalo2016stereotypes,bordalo2025imagining}. When forming heterogeneous expectations under varying conditions, economic agents tend to selectively retrieve different types of relevant information from memory (such as news, knowledge, and experiences) \citep{andre2022subjective}. This motivates us to analyze responses to open-ended questions, investigating whether LLM Agents can simulate the pattern of selective recall.

For the hypothetical vignette experiment, we first follow the approach of \citet{andre2022subjective} to focus on and quantify the proportions of words related to four distinct channels (topics)\footnote{{} Specifically, the four channels are defined as follows: Cost words include the word (stem) ``cost''. Demand words include the words (stems) ``demand'', ``buy'', ``purchas'', ``invest'', ``spend'', ``consum''. Labor words include the words (stems) ``layoff'', ``lay-off'', ``lay off'', ``fire'', ``hire'', ``labor'', ``work'', ``job''. Central bank words (phrases) include ``monetary policy'', ``federal funds rate'', ``fed funds rate'', ``federal funds target rate''.} mentioned by LLM Agents in their open-ended responses when generating expectations under each vignette.

\begin{figure}[!htbp]
\centering
\includegraphics[width=\linewidth,keepaspectratio]{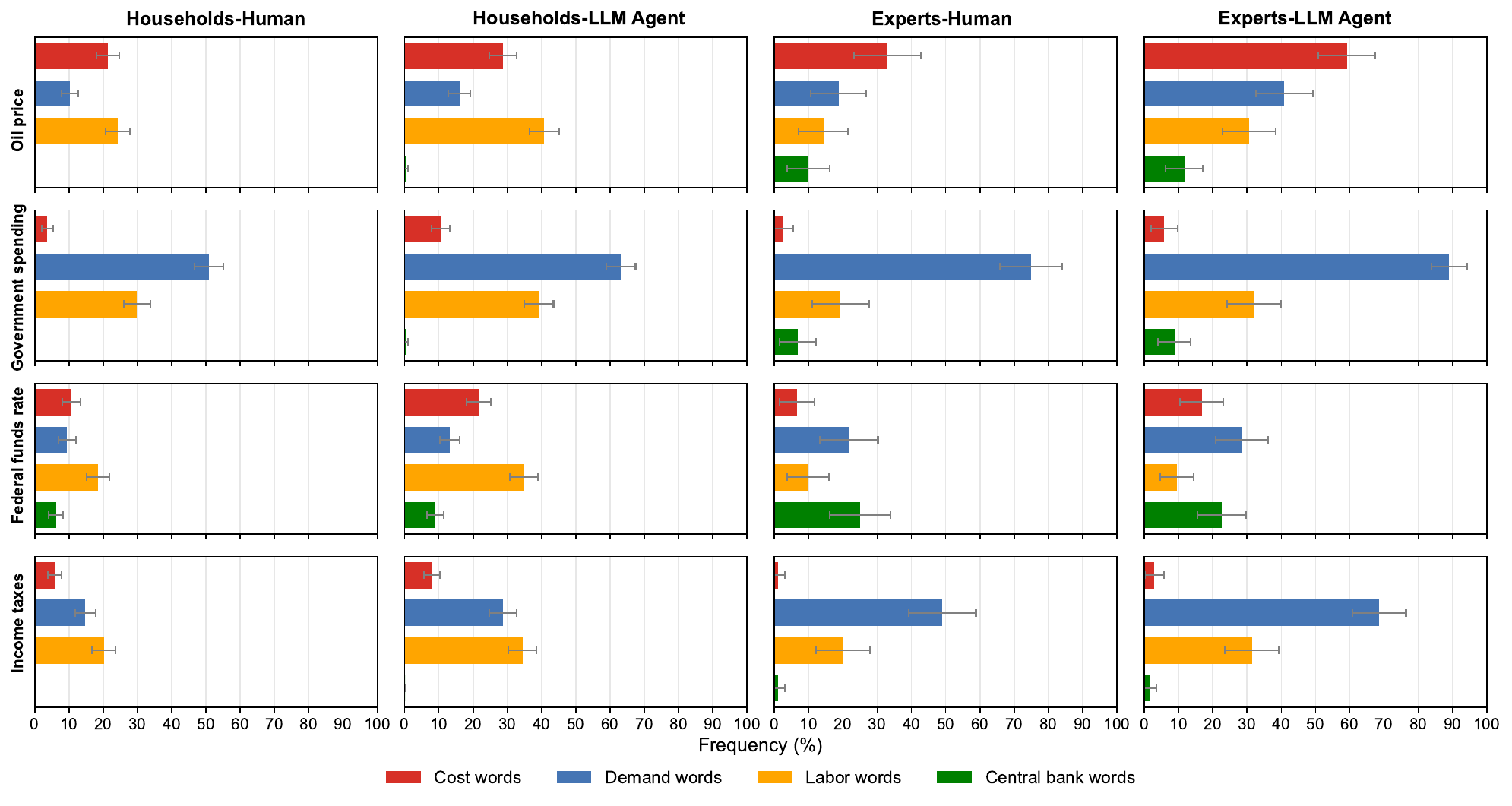}
\caption{Word usage for open-ended responses of humans and LLM Agents in Experiment 1}\label{fig:7}
\notepar{%
Notes: This figure presents the proportions of human households (i.e., Households-Human in Column 1), Household Agents (i.e., Households-LLM Agent in Column 2), Human Experts (i.e., Experts-Human in Column 3), and Expert Agents (i.e., Experts-LLM Agent in Column 4) mentioning words from four word groups in their open-ended responses under four different vignettes. The error bars indicate 95\% confidence intervals.
}
\end{figure}

As shown in Figure~\ref{fig:7}, both Household Agents and Expert Agents are able to capture the key heterogeneity of thoughts within and between human households and experts: experts tend to concentrate their reasoning within each vignette on channels that are recognized by the mainstream literature or textbooks as playing a central role in real-world shocks, whereas households often overlook mechanisms that may be dominant in reality. For example, across all four vignettes, whether facing supply or demand shocks, a considerable number of households refer to cost-related, particularly labor-related, supply-side channels. In contrast, for experts, cost-related supply-side mechanisms predominate in the case of an oil price shock (a supply shock), whereas demand-side channels dominate in the latter three vignettes, which involve demand shocks. Moreover, experts make more frequent references to central banks (Federal Reserve), further illustrating the professional nature of their recall content.

Further comparison reveals that while LLM Agents can qualitatively simulate the various channels mentioned by humans in forming expectations, there are quantitative differences: specifically, LLM Agents recall these types of channels at a slightly higher frequency than humans, indicating greater homogeneity in the content recalled by LLM Agents. These patterns are also echoed in the responses of LLM Agents to structured questions, as shown in Supplementary Appendix Figure~\ref{fig:a-10}.

Second, we need to focus on the overall semantics of the open-ended responses from LLM Agents rather than merely their lexical features. Specifically, we examine what kinds of content their selective recall draws upon when providing explanations (for instance, whether it is mere conjecture or well-reasoned analysis), as well as the similarities and differences relative to the recall of their human counterparts. Following the coding scheme defined by \citet{andre2022subjective}, we design and implement an agentic workflow (see Supplementary Appendix Figure~\ref{fig:a-11}) that leverages two different types of LLMs to simulate the process of two human annotators independently labeling responses and reaching consensus through multiple rounds of discussion. This procedure categorizes open-ended responses from LLM Agents into nine distinct categories\footnote{{} We adopt the following categories as defined by \citet{andre2022subjective}: i) ``Mechanism'' encompasses all responses addressing how shocks transmit through economic channels; ii) ``Model'' covers statements invoking a particular economic framework or theory; iii) ``Guess'' flags any expressions of uncertainty or admissions that the forecast is speculative; iv) ``Politics'' gathers broad political or normative commentary; v) ``Historical'' captures references to past developments or typical evolutionary patterns; vi) ``Misunderstanding'' marks instances where respondents misinterpret aspects of the scenario; vii) ``Restates prediction'' identifies replies that merely reiterate or paraphrase the provided inflation and unemployment forecasts; viii) ``Endogenous shock'' refers to understanding an exogenous shock as an endogenous response, such as mentioning that interest-rate adjustments are responses by the Fed to other economic changes; and ix) ``Other'' serves as a residual category. Each response is allowed to fall into more than one category.}. These results are independently verified by two graduate students in economics, who ultimately reach a consensus on any necessary corrections.

\begin{figure}[h]
\centering
\includegraphics[width=\linewidth,keepaspectratio]{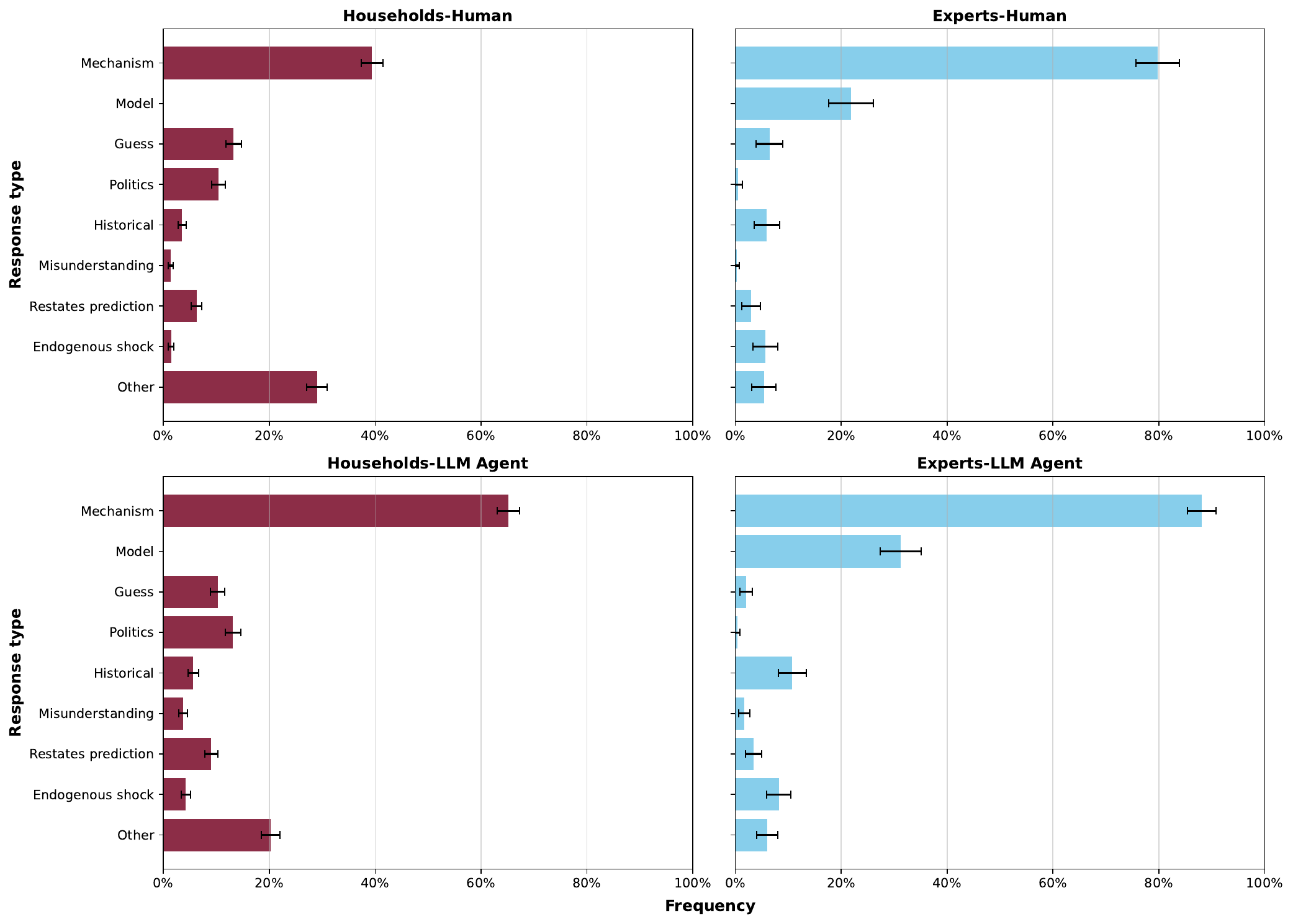}
\caption{Response types in open-ended responses of humans and LLM Agents in Experiment 1}\label{fig:8}
\notepar{%
Notes: This figure presents ``response type'' classification of open-ended responses generated by human households (i.e., Households-Human), human experts (i.e., Experts-Human), Household Agents (i.e., Households-LLM Agent) and Expert Agents (i.e., Experts-LLM Agent), averaged across all four vignettes. The human data annotations are directly obtained from \citet{andre2022subjective}, while the open-ended responses from LLM Agents are automatically classified by an agentic workflow and manually verified. Error bars display 95\% confidence intervals.
}
\end{figure}

As shown in Figure~\ref{fig:8}, LLM Agents qualitatively recapitulate the main differences between households and experts in selectively retrieving various types of content from memory when forming expectations: when making predictions, households tend to rely more on guesses and are more susceptible to politics. Their reasoning may be simpler, often merely restating predictions, and is more diverse, falling largely into the ``Other'' category. In contrast, experts more frequently recall and refer to ``Mechanism'' and ``Model,'' and are more inclined to cite ``Historical'' content. This also explains why changes in experts' expectations are more concentrated and generally align with textbook theories. However, quantitative differences exist between the thoughts of LLM Agents and humans: responses from LLM Agents more frequently mention ``Mechanism'' and are less frequently categorized as ``Other''.

Similarly, for Sub-Experiment 2 of the randomized information experiment, we follow the coding scheme defined by \citet{chopra2025home} and use the previously constructed agentic workflow to categorize the open-ended responses of LLM Agents into nine distinct mechanisms\footnote{{} We adopt the following mechanisms as defined by \citet{chopra2025home}: i) ``Wealth effects,'' referring to changes in the value of housing currently owned by the respondent's household. ii) ``Income effects (cost of buying),'' referring to changes in the cost of buying a home. iii) ``Home price growth irrelevant,'' meaning that home price growth is irrelevant because the respondent does not plan to buy, sell, or move. iv) ``Income effects (rental prices),'' referring to changes in the rental prices of homes. v) ``Collateral effects,'' referring to changes in the ease of borrowing against home equity. vi) ``Endogenous adjustments to housing,'' referring to endogenous up-/downsizing, buying/selling, or changes in timing---for example, due to substitution effects, the investment channel, or purchase timing considerations. vii) ``Inflation,'' referring to inflation and changes in the overall price level. viii) ``Household income,'' referring to changes in the household's overall income. ix) ``Interest rates,'' referring to changes in interest rates. Specific examples for each mechanism can be found in Supplemental Appendix Table A.21 of \citet{chopra2025home}. Responses are allowed to correspond to more than one mechanism.}, with the results manually verified and corrected.

\begin{figure}[!htbp]
\centering
\includegraphics[width=\linewidth,keepaspectratio]{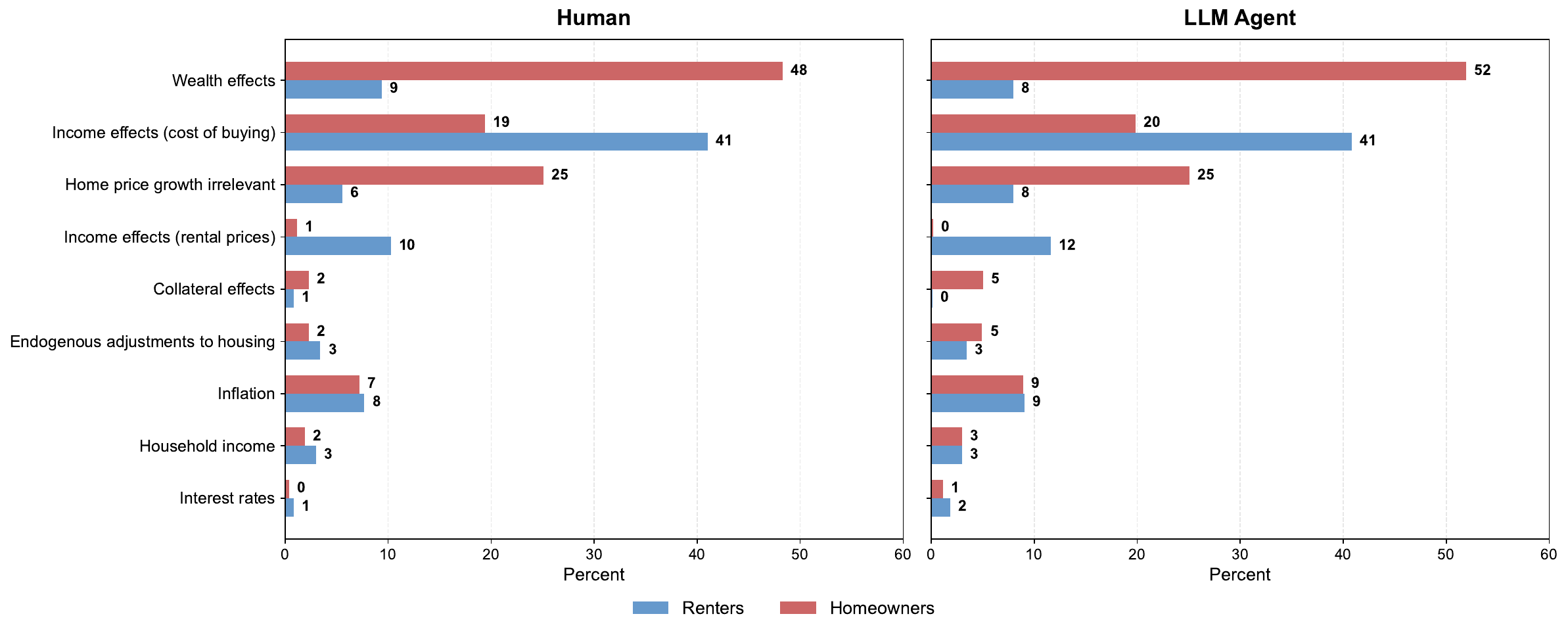}
\caption{Open-ended responses on how higher expected home price growth affects humans' and LLM Agents' expectations about economic situation in Sub-Experiment 2 of Experiment 2}\label{fig:9}
\notepar{%
Notes: The figure shows the proportion of human respondents and LLM Agents who invoke different arguments to explain why an increase in their expectations about home price growth over the next 10 years would affect their household economic outlook. The human data annotations are directly obtained from \citet{chopra2025home}, while the open-ended responses from LLM Agents are automatically classified by an agentic workflow and manually verified.
}
\end{figure}

Figure~\ref{fig:9} shows that the LLM Agents capture a key heterogeneity between the thoughts of renters and homeowners: most homeowners believe that an expected rise in home prices will increase the value of their housing via wealth effects, thereby improving their outlook on future economic situation; alternatively, they consider the home price growth irrelevant since they have no plans to buy or sell homes. In contrast, most renters believe that an expected rise in home prices will increase their future costs of buying or renting via income effects, thereby worsening their expectations about economic situation. However, similar to the earlier findings, compared to humans, the thought processes of LLM Agents more frequently refer to certain specific mechanisms, making them more homogeneous.

For the large-scale expectations survey (MSC), similar to the method by \citet{andre2022subjective} used in Experiment 1, we focus on and quantify the proportions of words related to seven distinct channels (topics)\footnote{{} Specifically, the seven channels are defined as follows: ``Cost'' channel includes the word (stem) ``cost'', ``expense'', ``fee''. ``Demand'' channel includes the word (stem) ``demand'', ``buy'', ``purchas'', ``invest'', ``spend'', ``consum''. ``Borrowing \& Lending'' channel includes the word (stem) ``loan'', ``lend'', ``borrow'', ``debt'', ``credit'', ``interest'', ``mortgage''. ``Politics'' channel includes the word (stem) ``republic'', ``democratic'', ``trump'', ``biden'', ``harris'', ``elect''. ``Policy'' channel includes the word (stem) ``government'', ``fed'', ``monetary'', ``fiscal'', ``tax'', ``tariff''. ``Energy'' channel includes the word (stem) ``oil'', ``gas'', ``fuel'', ``electricity'', ``energy''. ``Black Swan Event'' channel includes the word (stem) ``russia'', ``ukraine'', ``war'', ``invasion'', ``sanction'', ``pandemic'', ``covid'', ``lockdown'', ``crisis'', ``disaster'', ``collapse'', ``crash'', ``breaking'', ``recession'', ``bubble''.} mentioned by Household Agents in their open-ended explanations when generating inflation and home price expectations for the period beyond the knowledge cutoff.

\begin{figure}[!htbp]
\centering
\includegraphics[width=\linewidth,keepaspectratio]{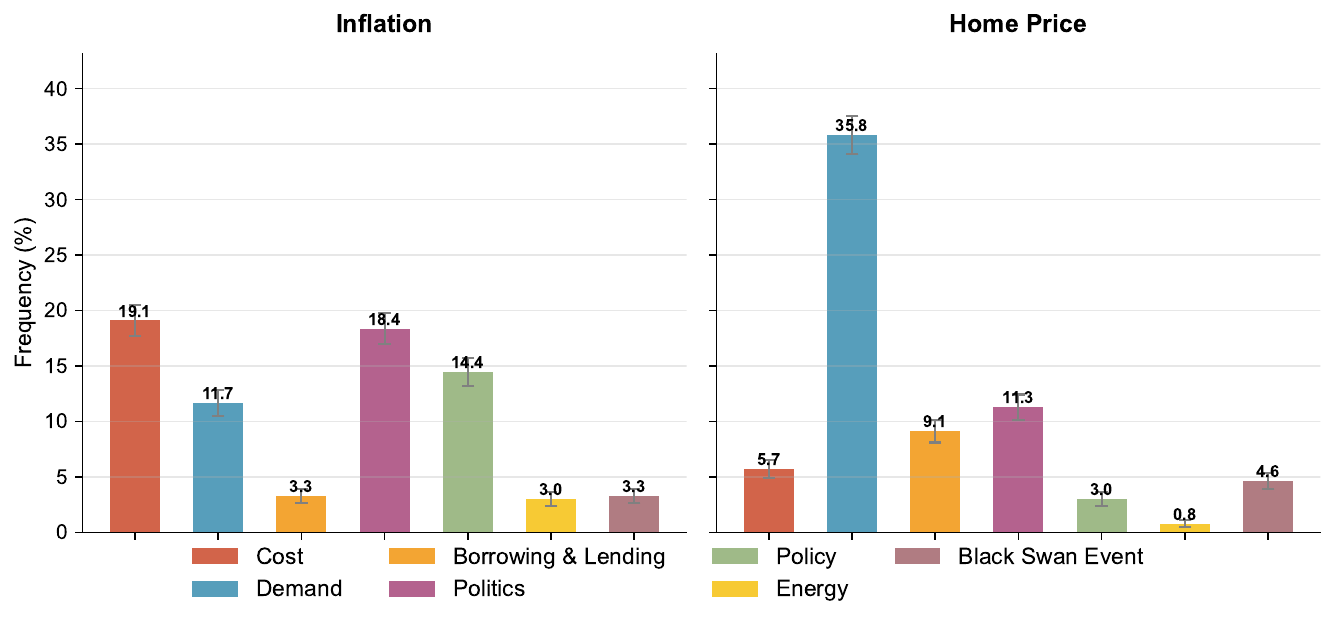}
\caption{Proportion of various channels recalled by Household Agents in Experiment 3}\label{fig:10}
\notepar{%
Notes: This figure displays the proportions of seven channels recalled by Household Agents when simulating inflation and home price expectations for 2025. The left panel presents the results for inflation expectations, while the right panel shows those for home price expectations. Error bars display 95\% confidence intervals.
}
\end{figure}

As shown in Figure~\ref{fig:10}, Household Agents primarily recall channels related to Cost, Politics, and Policy when simulating inflation expectations, whereas for home price expectations, they focus more on Demand, Politics, and Borrowing \& Lending. Although direct comparison with real households is limited due to the absence of open-ended responses in the MSC, the results in Figure~\ref{fig:10} still reflect key characteristics of selective recall among real-world households: (1) Households' inflation expectations are mainly influenced by cost-related (or supply-side) factors, leading them to recall cost-related channels more frequently \citep{dacunto2021exposure,coibion2022monetary,andre2026narratives}, while their home price expectations are driven more by demand-side factors, making them more likely to recall demand-related channels \citep{binder2026central,gohl2024house,bro2025subjective}. (2) Households tend to consider politics-related narratives when forming macroeconomic expectations, a finding consistent with Figure~\ref{fig:8}. (3) Households' inflation expectations are susceptible to government policies \citep{dacunto2024household,weber2022subjective}, as reflected in Household Agents' references to Federal Reserve monetary policy and the 2025 Trump tariffs. Meanwhile, research shows that households often factor in Borrowing \& Lending considerations, such as mortgage rates, when forming home price expectations \citep{binder2026central}.

In addition, Household Agents mention several major black swan events in recent years, such as the Russia--Ukraine war and COVID-19. This pattern aligns with the general qualitative findings from several expectation surveys of households \citep{binetti2024peoples,andre2026narratives}. The evidence in Supplementary Appendix Figure~\ref{fig:a-26} shows that the content related to black swan events in the selective recall of Household Agents without SMIM almost disappears, indicating that this phenomenon is not directly retrieved by the foundation models from their training data, but rather arises because Household Agents actively perceive external information through SMIM.

These results in this subsection indicate that the open-ended responses generated by LLM Agents capture a key feature underlying human expectation formation, namely selective recall. However, the recapitulation of these patterns is primarily qualitative; quantitative differences remain. Specifically, the content recalled by LLM Agents is more concentrated on certain channels or categories than that of humans, and these are typically the dominant channels or categories that already account for a high share in the corresponding human samples.

\subsubsection{The Mental Models of LLM Agents vs. Humans}\label{the-mental-model-of-llm-agents}\label{sec:5-2-2}

Section~\ref{sec:5-2-1} analyzes the channels mentioned in the open-ended responses of LLM Agents, or the categories to which these responses belong, but does not examine how the underlying causal reasoning pathways compare to those of humans. Therefore, this subsection investigates this issue by identifying and comparing the mental models reflected in the open-ended responses of both LLM Agents and humans. According to \citet{andre2023mental}, a \emph{Mental Model} represents an individual's beliefs about the relationships between different variables, such as the reasoning process underlying the connection between rising oil prices and expected future inflation. Following this definition and the methodology of \citet{andre2026narratives}, we extract causal DAGs\footnote{{} A causal Directed Acyclic Graph is a graphical model composed of nodes representing variables and directed edges that signify causal relationships between them. The direction of each edge reflects the flow of causality, while the acyclic structure ensures that no variable can be a cause of itself, either directly or through a sequence of causal links. Causal DAGs have become a fundamental tool for formalizing and analyzing causal inference across diverse disciplines, including statistics, computer science, and the social sciences \citep{pearl2009causality,sloman2015causality}. More recently, this framework has been extended to the analysis of narrative or mental model structures in economic theory, wherein causal reasoning and story-based explanations play a central role \citep{eliaz2020model,spiegler2016bayesian,spiegler2020behavioral}.} over the mentioned variables from the open-ended responses, using these as representations of mental models. For comparison, we include the results of foundation models with only naive personas (i.e., the naive persona approach, hereafter ``Naive Persona") as a baseline. To simplify the analysis, we take four sub-experiments from the hypothetical vignette experiment as examples.

First, we construct an agentic workflow (see Supplementary Appendix Figure~\ref{fig:a-12}) that fully automatically identifies and labels the DAGs for each open-ended response. The labeling results are reviewed and corrected by two graduate students. Supplementary Appendix Section~\ref{sec:c-1} provides a detailed explanation of how DAGs are identified.

After converting each open-ended response into a DAG, we treat all DAGs from the same group of agents (households or experts) regarding the same type of expectations (inflation or unemployment) under the same vignette as a set of mental models. We then compute the Jaccard similarity of mental model sets underlying the expectations of LLM Agents and Naive Persona to those of humans for each vignette (see Supplementary Appendix Section~\ref{sec:c-2} for the detailed calculation method). The results are presented in Table~\ref{tab:1}.

\begin{table}[!htbp]
\centering
\caption{Similarity of mental models of LLM Agents and Naive Persona to those of humans}\label{tab:1}
\small
\begin{tabular}{@{}>{}p{(\linewidth - 8\tabcolsep) * \real{0.1940}}
  >{}p{(\linewidth - 8\tabcolsep) * \real{0.1874}}
  >{}p{(\linewidth - 8\tabcolsep) * \real{0.1876}}
  >{}p{(\linewidth - 8\tabcolsep) * \real{0.1874}}
  >{}p{(\linewidth - 8\tabcolsep) * \real{0.2036}}@{}}
\toprule
\multirow{2}{*}{Vignettes} & \multicolumn{2}{>{}p{(\linewidth - 8\tabcolsep) * \real{0.3700} + 2\tabcolsep}}{%
\textbf{Panel A: Inflation (Households)}} & \multicolumn{2}{>{}p{(\linewidth - 8\tabcolsep) * \real{0.4410} + 2\tabcolsep}@{}}{%
\textbf{Panel B: Unemployment (Households)}} \\
& LLM Agent & Naive Persona & LLM Agent & Naive Persona \\
\midrule
Oil price & 0.87 & 0.52 & 0.68 & 0.43 \\
Government spending & 0.65 & 0.31 & 0.78 & 0.34 \\
Federal funds rate & 0.68 & 0.46 & 0.70 & 0.48 \\
Income taxes & 0.85 & 0.35 & 0.78 & 0.33 \\
\midrule
\multirow{2}{=}{ Vignettes} & \multicolumn{2}{>{}p{(\linewidth - 8\tabcolsep) * \real{0.3700} + 2\tabcolsep}}{%
\textbf{Panel C: Inflation (Experts)}} & \multicolumn{2}{>{}p{(\linewidth - 8\tabcolsep) * \real{0.4410} + 2\tabcolsep}@{}}{%
\textbf{Panel D: Unemployment (Experts)}} \\
& LLM Agent & Naive Persona & LLM Agent & Naive Persona \\
\midrule
Oil price & 0.73 & 0.46 & 0.77 & 0.51 \\
Government spending & 0.76 & 0.40 & 0.82 & 0.46 \\
Federal funds rate & 0.92 & 0.42 & 0.63 & 0.53 \\
Income taxes & 0.70 & 0.28 & 0.64 & 0.44 \\
\bottomrule
\end{tabular}
\notepar{%
Notes: This table presents the Jaccard similarity of the mental models underlying expectation formation of LLM Agents and foundation models with only naive personas (Naive Persona) to those of humans in each vignette. Panels A and B present the results for Household Agents and Naive Persona regarding inflation and unemployment expectations, respectively. Panels C and D present the results for Expert Agents and Naive Persona regarding inflation and unemployment expectations, respectively.
}
\end{table}

Table~\ref{tab:1} shows that across all vignettes, both Household Agents and Expert Agents exhibit mental models more closely aligned with those of humans (with a minimum similarity of 0.63). In contrast, the mental models of Naive Persona are significantly less aligned with those of humans (with a maximum similarity of only 0.53).

\begin{figure}[h]
\centering
\includegraphics[width=\linewidth,keepaspectratio]{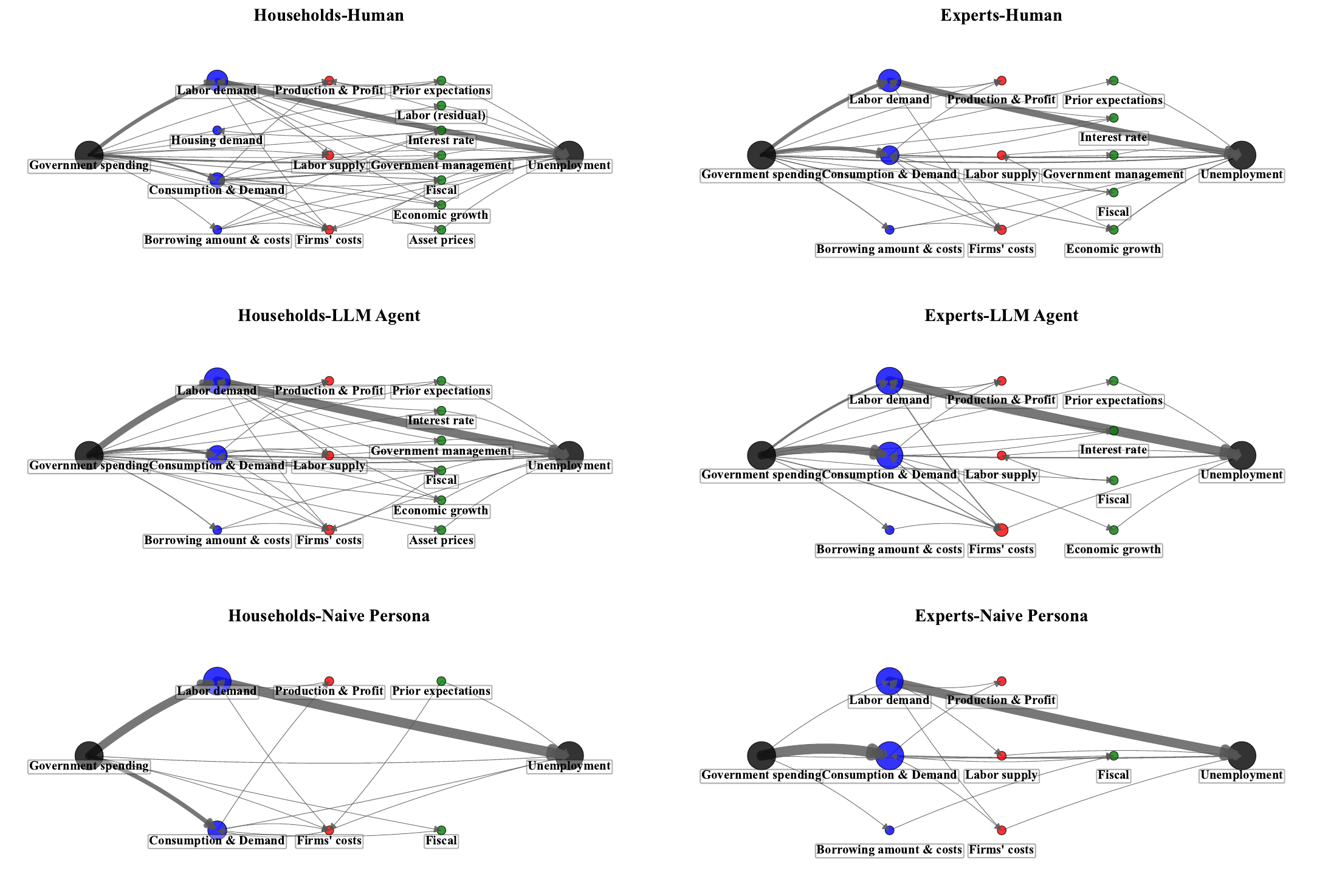}
\caption{The ``average'' DAGs underlying the formation of unemployment expectations in the government spending vignette}\label{fig:11}
\notepar{%
Notes: The figure presents the ``average'' DAGs underlying unemployment expectation formation for humans (i.e., ``Households-Human'' and ``Experts-Human''), LLM Agents (i.e., ``Households-LLM Agent'' and ``Experts-LLM Agent''), and foundation models with only naive personas (i.e., ``Households-Naive Persona'' and ``Experts-Naive Persona'') in the government spending vignette. The nodes represent categories of intermediate variables, whose definitions and classifications are provided in Supplementary Appendix Table~\ref{tab:a-3}. The aggregated DAGs reveal the most relevant variables (nodes) and causal links in the responses of humans, LLM Agents, and Naive Persona. Node size: The size of the nodes is proportional to the share of responses that refer to the nodes. Node color: Red indicates supply-side variables, blue indicates demand-side variables, green indicates miscellaneous variables, black indicates start and end nodes. Edge thickness: The thickness of the edges is proportional to the share of responses that refer to the causal connections (among humans, LLM Agents and Naive Persona, respectively).
}
\end{figure}

Furthermore, we construct ``average DAGs'' to visualize the aggregated mental models underlying inflation or unemployment expectations among humans, LLM Agents, and Naive Persona under each vignette. Using shifts in unemployment expectations following a government spending shock as an example, the results are presented in Figure~\ref{fig:11}. In this figure, variables (nodes) more frequently cited in respondents' mental models are depicted as larger circles, while more common causal relationships are represented by thicker edges. This approach intuitively reveals the most prevalent variables and causal links in the mental models of both households and experts.

As shown in Figure~\ref{fig:11}, both Household Agents and Expert Agents capture most of the nodes and their relationships within the mental models of households and experts, respectively. Compared to experts, households exhibit greater diversity in the nodes and cognitive pathways within their mental models, leading to more dispersed expectation distributions. However, the mental models of LLM Agents still differ from those of humans in several respects. They concentrate more heavily on certain prevalent nodes and causal links, while omitting less common but distinctive nodes or links, particularly those associated with miscellaneous variables. This pattern mirrors the more concentrated distribution of expectations generated by LLM Agents. By contrast, the mental models of Naive Persona lack many key nodes and edges, resulting in a highly distorted structure. Similar patterns are observed across the other vignettes in Supplementary Appendix Figure~\ref{fig:a-13} to Figure~\ref{fig:a-19} for inflation and unemployment expectations.

In addition, we calculate the average number of causal links and unique nodes in the causal DAGs of humans, LLM Agents, and Naive Persona to assess the complexity of their mental models. The results are presented in Supplementary Appendix Table~\ref{tab:a-2}. We find that, in most cases, the average number of variables and causal links in the mental models of LLM Agents exceeds that of humans, with this pattern being even more pronounced in the results of Naive Persona. This reveals a key quantitative difference between the open-ended responses of GenAI and those of humans. At the individual level, the causal chains through which humans explain expectation formation are relatively simpler, with the average number of causal links in their responses falling below 3.5 in most cases. By contrast, the causal narratives generated by GenAI, whether by LLM Agents or Naive Persona, are typically more complete and complex, with the average number of causal links exceeding 3.5 in most cases.

The results in Sections~\ref{sec:5-2-1} and \ref{sec:5-2-2} jointly demonstrate that LLM Agents calibrated with expectation survey data and text data from human society can qualitatively recapitulate key patterns underlying human expectation formation at the population level, whereas Naive Persona fails to do so. However, significant quantitative discrepancies remain, which delineate the application boundaries of LLM Agents and highlight the uniqueness of human samples. Humans possess unique personal experiences, richer information sources, and unobservable heterogeneity that are difficult for LLM Agents to fully capture. The modules we construct and the theory-driven rules we introduce map to the key determinants and mechanisms of expectation formation, serving as a high-level abstraction of real-world agents. Our LLM Agents therefore function as ``distilled'' humans: they capture stylized qualitative features of human expectations but cannot precisely generate quantitative results comparable to those of humans.

\section{Contributions of Components in LLM Agents}\label{evaluation-of-contributions-of-components-in-llm-agents}\label{sec:6}

In this section, we conduct Step 4 of our framework to investigate the origin of LLM Agents' ability to simulate human-like expectation distributions and capture underlying thinking patterns (e.g., selective recall and mental models). We remove one component from the LLM Agents at a time while holding the others fixed\footnote{{} {Removing only one component at a time yields the following five types of Household Agents: (1) those without Random Disturbances of hyperparameters (i.e., setting \texttt{temperature} and \texttt{top-p} as constants equal to 1 and 0.5, respectively), denoted as ``w/o RD''; (2) those without SMIM, denoted as ``w/o SMIM''; (3) those without PEPM, denoted as ``w/o PEPM''; (4) those without PCM, denoted as ``w/o PCM''; and (5) those without initialization prompts, denoted as ``w/o INITIAL.'' Similarly, removing each component individually produces five types of Expert Agents. Types (1), (3), and (5) correspond to those of Household Agents, while the other two are: (2) Expert Agents without KAM, denoted as ``w/o KAM''; and (4) Expert Agents without PBM, denoted as ``w/o PBM.''}}, and then compare their simulated distributions and open-ended responses with those of humans one by one, so as to evaluate each component's contribution to simulation performance across different dimensions. In these comparisons, we include the simulation results of Naive Persona (that is, removing all components at once except for the initialization prompts) as a baseline.

\begin{figure}[!h]
\centering
\includegraphics[width=\linewidth,keepaspectratio]{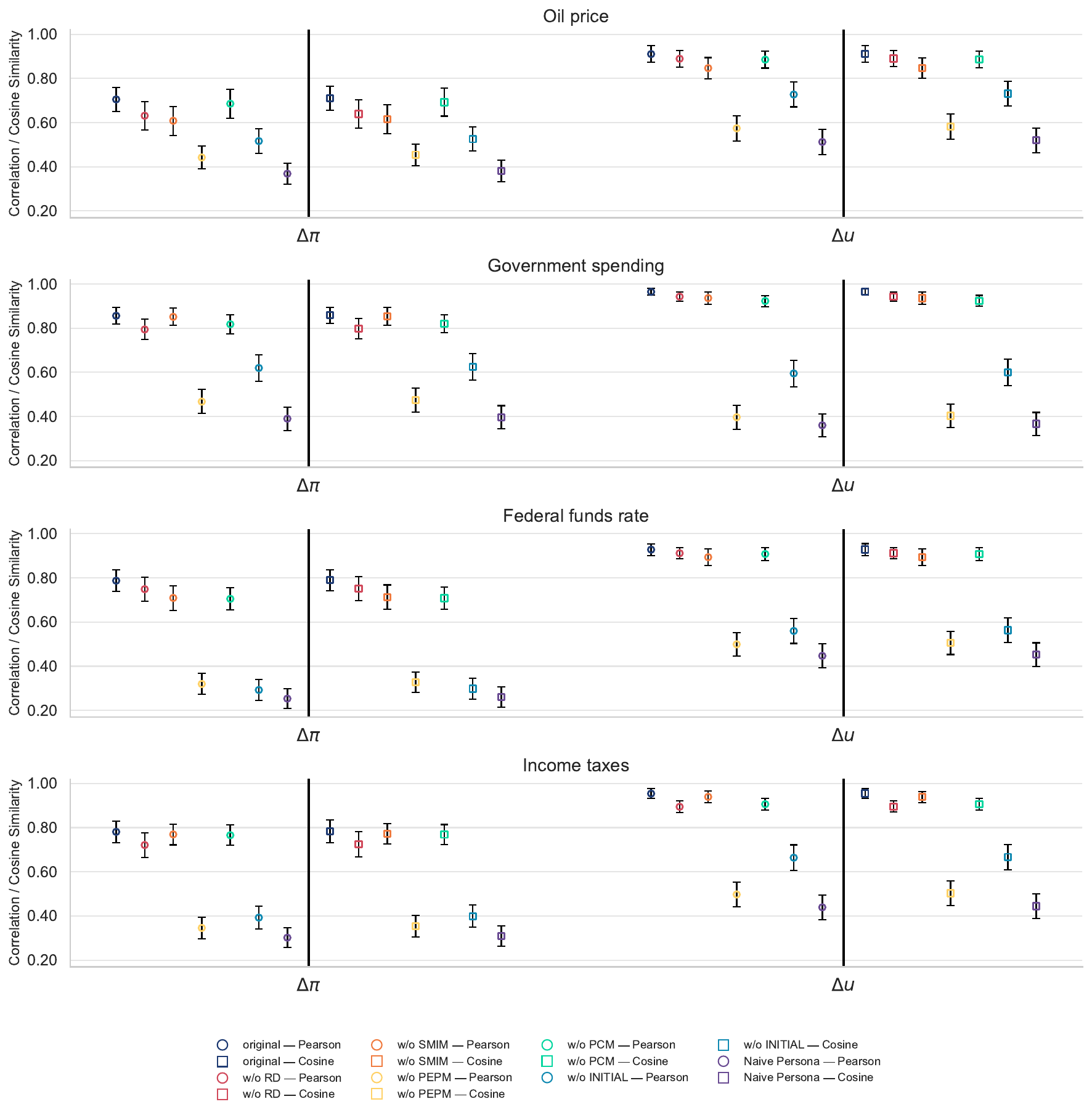}
\caption{Shape similarity between the expectation distributions generated by different Household Agents and those generated by humans in Experiment 1}\label{fig:12}
\notepar{%
Notes: This figure displays the distributional shape similarity, as measured by Pearson correlation (displayed to the left of the bold vertical lines) and cosine similarity (displayed to the right of the bold vertical lines), between the changes in inflation expectations (\(\Delta\ \pi\)) and unemployment expectations (\(\Delta\ u\)) generated by Household Agents (original and those without different components) and those of households under four different vignettes. Error bars present two-sided 95\% confidence intervals for the similarity metrics, obtained by bootstrap over histogram-based probability vectors.
}
\end{figure}

First, to evaluate the simulation performance along the distributional dimension, we compute the shape similarity between the human distributions and those generated by the original LLM Agents, LLM Agents without a certain component, and Naive Persona, respectively. Taking the Household Agents in the hypothetical vignette experiment as an example, Figure~\ref{fig:12} shows that: (1) Compared with the original Household Agents, removing any single component reduces the similarity of the simulated distributions, though the magnitude of the decline varies. (2) Removing the PEPM or the initialization prompts leads to the most pronounced decline in distributional similarity. (3) Naive Persona yields the lowest distributional similarity, averaging about 0.4, far below that of the original Household Agents. Similar patterns are also observed in the Expert Agents and two other survey experiments (see Supplementary Appendix Figure~\ref{fig:a-20} to Figure~\ref{fig:a-23}).

\begin{figure}[h]
\centering
\includegraphics[width=\linewidth,keepaspectratio]{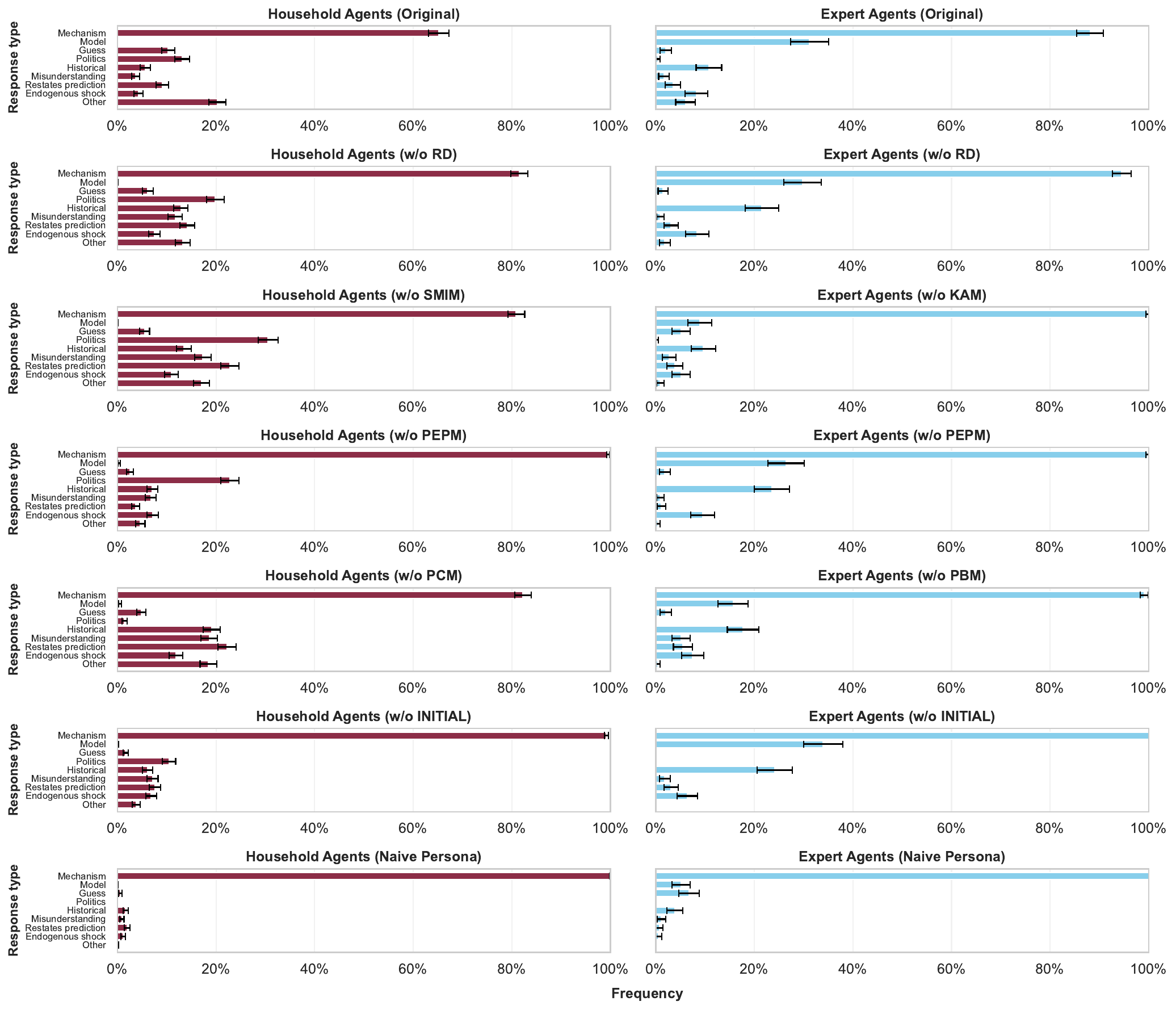}
\caption{Response types in open-ended responses of different LLM Agents in Experiment 1}\label{fig:13}
\notepar{%
Notes: This figure presents ``response type'' classification of open-ended responses generated by LLM Agents (original and those without different components) across all four vignettes. The open-ended responses from all LLM Agents are automatically classified by an agentic workflow and manually verified. Error bars display 95\% confidence intervals.
}
\end{figure}

Second, unlike the prior expectations from the PEPM, which are important for distributional simulation, the personal information from the PCM and PBM and the human social textual data extracted by the SMIM and KAM contribute less to the simulation. Instead, they play a major role in recapitulating certain key patterns in the thoughts underlying expectation formation. Taking the hypothetical vignette experiment as an example, as shown in Figure~\ref{fig:13}, the overall results show that removing any component leads to an increase in the proportion of recalled content categorized as ``Mechanism,'' while the proportion categorized as ``Other'' decreases. Specifically, for Expert Agents, the removal of KAM or PBM results in a significant reduction in their recall of ``Model''-related content, suggesting that information such as expertise and professional background aids in establishing selective recall regarding models or theories. For Household Agents, the removal of SMIM leads to a loss of response diversity, with a stronger focus on personal characteristics or prior predictions, thereby increasing the proportions of ``Politics'' and ``Restates prediction.'' Household Agents without PCM lose access to personal information, resulting in a decline in the proportion of recalled content categorized as ``Politics.''

Additionally, supplementary evidence yields similar findings. Supplementary Appendix Figure~\ref{fig:a-24} reveals that, compared to the original LLM Agents, Household Agents without SMIM and Expert Agents without KAM recall highly homogeneous channels. Specifically, Expert Agents without KAM fail to recall any professional content related to central banks. As shown in Supplementary Appendix Figure~\ref{fig:a-25}, in the randomized information experiment, the removal of any component leads to varying degrees of increased homogeneity in the channels recalled by LLM Agents. For instance, after removing SMIM, the proportion of mechanisms recalled by Homeowner Agents that fall under ``Home price growth irrelevant'' increases, which explains the rise in the proportion of agents expecting unchanged future economic situations. Meanwhile, more Renter Agents recall mechanisms centered on income effects, accounting for the significant increase in those expecting worsened future economic situations. Supplementary Appendix Figure~\ref{fig:a-26} indicates that, in simulating the MSC, Household Agents without SMIM are almost unable to recall certain recent black swan events and exhibit reduced perception of policy-related information. Household Agents without PCM fail to recall channels associated with politics.

Furthermore, we measure and compare the diversity of thoughts underlying expectation formation among LLM Agents with different components removed and humans across three survey experiments (see Supplementary Appendix Section~\ref{sec:d} for details). The results are presented in Supplementary Appendix Table~\ref{tab:a-4} to Table~\ref{tab:a-6}. We find that: (1) Removing any component reduces the diversity of thoughts generated by LLM Agents, suggesting that homogenization of thoughts may explain their diminished simulation performance. (2) Removing the initialization prompts reduces the diversity of the generated thoughts more substantially than removing any other component, indicating that simply feeding in rich contextual information does not enable LLM Agents to generate more diverse reasoning. (3) Although LLM Agents generate significantly more diverse thoughts than Naive Persona, a pronounced quantitative gap remains, as their diversity is substantially lower than that of human samples.

From a qualitative perspective, these findings collectively show that: (1) The naive persona approach commonly used in the existing literature severely distorts and homogenizes both the expectation distributions and the open-ended responses generated by foundation models, whereas LLM Agents calibrated with real-world information and expectation-formation theories mitigate these biases and achieve effective simulations. (2) Different modules contribute to distinct dimensions of the simulation. For distributional simulation, priors from PEPM contribute the most. A similar pattern appears in human samples: priors (particularly the most recent perceptions) are among the most critical factors in expectation formation \citep{jonung1981perceived,coibion2020inflation}. For recapitulating certain key qualitative results in open-ended responses, the personal information from PCM and PBM, together with the textual data on human society extracted by SMIM and KAM, make important contributions. (3) Even when we feed the LLM Agents all information extracted by every module but omit the initialization prompts, the agents still fail to achieve the original simulation performance. This indicates that their strong performance does not simply stem from foundation models' recall triggered by directly inputting rich contextual information, but rather from the agents' deeper understanding of roles, tasks, and Bayesian updating mechanisms, combined with the effective integration and utilization of information across modules. These findings provide a roadmap for designing modules grounded in empirical evidence to construct LLM Agents calibrated to real-world information and theories of expectation updating, and offer insights into narrowing the systematic belief gap between GenAI and human beliefs at the aggregate level.

Combined with the findings in Section~\ref{sec:5}, a quantitative perspective indicates that the primary difference between GenAI simulations and human samples lies in the degree of homogenization. While LLM Agents improve upon the naive persona approach by utilizing modules and randomized hyperparameters to capture key heterogeneity, they still fail to fully reflect the unobservable heterogeneity and random factors inherent in human samples. This homogenization of LLM Agents is particularly evident in open-ended responses and manifests in four dimensions: (1) the channels mentioned in selective recall are highly singular and concentrated; (2) the content of selective recall skews heavily toward mechanism explanations, lacking diverse elements; (3) mental models primarily focus on prevalent causal chains, omit nodes or links related to miscellaneous variables, and exhibit longer average causal chains; and (4) the semantic diversity of open-ended responses is substantially lower than in human samples. These differences highlight the potential risks of applying LLM Agents to quantitative research and clarify their boundary of application. In addition, many recent online survey participants use LLMs to generate open-ended responses \citep{haaland2025understanding}, and several commercial platforms directly generate survey samples using GenAI\footnote{{} Three illustrative examples include: (1) HumanAI: \url{https://www.syntheticrespondents.io/}; (2) Qualtrics: \url{https://www.qualtrics.com/articles/strategy-research/qualtrics-ai-outperforms-general-use-llms/}; (3) BlockSurvey: \url{https://blocksurvey.io/ai-sample-response-generator}}. If these AI-generated data contaminate human samples, they threaten to severely compromise the diversity and authenticity of the data. Consequently, the quantitative empirical evidence in this paper offers insights for distinguishing GenAI-generated open-ended responses from those produced by humans.

\section{Concluding Remarks}\label{concluding-remarks}\label{sec:7}

This paper develops and validates a framework for simulating the macroeconomic expectations of heterogeneous respondents in survey experiments. Grounded in economic theory and the empirical determinants of expectation formation, we construct LLM Agents that use a foundation model as their reasoning core and a set of task-specific modules to incorporate personal characteristics, prior beliefs, and dynamic information drawn from human society, with module invocation disciplined by a Bayesian updating rule. We deploy Household Agents and Expert Agents to recapitulate three representative designs spanning a hypothetical vignette experiment, a randomized information experiment, and a large-scale recurring survey, covering inflation, unemployment, and home price expectations across distinct agent types. The simulated distributions track the human ones closely and preserve the key heterogeneity observed within and across groups, while the accompanying open-ended responses recapitulate the qualitative signatures of human expectation formation, namely selective recall and mental models. Component ablation reveals that prior expectations are decisive for matching distributions, whereas personal information and externally retrieved text are what allow the agents to recover human-like reasoning; absent a theory-grounded initialization, even agents endowed with all available information fail to simulate effectively.

These results carry implications along several dimensions. Methodologically, we establish that calibrating foundation models with real-world information and expectation-formation theory, rather than relying on naive persona prompts alone, is what converts a foundation model into a \emph{Homo silicus}. The framework is accordingly best understood as a complement to traditional surveys rather than a substitute for them, offering low-cost pre-experimental simulation and the imputation of scarce observations once its capabilities have been validated. Conceptually, it demonstrates that LLM Agents can serve as an abstraction of human agents, reflecting aggregate behavioral regularities rather than merely mimicking numerical outputs.

The framework nonetheless has clear limitations. While LLM Agents qualitatively recapitulate the mental patterns underlying expectation formation, they remain quantitatively more homogeneous than human samples, concentrating recall on dominant channels, simplifying mental models, and exhibiting lower semantic diversity. These agents smooth away the idiosyncratic experience and unobserved heterogeneity of real respondents, and researchers with strict quantitative requirements should treat their output with corresponding caution.

Two directions appear especially promising but, as this paper adopts the perspective of experimental and behavioral economics, both lie beyond our scope and are left for future research. The first concerns mechanism and interpretability. We have characterized the qualitative patterns in LLM Agents' open-ended explanations, but have not opened the foundation models' internal black box. The micro-level mechanisms of their expectation formation and the interpretability of their outputs remain unresolved at the frontier of AI research and central to AI behavioral science. The second concerns correcting the quantitative biases that current GenAI tools still display, an issue of statistical theory and econometrics beyond our present remit. We hope future work will develop an integrated paradigm combining GenAI as the structural backbone, a small but carefully curated sample of real survey data for calibration, and rigorous econometric methods as the disciplining standard. Once such bias is corrected, the approach could extend to far wider survey experiments.

\newpage
\bibliographystyle{apalike}
\bibliography{bib/refs}

\clearpage
\renewcommand{\thepage}{SA-\arabic{page}}
\setcounter{page}{1}
\thispagestyle{plain}
\addtocontents{toc}{\protect\setcounter{tocdepth}{2}}

\appendix
\renewcommand{\thesection}{\Alph{section}}
\renewcommand{\thesubsection}{\Alph{section}.\arabic{subsection}}
\setcounter{figure}{0}
\setcounter{table}{0}
\setcounter{footnote}{0}
\renewcommand{\thefigure}{A.\arabic{figure}}
\renewcommand{\thetable}{A.\arabic{table}}
\begin{center}
\section*{Supplementary Appendix For \\ ``Simulating Macroeconomic Expectations in Survey Experiments with LLM-based Economic Agents''}
\end{center}
\renewcommand{\contentsname}{Contents}
\setcounter{tocdepth}{2}
\tableofcontents
\bigskip
\clearpage
\thispagestyle{plain}
\section{Additional Results}\label{additional-results}\label{sec:a}

\subsection{Tables}\label{tables}\label{sec:a-1}

\begin{table}[H]
\centering
\caption{Abbreviations used in this paper}\label{tab:abbrev}
\small
\begin{tabular}{@{}>{\raggedright\arraybackslash}p{(\linewidth - 4\tabcolsep) * \real{0.0855}}
  >{\raggedright\arraybackslash}p{(\linewidth - 4\tabcolsep) * \real{0.2300}}
  >{\raggedright\arraybackslash}p{(\linewidth - 4\tabcolsep) * \real{0.6700}}@{}}
\toprule
\textbf{No.} & \textbf{Abbreviation} & \textbf{Full Term} \\
\midrule
1 & ACS & American Community Survey \\
2 & DAG & Directed Acyclic Graph \\
3 & Expert Agents & LLM Agents for simulating expert expectations \\
4 & GenAI & generative AI \\
5 & Homeowner Agents & Household Agents for simulating homeowners \\
6 & Household Agents & LLM Agents for simulating household expectations \\
7 & i.i.d. & independent and identically distributed \\
8 & INITIAL & initialization prompts \\
9 & KAM & Knowledge Acquisition Module \\
10 & LLM & Large Language Model \\
11 & LLM Agents & LLM-based economic agents \\
12 & MSC & Michigan Surveys of Consumers \\
13 & Naive Persona & foundation models with only naive personas, equivalent to LLM Agents with only initialization prompts \\
14 & PBM & Professional Background Module \\
15 & PCM & Personal Characteristics Module \\
16 & PEPM & Prior Expectations \& Perceptions Module \\
17 & RAG & Retrieval-Augmented Generation \\
18 & RCT & randomized controlled trial \\
19 & RD & Random Disturbances \\
20 & Renter Agents & Household Agents for simulating renters \\
21 & SBERT & Sentence-BERT \\
22 & SMIM & Social Media Information Module \\
23 & SPF & Survey of Professional Forecasters \\
24 & w/o INITIAL & Household (or Expert) Agents without initialization prompts \\
25 & w/o KAM & Expert Agents without Knowledge Acquisition Module \\
26 & w/o PBM & Expert Agents without Professional Background Module \\
27 & w/o PCM & Household Agents without Personal Characteristics Module \\
28 & w/o PEPM & Household (or Expert) Agents without Prior Expectations \& Perceptions Module \\
29 & w/o RD & Household (or Expert) Agents without Random Disturbances \\
30 & w/o SMIM & Household Agents without Social Media Information Module \\
\bottomrule
\end{tabular}
\notepar{%
Notes: This table lists all abbreviations used throughout this paper and their corresponding full terms, sorted alphabetically (A--Z) by abbreviation for ease of reference. The abbreviations and full terms are kept consistent with the wording used in the main text. Plural initialisms (LLMs, DAGs, RCTs) are listed in their singular base form.
}
\end{table}

\begin{table}[H]
\centering
\caption{Information on the foundation models in our paper}\label{tab:a-1}
\small
\begin{tabular}{@{}>{}p{(\linewidth - 6\tabcolsep) * \real{0.3438}}
  >{}p{(\linewidth - 6\tabcolsep) * \real{0.1940}}
  >{}p{(\linewidth - 6\tabcolsep) * \real{0.1793}}
  >{}p{(\linewidth - 6\tabcolsep) * \real{0.2693}}@{}}
\toprule
Foundation Model (LLM) & Developer & Release Date & Knowledge Cutoff \\
\midrule
Qwen3-235B-A22B-Thinking-2507 & Alibaba & July 25, 2025 & October 2024 (or earlier)\textsuperscript{*} \\
DeepSeek-R1-0528 & DeepSeek & May 28, 2025 & July 2024 (or earlier)\textsuperscript{*} \\
DeepSeek-V3-0324 & DeepSeek & March 24, 2025 & July 2024 (or earlier)\textsuperscript{*} \\
GPT-o4-mini & OpenAI & April 16, 2025 & June 2024 \\
GPT-4.1-mini & OpenAI & April 14, 2025 & June 2024 \\
Gemini-2.5-Pro & Google DeepMind & June 17, 2025 & January 2025 \\
\bottomrule
\end{tabular}
\notepar{%
Notes: This table presents information on the developers, release dates, and knowledge cutoffs of the six advanced foundation models discussed in this paper. The latest knowledge cutoff among these models is no later than January 2025. The knowledge cutoffs marked with an asterisk (*) are not officially released dates---as the developers do not disclose them in their technical reports---but are inferred by querying the LLMs with a series of questions, including: (1) What is your knowledge cutoff? (2) What is today's date? (3) What happened in January 2025? These questions are designed to elicit responses revealing the models' actual knowledge cutoffs.
}
\end{table}

\begin{table}[H]
\centering
\caption{Complexity of the mental models underlying the expectations of humans, LLM Agents, and Naive Persona under each vignette}\label{tab:a-2}
\small
\begin{tabular}{@{}>{}p{(\linewidth - 12\tabcolsep) * \real{0.1493}}
  >{}p{(\linewidth - 12\tabcolsep) * \real{0.1106}}
  >{}p{(\linewidth - 12\tabcolsep) * \real{0.1406}}
  >{}p{(\linewidth - 12\tabcolsep) * \real{0.1612}}
  >{}p{(\linewidth - 12\tabcolsep) * \real{0.1106}}
  >{}p{(\linewidth - 12\tabcolsep) * \real{0.1406}}
  >{}p{(\linewidth - 12\tabcolsep) * \real{0.1871}}@{}}
\toprule
\multirow{2}{*}{Vignettes} & \multicolumn{3}{>{}p{(\linewidth - 12\tabcolsep) * \real{0.4124} + 4\tabcolsep}}{%
\textbf{Panel A: Inflation (Households)}} & \multicolumn{3}{>{}p{(\linewidth - 12\tabcolsep) * \real{0.4383} + 4\tabcolsep}@{}}{%
\textbf{Panel B: Unemployment (Households)}} \\
& Human & LLM Agent & Naive Persona & Human & LLM Agent & Naive Persona \\
\midrule
Oil price & 2.2 (3.0) & 3.0 (3.5) & 2.7 (3.4) & 3.0 (3.7) & 4.0 (4.3) & 5.7 (5.6) \\
Government spending & 3.2 (3.7) & 4.4 (4.5) & 5.6 (5.3) & 2.7 (3.5) & 3.8 (4.1) & 3.0 (3.6) \\
Federal funds rate & 2.8 (3.5) & 3.4 (3.8) & 5.7 (5.3) & 3.0 (3.7) & 3.6 (4.2) & 6.1 (6.0) \\
Income taxes & 3.7 (4.0) & 5.0 (5.1) & 5.8 (5.7) & 3.6 (4.0) & 5.3 (5.3) & 6.0 (6.0) \\
\midrule
\multirow{2}{=}{ Vignettes} & \multicolumn{3}{>{}p{(\linewidth - 12\tabcolsep) * \real{0.4124} + 4\tabcolsep}}{%
\textbf{Panel C: Inflation (Experts)}} & \multicolumn{3}{>{}p{(\linewidth - 12\tabcolsep) * \real{0.4383} + 4\tabcolsep}@{}}{%
\textbf{Panel D: Unemployment (Experts)}} \\
& Human & LLM Agent & Naive Persona & Human & LLM Agent & Naive Persona \\
\midrule
Oil price & 2.7 (3.2) & 3.5 (3.7) & 2.6 (3.3) & 3.7 (4.1) & 6.1 (5.5) & 7.6 (6.6) \\
Government spending & 3.2 (3.6) & 3.8 (4.1) & 3.8 (4.1) & 3.1 (3.7) & 4.2 (4.5) & 3.6 (4.3) \\
Federal funds rate & 3.1 (3.6) & 4.2 (4.3) & 5.4 (5.2) & 2.5 (3.3) & 4.2 (4.6) & 6.1 (6.0) \\
Income taxes & 3.9 (4.2) & 5.6 (5.4) & 5.7 (5.6) & 3.7 (4.2) & 5.8 (5.7) & 6.3 (6.2) \\
\bottomrule
\end{tabular}
\notepar{%
Notes: This table presents the average number of causal links (values outside parentheses) and unique nodes (values inside parentheses) in the mental models (DAGs) of humans, LLM Agents, and foundation models with only naive personas (Naive Persona) across all vignettes. Panels A and B present the results corresponding to the inflation and unemployment expectations of households, Household Agents, and Naive Persona, respectively. Panels C and D present the results corresponding to the inflation and unemployment expectations of experts, Expert Agents, and Naive Persona, respectively. Higher values indicate greater structural complexity of the causal pathways on average.
}
\end{table}

\newpage
{\small
\begin{longtable}[]{@{}
  >{\raggedright\arraybackslash}p{(\linewidth - 4\tabcolsep) * \real{0.2614}}
  >{\raggedright\arraybackslash}p{(\linewidth - 4\tabcolsep) * \real{0.2614}}
  >{\arraybackslash}p{(\linewidth - 4\tabcolsep) * \real{0.4627}}@{}}
\caption{Intermediate variables involved in mental models and their categories}\label{tab:a-3}\tabularnewline
\toprule\noalign{}
\textbf{Category (Node)} & \textbf{Intermediate variable} & \textbf{Explanation} \\
\midrule\noalign{}
\endfirsthead
\toprule\noalign{}
\textbf{Category (Node)} & \textbf{Intermediate variable} & \textbf{Explanation} \\
\midrule\noalign{}
\endhead
\bottomrule\noalign{}
\endlastfoot
\multicolumn{3}{@{}>{\raggedright\arraybackslash}p{(\linewidth - 4\tabcolsep) * \real{0.9855} + 4\tabcolsep}@{}}{\centering
\textbf{Demand}} \\
\midrule
\multirow{12}{=}{Borrowing amount \& costs} & borrowing firms & Amount borrowed (debt) by firms, or amount lent by banks to firms. \\
& borrowing household & Amount borrowed (debt) by households, or amount lent by banks to households. \\
& borrowing government & Amount borrowed (debt) by the government, or amount lent by banks to the government. \\
& costs borrowing household & Borrowing rates and/or access to credit faced by households. \\
& costs borrowing banks & Borrowing rates and/or access to credit faced by banks. \\
& costs borrowing government & Borrowing rates and/or access to credit faced by the government. \\
\midrule
\multirow{11}{=}{Consumption \& Demand} & demand & Demand for goods, spending and consumption by different groups. \\
& demand firms & Demand for goods, spending and consumption by firms. \\
& demand household & Demand for goods, spending and consumption by households. \\
& costs household & Costs of subsistence goods, e.g., heating, gasoline, ... \\
& demand government & Demand for goods, spending and consumption by the government. \\
& investment & Investment (expenditure) of firms. \\
\midrule
Housing demand & housing demand & Quantity of housing demanded. \\
\midrule
Labor demand & labor demand & ``Job creation'', firm's/government's demand for employees, ``Job opportunities''. \\
\midrule
\multirow{5}{=}{Income, Saving \& Money} & income & Household income, wages received, purchasing power. \\
& money & Overall amount of money in circulation, money printing by the central bank. \\
& saving & Amount saved by households. \\
\toprule
\multicolumn{3}{@{}>{\raggedright\arraybackslash}p{(\linewidth - 4\tabcolsep) * \real{0.9855} + 4\tabcolsep}@{}}{\centering
\textbf{Supply}} \\
\midrule
\multirow{6}{=}{Firms' costs} & costs firms & Production costs, including costs of input goods, wages paid; ``Firms need to cover''; ``firms need to make up for it'', \ldots{} \\
& costs borrowing firms & Borrowing rates and/or access to credit faced by firms. \\
& firm prices & Firms' decisions about pricing. \\
\midrule
Housing supply & housing supply & Quantity of housing supplied. \\
\midrule
Labor supply & labor supply & Changes in households' desired work hours. \\
\midrule
\multirow{5}{=}{Production \& Profit} & production & Firms' production / supply of goods and services. \\
& profit & Firms' profits or profit margin, including firms facing pressure to take actions to keep the profit margin at a certain level. \\
\toprule
\multicolumn{3}{@{}>{\raggedright\arraybackslash}p{(\linewidth - 4\tabcolsep) * \real{0.9855} + 4\tabcolsep}@{}}{\centering
\textbf{Miscellaneous}} \\
\midrule
\multirow{6}{=}{Prior expectations} & expected inflation & Expectations of future realizations of inflation as intermediate causes (propagation mechanisms). \\
& expected unemployment & Expectations of future realizations of unemployment as intermediate causes (propagation mechanisms). \\
\midrule
\multirow{4}{=}{Fiscal} & government taxes & Tax revenue collected by the government. \\
& government finances & Residual category referring to unspecified improvements or deterioration in the government's budget. \\
\midrule
Economic growth & growth & GDP growth, overall growth of the economy. \\
\midrule
Housing (residual) & housing & The quantity of housing is mentioned, but it is unclear whether demand or supply is being referred to. \\
\midrule
Labor (residual) & labor & Residual category for cases where it is unclear whether the respondent is thinking about labor demand or supply, e.g., ``more people work''. \\
\midrule
\multirow{2}{=}{Asset prices} & prices stock & Stock prices. \\
& prices house & ‌House prices. \\
\midrule
\multirow{5}{=}{Interest rate} & interest & General interest rate category if agent not specified or if not specified whether households' rates on borrowing vs saving are meant. \\
& saving rate & Interest rate earned on savings. \\
\midrule
Government management & government management & Explicit reference to policy successes (failures), good (bad) management by policymakers, or politicized positive (negative) evaluations of policies. \\
\end{longtable}
}
\vspace{-4mm}
\notepar{%
Notes: This table presents all intermediate variables potentially mentioned in open-ended responses within the hypothetical vignette experiment, along with their categorization and corresponding explanations. The definitions and categorization of these variables are primarily based on \citet{andre2022subjective,andre2026narratives}.
}

\newpage
{\small
\begin{longtable}[]{@{}
  >{}p{(\linewidth - 10\tabcolsep) * \real{0.1681}}
  >{}p{(\linewidth - 10\tabcolsep) * \real{0.1681}}
  >{}p{(\linewidth - 10\tabcolsep) * \real{0.1288}}
  >{}p{(\linewidth - 10\tabcolsep) * \real{0.1681}}
  >{}p{(\linewidth - 10\tabcolsep) * \real{0.1681}}
  >{}p{(\linewidth - 10\tabcolsep) * \real{0.1288}}@{}}
\caption{The diversity of thoughts generated by LLM Agents (original and those without different components) and those generated by humans in Experiment 1}\label{tab:a-4}\tabularnewline
\toprule\noalign{}
\multicolumn{6}{@{}>{}p{(\linewidth - 10\tabcolsep) * \real{0.9901} + 10\tabcolsep}@{}}{%
\textbf{Panel A: Households}} \\
\midrule\noalign{}
\endfirsthead
\toprule\noalign{}
\multicolumn{6}{@{}>{}p{(\linewidth - 10\tabcolsep) * \real{0.9901} + 10\tabcolsep}@{}}{%
\textbf{Panel A: Households}} \\
\midrule\noalign{}
\endhead
\bottomrule\noalign{}
\endlastfoot
\textbf{Vignette} & \textbf{Agent} & \textbf{Semantic Diversity} & \textbf{Vignette} & \textbf{Agent} & \textbf{Semantic Diversity} \\
\multirow{8}{=}{ Oil price} & Human & 0.5335 & \multirow{8}{=}{ Government spending} & Human & 0.5564 \\
& Original & 0.3158 & & Original & 0.3355 \\
& w/o RD & 0.3123 & & w/o RD & 0.3261 \\
& w/o SMIM & 0.2921 & & w/o PCM & 0.3116 \\
& w/o PCM & 0.2836 & & w/o SMIM & 0.2970 \\
& w/o PEPM & 0.2605 & & w/o PEPM & 0.2646 \\
& w/o INITIAL & 0.2398 & & w/o INITIAL & 0.2530 \\
& Naive Persona & 0.1588 & & Naive Persona & 0.1783 \\
\midrule
\multirow{8}{=}{ Federal funds rate} & Human & 0.5913 & \multirow{8}{=}{ Income taxes} & Human & 0.5771 \\
& Original & 0.3449 & & Original & 0.3268 \\
& w/o RD & 0.3342 & & w/o RD & 0.3240 \\
& w/o PCM & 0.3156 & & w/o SMIM & 0.3004 \\
& w/o SMIM & 0.3103 & & w/o PCM & 0.2909 \\
& w/o PEPM & 0.2647 & & w/o PEPM & 0.2781 \\
& w/o INITIAL & 0.2318 & & w/o INITIAL & 0.2640 \\
& Naive Persona & 0.1429 & & Naive Persona & 0.1638 \\
\toprule\noalign{}
\multicolumn{6}{@{}>{}p{(\linewidth - 10\tabcolsep) * \real{0.9901} + 10\tabcolsep}@{}}{%
\textbf{Panel B: Experts}} \\
\bottomrule\noalign{}
\textbf{Vignette} & \textbf{Agent} & \textbf{Semantic Diversity} & \textbf{Vignette} & \textbf{Agent} & \textbf{Semantic Diversity} \\
\multirow{8}{=}{ Oil price} & Human & 0.5309 & \multirow{8}{=}{ Government spending} & Human & 0.5980 \\
& Original & 0.3632 & & Original & 0.3496 \\
& w/o KAM & 0.3586 & & w/o KAM & 0.3415 \\
& w/o RD & 0.3533 & & w/o RD & 0.3361 \\
& w/o PEPM & 0.3132 & & w/o PEPM & 0.3118 \\
& w/o PBM & 0.3103 & & w/o PBM & 0.3024 \\
& w/o INITIAL & 0.2848 & & w/o INITIAL & 0.2812 \\
& Naive Persona & 0.2333 & & Naive Persona & 0.2269 \\
\midrule
\multirow{8}{=}{ Federal funds rate} & Human & 0.5822 & \multirow{8}{=}{ Income taxes} & Human & 0.5875 \\
& Original & 0.4501 & & Original & 0.3389 \\
& w/o RD & 0.4383 & & w/o KAM & 0.3344 \\
& w/o KAM & 0.4356 & & w/o RD & 0.3129 \\
& w/o PEPM & 0.3718 & & w/o PEPM & 0.2797 \\
& w/o PBM & 0.3571 & & w/o PBM & 0.2651 \\
& w/o INITIAL & 0.3365 & & w/o INITIAL & 0.2526 \\
& Naive Persona & 0.2754 & & Naive Persona & 0.1820 \\
\end{longtable}
}
\vspace{-4mm}
\notepar{%
Notes: This table presents the semantic diversity of thoughts generated by LLM Agents (original and those without different components) and those generated by humans under each vignette in the hypothetical vignette experiment, respectively. Panel A compares the results of Household Agents with those of households, while Panel B compares Expert Agents with experts.
}

\begin{table}[H]
\centering
\caption{The diversity of thoughts generated by LLM Agents (original and those without different components) and those generated by humans in Sub-Experiment 2 of Experiment 2}\label{tab:a-5}
\small
\begin{tabular}{@{}>{}p{(\linewidth - 10\tabcolsep) * \real{0.1781}}
  >{}p{(\linewidth - 10\tabcolsep) * \real{0.1781}}
  >{}p{(\linewidth - 10\tabcolsep) * \real{0.1388}}
  >{}p{(\linewidth - 10\tabcolsep) * \real{0.1781}}
  >{}p{(\linewidth - 10\tabcolsep) * \real{0.1781}}
  >{}p{(\linewidth - 10\tabcolsep) * \real{0.1388}}@{}}
\toprule
\textbf{Respondent} & \textbf{Agent} & \textbf{Semantic Diversity} & \textbf{Respondent} & \textbf{Agent} & \textbf{Semantic Diversity} \\
\midrule
\multirow{8}{=}{ Homeowners} & Human & 0.4996 & \multirow{8}{=}{ Renters} & Human & 0.5375 \\
& Original & 0.2871 & & Original & 0.2792 \\
& w/o RD & 0.2852 & & w/o PEPM & 0.2788 \\
& w/o PEPM & 0.2711 & & w/o RD & 0.2763 \\
& w/o PCM & 0.2626 & & w/o PCM & 0.2522 \\
& w/o SMIM & 0.2598 & & w/o SMIM & 0.2442 \\
& w/o INITIAL & 0.2281 & & w/o INITIAL & 0.2420 \\
& Naive Persona & 0.1575 & & Naive Persona & 0.1845 \\
\bottomrule
\end{tabular}
\notepar{%
Notes: This table presents the semantic diversity of thoughts generated by LLM Agents (original and those without different components) and those generated by humans in Sub-Experiment 2 of the randomized information experiment, respectively. The left panel compares the results of Homeowner Agents with those of homeowners, while the right panel compares Renter Agents with renters.
}
\end{table}

\begin{table}[H]
\centering
\caption{The diversity of thoughts generated by LLM Agents (original and those without different components) in Experiment 3}\label{tab:a-6}
\small
\begin{tabular}{@{}>{}p{(\linewidth - 10\tabcolsep) * \real{0.1781}}
  >{}p{(\linewidth - 10\tabcolsep) * \real{0.1781}}
  >{}p{(\linewidth - 10\tabcolsep) * \real{0.1388}}
  >{}p{(\linewidth - 10\tabcolsep) * \real{0.1781}}
  >{}p{(\linewidth - 10\tabcolsep) * \real{0.1781}}
  >{}p{(\linewidth - 10\tabcolsep) * \real{0.1388}}@{}}
\toprule
\textbf{Expectation} & \textbf{Agent} & \textbf{Semantic Diversity} & \textbf{Expectation} & \textbf{Agent} & \textbf{Semantic Diversity} \\
\midrule
\multirow{7}{=}{ Inflation} & Original & 0.3097 & \multirow{7}{=}{ Home price} & Original & 0.3214 \\
& w/o RD & 0.3097 & & w/o RD & 0.3187 \\
& w/o PCM & 0.2938 & & w/o PCM & 0.3076 \\
& w/o SMIM & 0.2674 & & w/o SMIM & 0.2670 \\
& w/o PEPM & 0.2102 & & w/o PEPM & 0.2366 \\
& w/o INITIAL & 0.1801 & & w/o INITIAL & 0.1709 \\
& Naive Persona & 0.0727 & & Naive Persona & 0.0921 \\
\bottomrule
\end{tabular}
\notepar{%
Notes: This table presents the semantic diversity of thoughts generated by Household Agents (original and those without different components) when simulating inflation and home price expectations in 2025 Michigan Surveys of Consumers, respectively. The left panel compares the results for inflation expectations, and the right panel for home price expectations.
}
\end{table}

\newpage
\subsection{Figures}\label{figures}\label{sec:a-2}

\begin{figure}[!htbp]
\centering
\includegraphics[width=\linewidth,keepaspectratio]{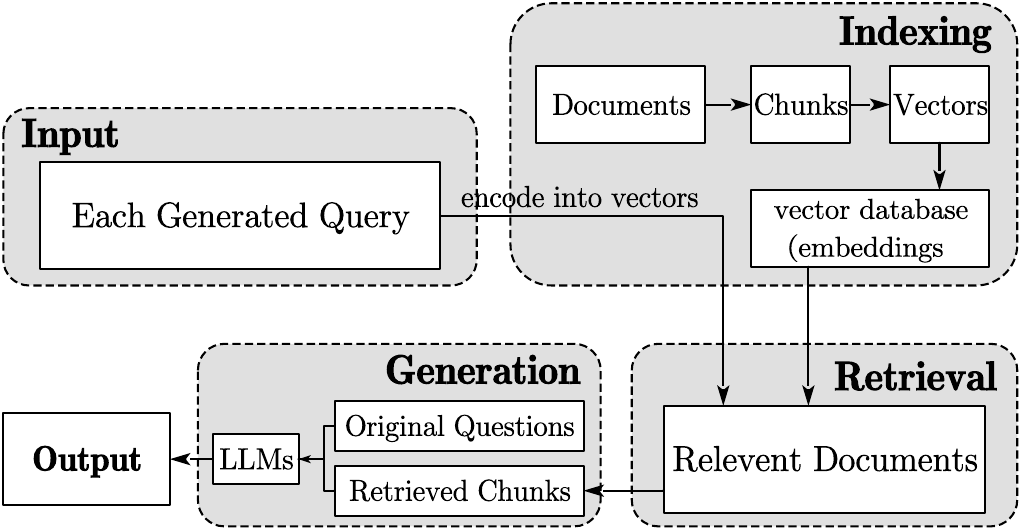}
\caption{The framework of our RAG workflow}\label{fig:a-1}
\notepar{%
Notes: This figure illustrates in detail the framework of the RAG workflow, specifically how KAM retrieves relevant knowledge \& information. It comprises three core steps: (1) Indexing: documents in the personalized knowledge base are segmented into chunks, encoded into vector representations, and stored in a vector database; (2) Retrieval: based on semantic similarity, the top k most relevant chunks are retrieved for each query; (3) Generation: the original survey question and the retrieved chunks are jointly input into the LLMs to generate the answer.
}
\end{figure}

\begin{figure}[!htbp]
\centering
\includegraphics[width=\linewidth,keepaspectratio]{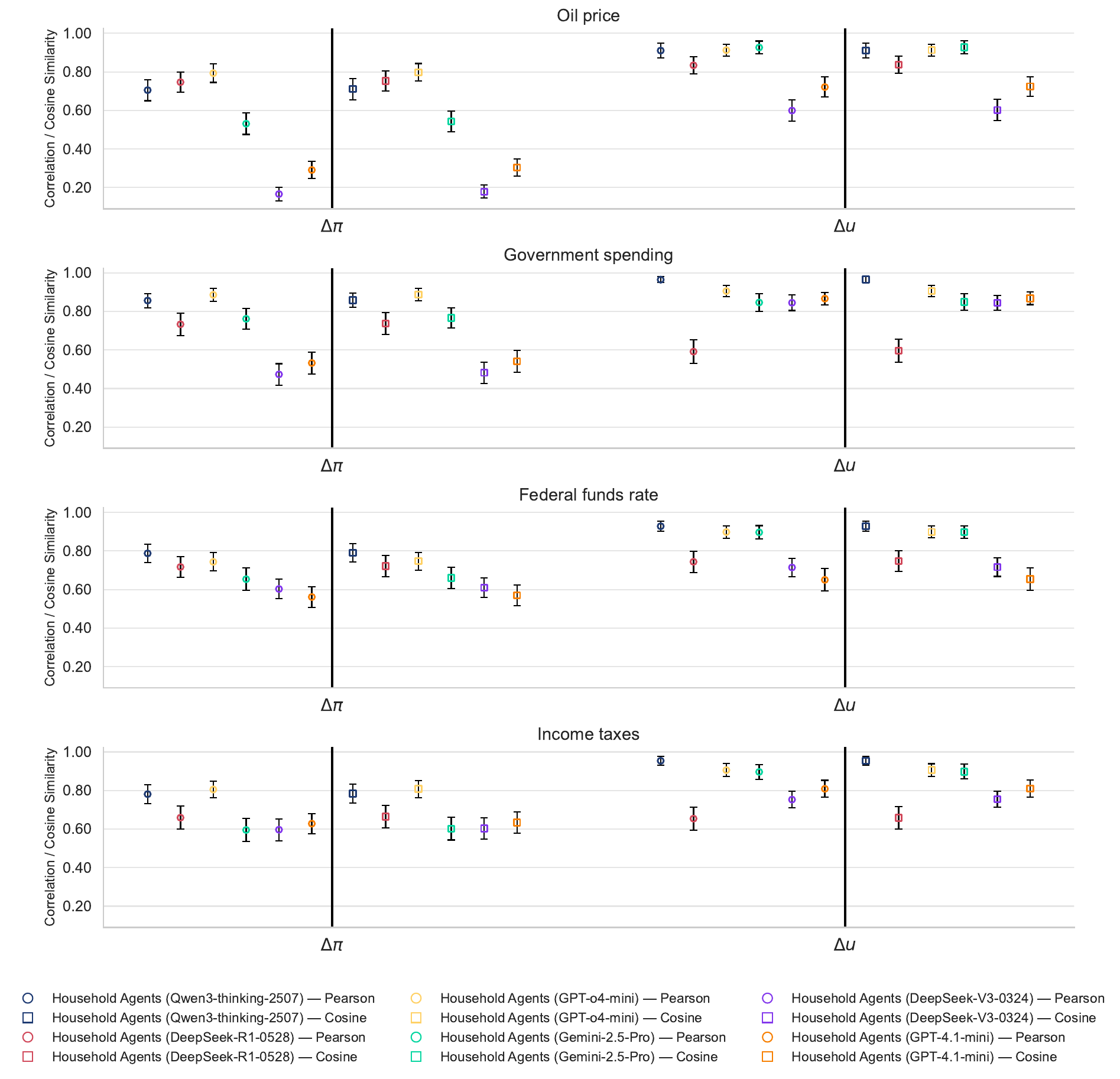}
\caption{Shape similarity between the expectation distributions generated by Household Agents based on different foundation models and those generated by humans in Experiment 1}\label{fig:a-2}
\notepar{%
Notes: This figure displays the distributional shape similarity, as measured by Pearson correlation (displayed to the left of the bold vertical lines) and cosine similarity (displayed to the right of the bold vertical lines), between the changes in inflation expectations (\(\Delta\ \pi\)) and unemployment expectations (\(\Delta\ u\)) generated by Household Agents based on six different types of foundation models (reasoning models: Qwen3-235B-A22B-Thinking-2507, DeepSeek-R1-0528, GPT-o4-mini, and Gemini-2.5-Pro; non-reasoning models: DeepSeek-V3-0324 and GPT-4.1-mini) and those of households under four different vignettes. Error bars present two-sided 95\% confidence intervals for the similarity metrics, obtained by bootstrap over histogram-based probability vectors.
}
\end{figure}

\begin{figure}[!htbp]
\centering
\includegraphics[width=\linewidth,keepaspectratio]{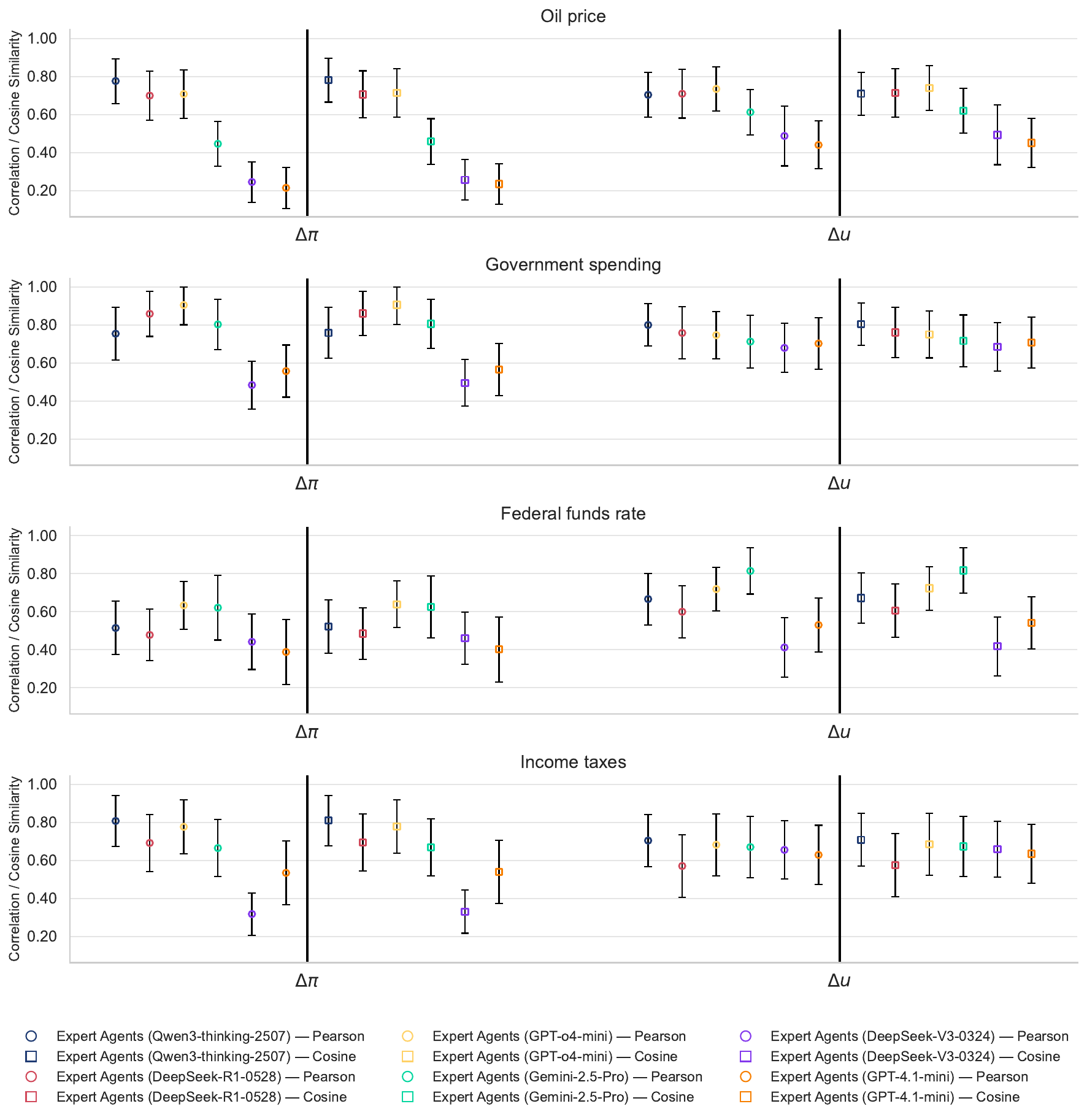}
\caption{Shape similarity between the expectation distributions generated by Expert Agents based on different foundation models and those generated by humans in Experiment 1}\label{fig:a-3}
\notepar{%
Notes: This figure displays the distributional shape similarity, as measured by Pearson correlation (displayed to the left of the bold vertical lines) and cosine similarity (displayed to the right of the bold vertical lines), between the changes in inflation expectations (\(\Delta\ \pi\)) and unemployment expectations (\(\Delta\ u\)) generated by Expert Agents based on six different types of foundation models (reasoning models: Qwen3-235B-A22B-Thinking-2507, DeepSeek-R1-0528, GPT-o4-mini, and Gemini-2.5-Pro; non-reasoning models: DeepSeek-V3-0324 and GPT-4.1-mini) and those of experts under four different vignettes. Error bars present two-sided 95\% confidence intervals for the similarity metrics, obtained by bootstrap over histogram-based probability vectors.
}
\end{figure}

\begin{figure}[!htbp]
\centering
\includegraphics[width=\linewidth,keepaspectratio]{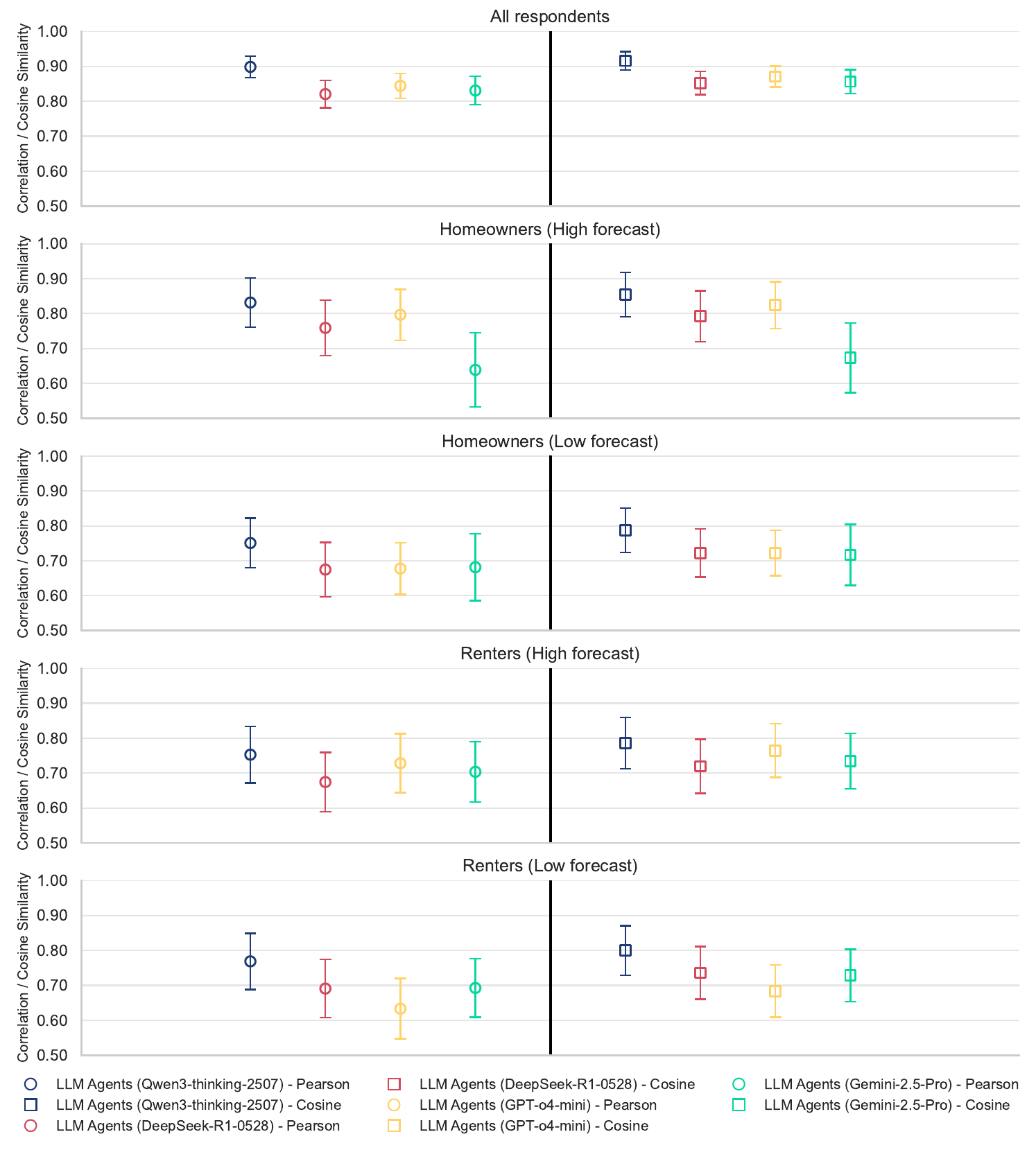}
\caption{Shape similarity between the expectation distributions generated by LLM Agents based on different foundation models and those generated by humans in Sub-Experiment 1 of Experiment 2}\label{fig:a-4}
\notepar{%
Notes: This figure displays the distributional shape similarity, as measured by Pearson correlation (displayed to the left of the bold vertical lines) and cosine similarity (displayed to the right of the bold vertical lines), between the home price expectations generated by LLM Agents based on four different types of reasoning models (Qwen3-235B-A22B-Thinking-2507, DeepSeek-R1-0528, GPT-o4-mini, and Gemini-2.5-Pro) and those of humans in different treatment groups. Error bars present two-sided 95\% confidence intervals for the similarity metrics, obtained by bootstrap over histogram-based probability vectors.
}
\end{figure}

\begin{figure}[!htbp]
\centering
\includegraphics[width=\linewidth,keepaspectratio]{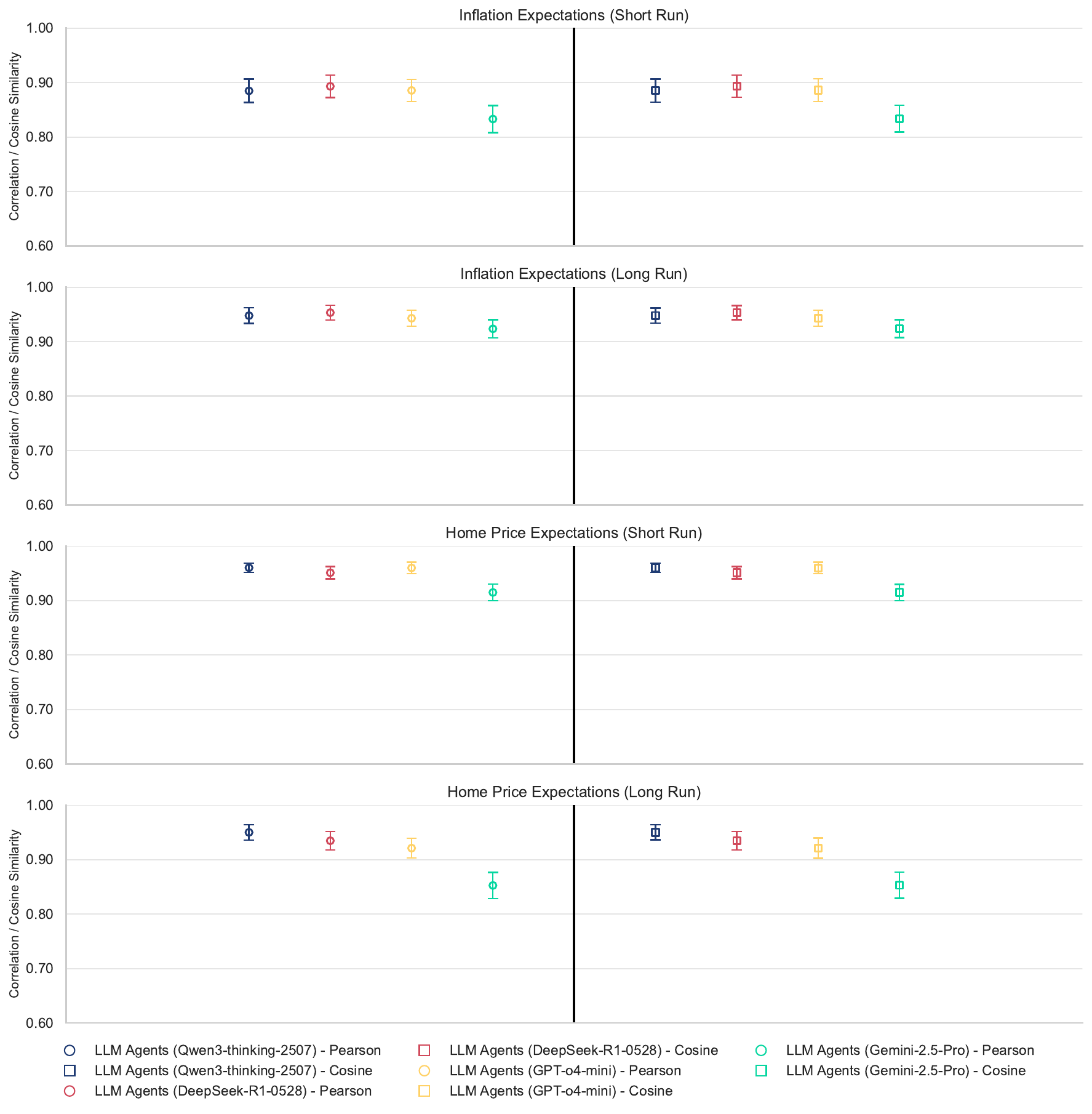}
\caption{Shape similarity between the expectation distributions generated by LLM Agents based on different foundation models and those generated by humans in Experiment 3}\label{fig:a-5}
\notepar{%
Notes: This figure displays the distributional shape similarity, as measured by Pearson correlation (displayed to the left of the bold vertical lines) and cosine similarity (displayed to the right of the bold vertical lines), between long- and short-term inflation expectations and home price expectations generated by LLM Agents based on four different types of reasoning models (Qwen3-235B-A22B-Thinking-2507, DeepSeek-R1-0528, GPT-o4-mini, and Gemini-2.5-Pro) and those of households in 2025 Michigan Surveys of Consumers. Error bars present two-sided 95\% confidence intervals for the similarity metrics, obtained by bootstrap over histogram-based probability vectors.
}
\end{figure}

\begin{figure}[!htbp]
\centering
\subcaptionbox{Humans\label{fig:sub-a-7-a}}{%
  \includegraphics[width=\linewidth,keepaspectratio]{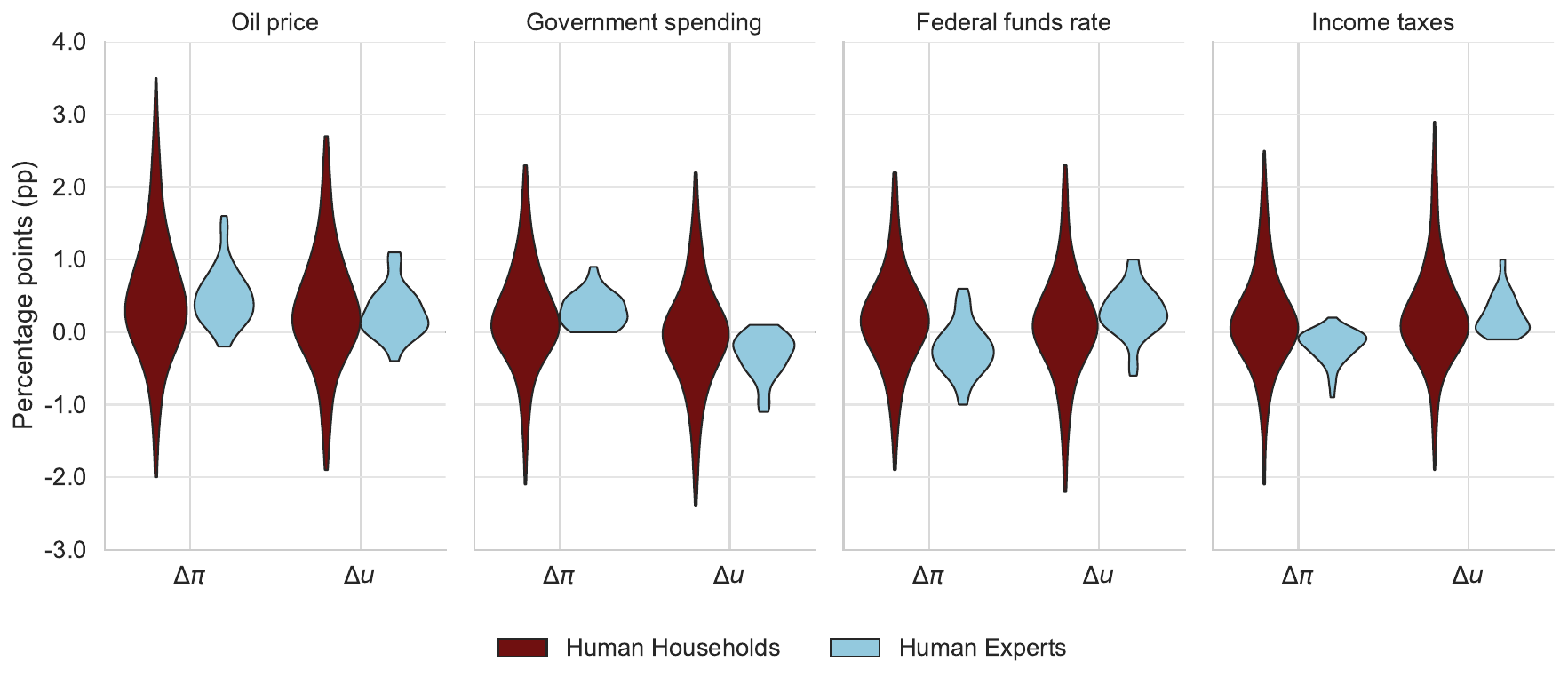}%
}\\[0.6em]
\subcaptionbox{LLM Agents\label{fig:sub-a-7-b}}{%
  \includegraphics[width=\linewidth,keepaspectratio]{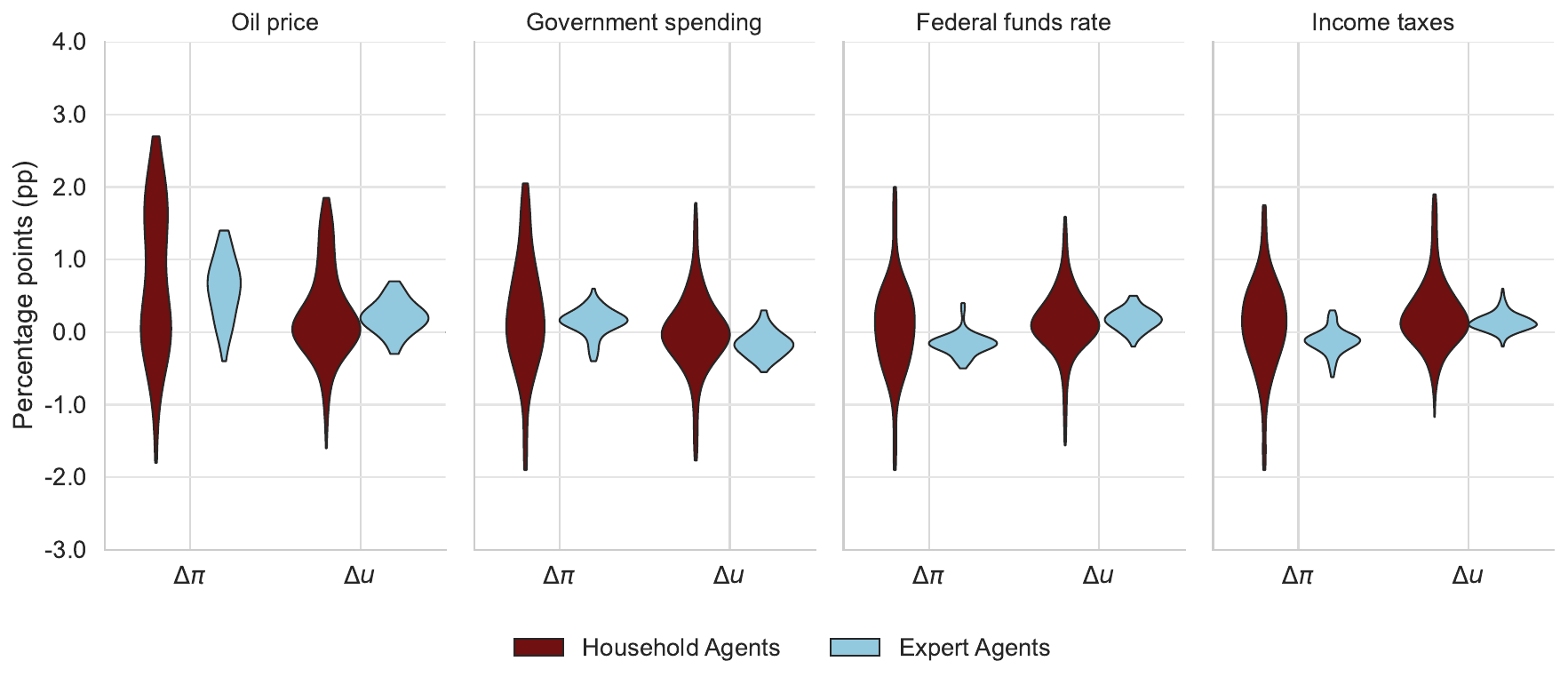}%
}
\caption{Forecast distributions of the quantitative effects of macroeconomic shocks (humans vs. LLM Agents) in Experiment 1}\label{fig:a-7}
\notepar{%
Notes: This figure presents the forecast distributions (with trimmed 5\% tails) of the quantitative effects of macroeconomic shocks on the inflation rate (\(\Delta\ \pi\)) and the unemployment rate (\(\Delta\ u\)) by humans (households and experts) and LLM Agents (Household Agents and Expert Agents), using kernel density estimators. This figure aggregates forecasts for the ``rise'' and ``fall'' scenarios, with fall predictions reversed to be comparable to rise predictions.
}
\end{figure}

\begin{figure}[!htbp]
\centering
\includegraphics[width=\linewidth,keepaspectratio]{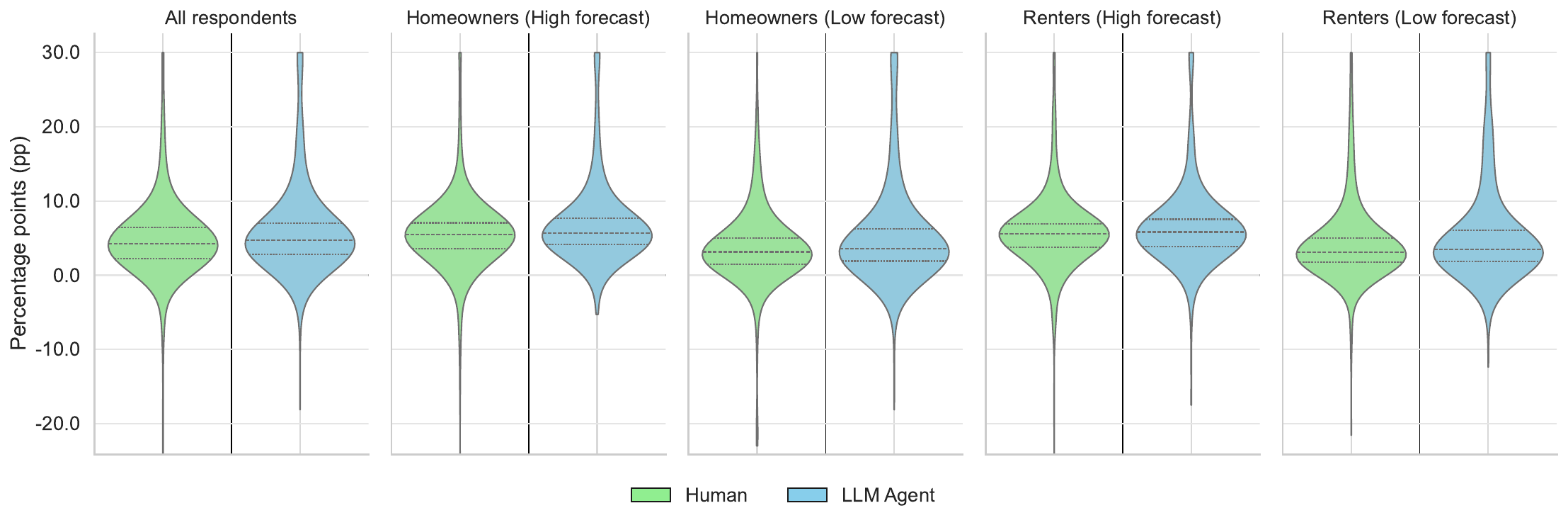}
\caption{Distributions of home price expectations generated by LLM Agents and humans in Sub-Experiment 1 of Experiment 2}\label{fig:a-8}
\notepar{%
Notes: This figure presents distributions of home price expectations generated by LLM Agents (Homeowner Agents and Renter Agents) and humans (homeowners and renters) in different treatment groups, using kernel density estimators. The dashed lines in each violin plot represent the quartiles of each distribution.
}
\end{figure}

\begin{figure}[!htbp]
\centering
\includegraphics[width=\linewidth,keepaspectratio]{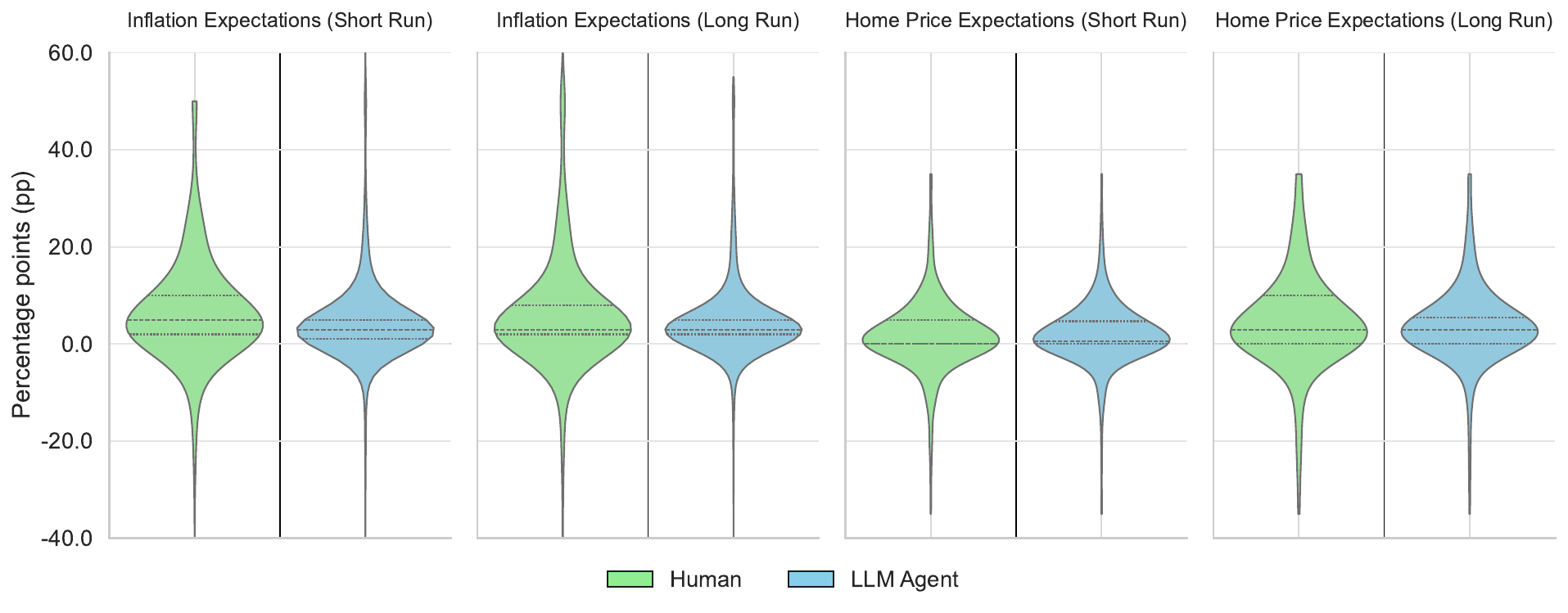}
\caption{Expectation distributions generated by LLM Agents and humans in Experiment 3}\label{fig:a-9}
\notepar{%
Notes: This figure presents distributions of long- and short-term inflation expectations and home price expectations generated by Household Agents and households in 2025 Michigan Surveys of Consumers, using kernel density estimators. The dashed lines in each violin plot represent the quartiles of each distribution.
}
\end{figure}

\begin{figure}[!htbp]
\centering
\includegraphics[width=\linewidth,keepaspectratio]{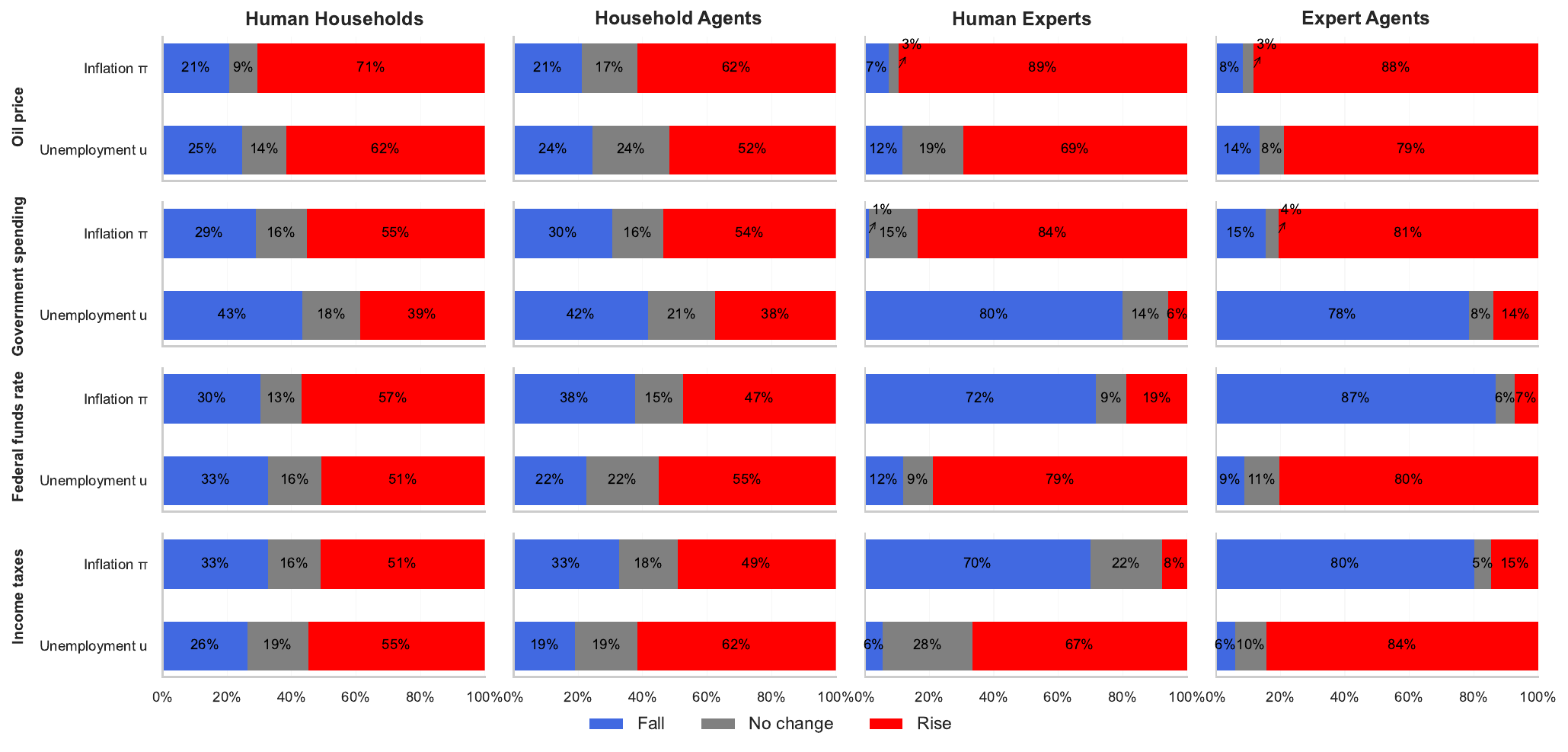}
\caption{Forecasts of the directional effects of macroeconomic shocks (humans v.s. LLM Agents) in Experiment 1}\label{fig:a-6}
\notepar{%
Notes: This figure shows the forecasts of the directional effects of macroeconomic shocks on the inflation rate and the unemployment rate by humans (households and experts) and LLM Agents (Household Agents and Expert Agents), using percentage bar charts. Predictions in the ``fall'' scenarios are reversed to make them comparable to rise predictions.
}
\end{figure}

\begin{figure}[!htbp]
\centering
\includegraphics[width=\linewidth,keepaspectratio]{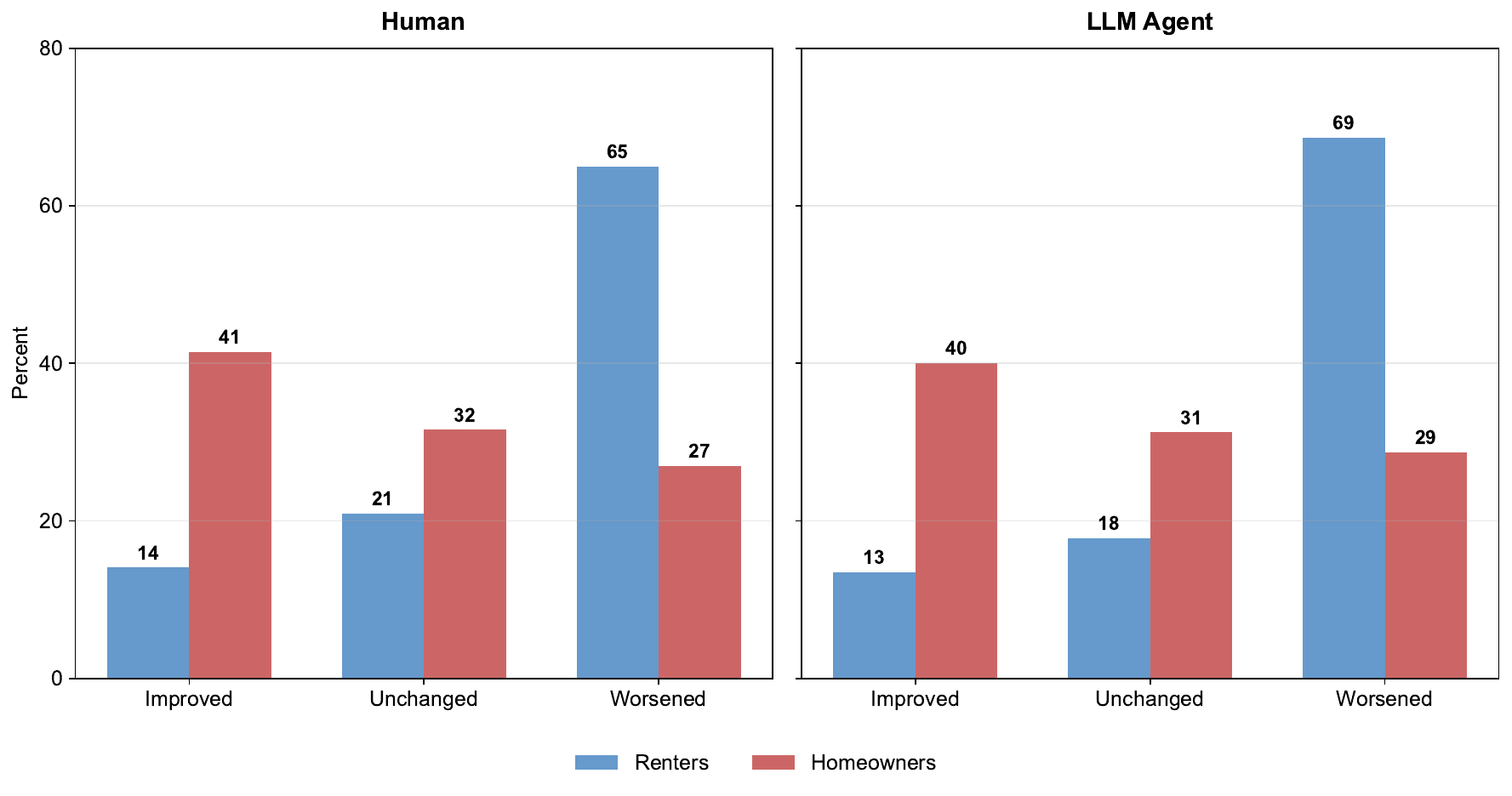}
\caption{Comparison of LLM Agents' simulated results with human data in Sub-Experiment 2 of Experiment 2}\label{fig:6}
\notepar{%
Notes: This figure compares the changes in expectations about their household's future economic situation generated by LLM Agents with those generated by humans in Sub-Experiment 2 of the randomized information experiment. The left panel presents the responses from human participants, while the right panel displays the simulation results from Homeowner Agents. The horizontal axis represents the three possible directions of changes in expectations (improved, unchanged, worsened), and the vertical axis indicates the percentage of respondents selecting each direction.
}
\end{figure}

\begin{figure}[!htbp]
\centering
\includegraphics[width=\linewidth,keepaspectratio]{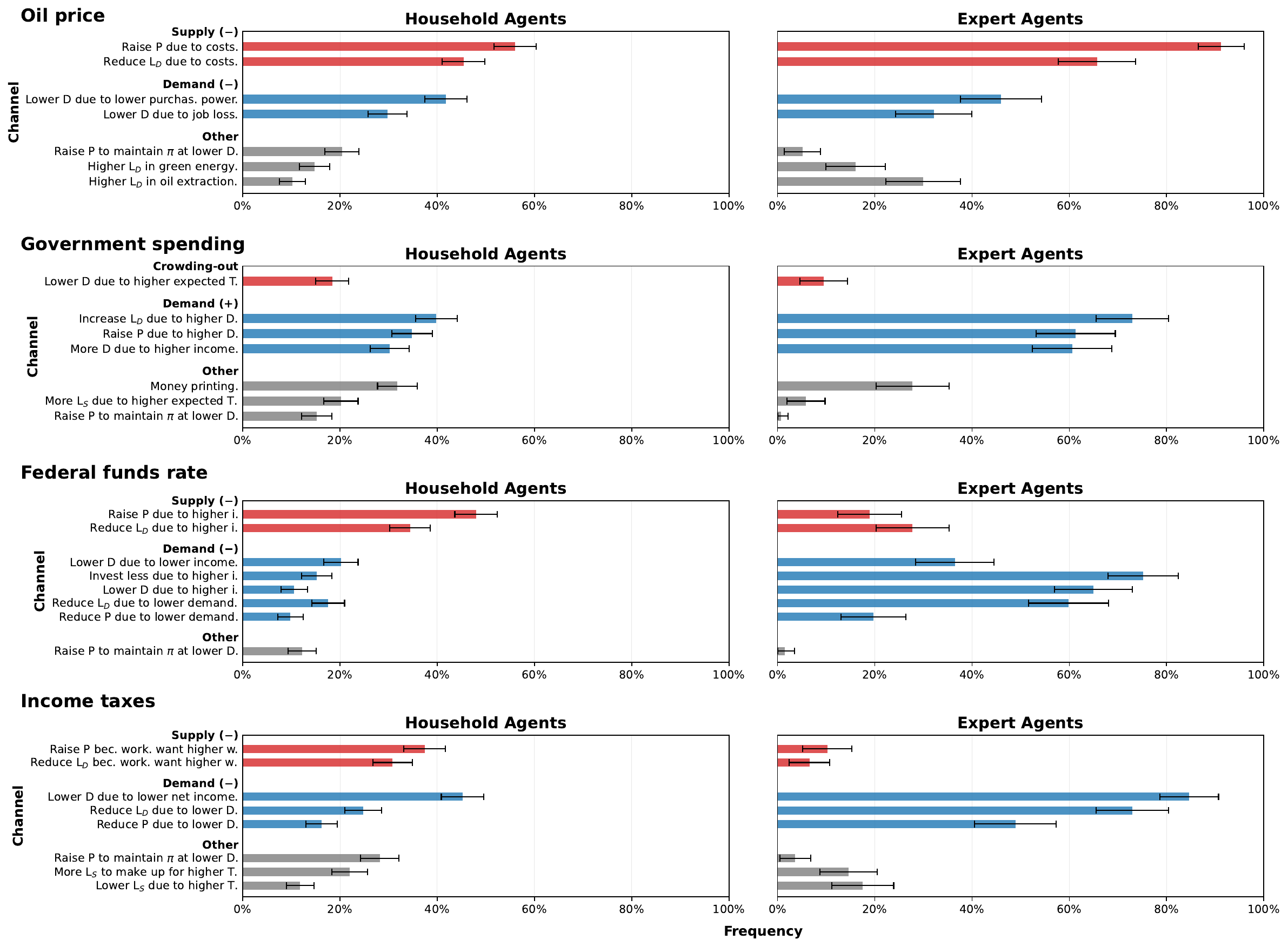}
\caption{LLM Agents' thoughts of propagation channels}\label{fig:a-10}
\notepar{%
Notes: This figure shows which propagation channels are selected by LLM Agents when they make their predictions. LLM Agents can select the channels from a list. The results are displayed separately for each vignette and for Household Agents (left panel) and Expert Agents (right panel). Error bars display 95\% confidence intervals. P abbreviates ``firm prices,'' L\emph{\textsubscript{D}} ``labor demand,'' D ``product demand,'' \(\pi\) ``firm profits,'' T ``taxes,'' i ``interest rates,'' w ``wages,'' and L\emph{\textsubscript{S}} ``labor supply.'' The format of this figure is consistent with Figure 3 in \citet{andre2022subjective}, making them comparable.
}
\end{figure}

\begin{figure}[!htbp]
\centering
\includegraphics[width=\linewidth,keepaspectratio]{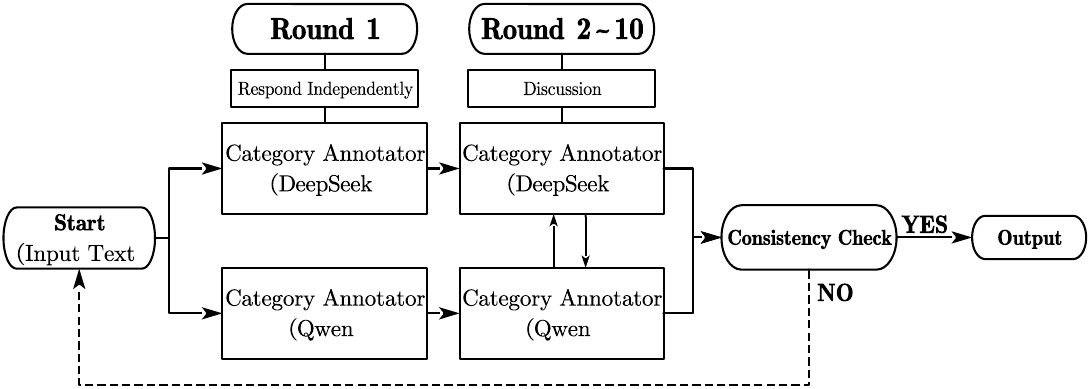}
\caption{Architecture of an agentic workflow for classifying open-ended responses}\label{fig:a-11}
\notepar{%
Notes: This figure illustrates the architecture of an agentic workflow for classifying or labeling open-ended responses. In Round 1, two distinct medium-scale LLMs (deepseek-r1-distill-qwen-32b and qwq-32b) serve as Category Annotators, independently labeling input text based on predefined criteria. After Round 1, the two Annotators discuss their initial results. When their outputs agree, the result passes consistency check and is output. If discrepancies arise, they engage in multiple discussion rounds until consensus is reached. Should no agreement be achieved by Round 10, the entire process repeats until consistent outputs are attained.
}
\end{figure}

\begin{figure}[!htbp]
\centering
\includegraphics[width=\linewidth,keepaspectratio]{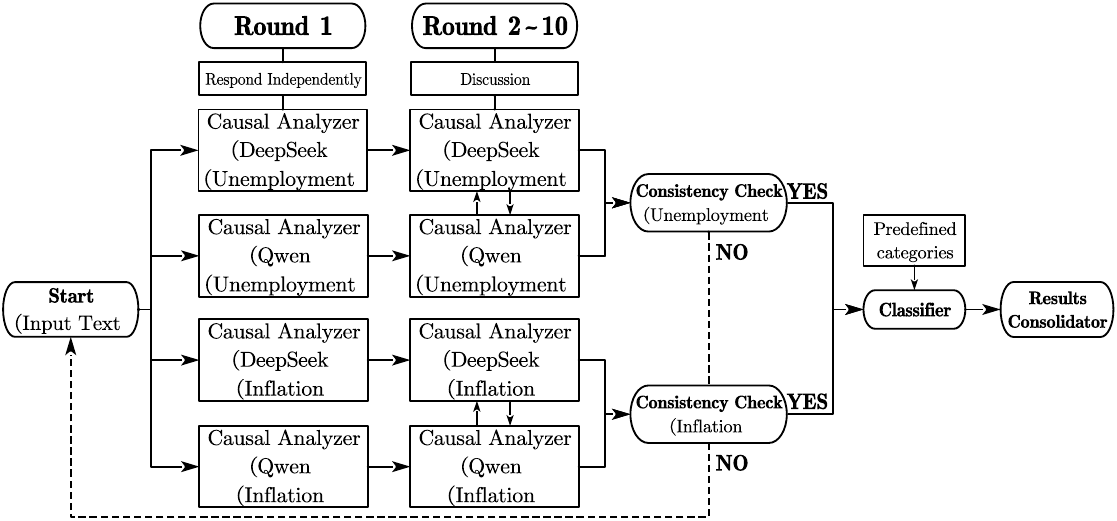}
\caption{Architecture of an agentic workflow for identifying DAGs in open-ended responses}\label{fig:a-12}
\notepar{%
Notes: This figure illustrates the architecture of an agentic workflow designed to identify Directed Acyclic Graphs (DAGs) in open-ended responses. Two medium-scale LLMs of different types---deepseek-r1-distill-qwen-32b and qwq-32b---serve as Causal Analyzers, each independently identifying causal pathways in open-text responses related to unemployment and inflation expectations. After Round 1, the two Analyzers discuss their initial results. If their outputs agree, they pass the consistency check; if not, they engage in multiple discussion rounds until consensus is reached. If no agreement is achieved after 10 rounds, the entire process is repeated. A Classifier then categorizes the intermediate variables in the causal pathways according to predefined categories. Finally, a Results Consolidator consolidates the categorized results. Further details are provided in Section~\ref{sec:c-1}.
}
\end{figure}

\begin{figure}[!htbp]
\centering
\includegraphics[width=\linewidth,keepaspectratio]{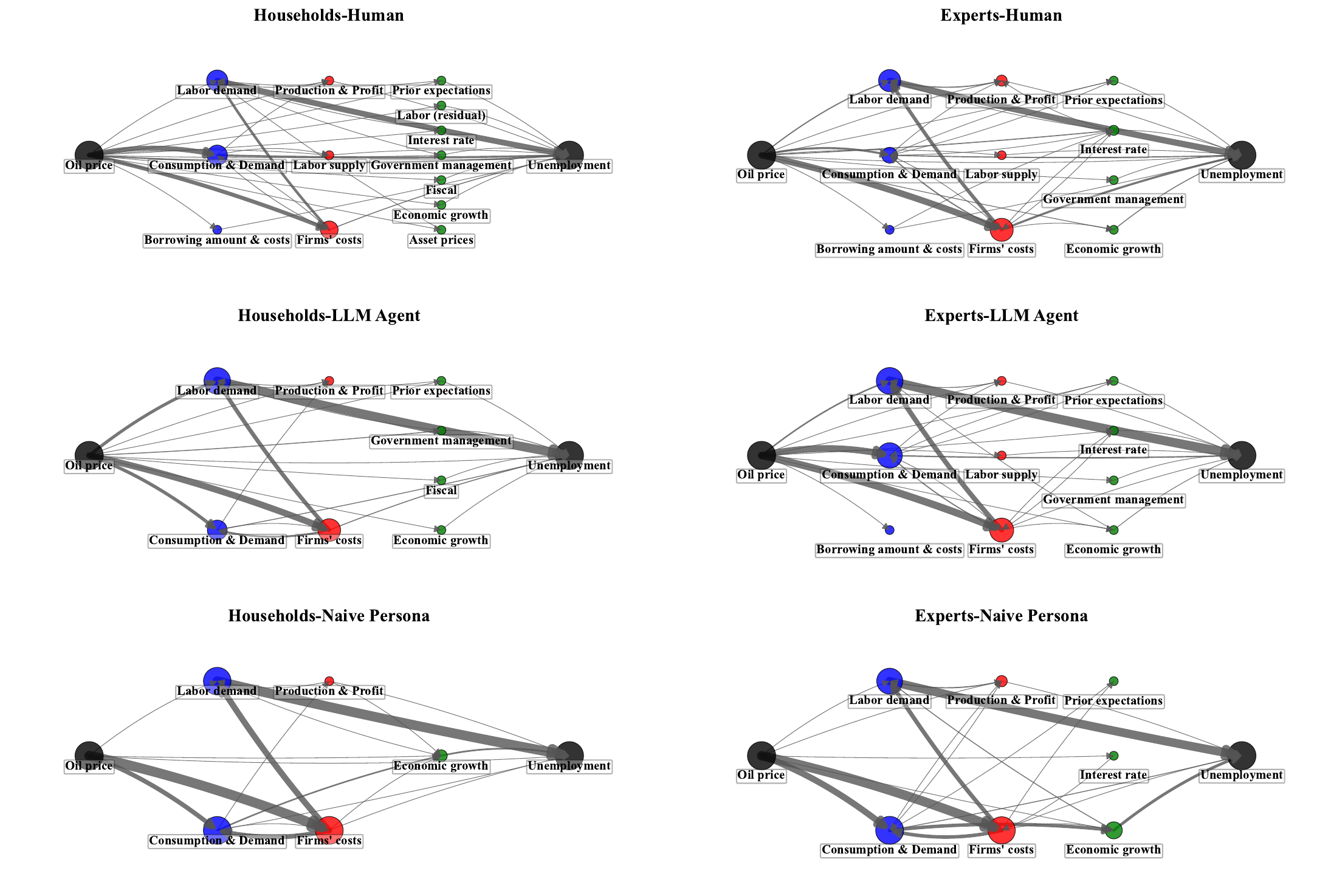}
\caption{The ``average'' DAGs underlying the formation of unemployment expectations in the oil price vignette}\label{fig:a-13}
\notepar{%
Notes: The figure presents the ``average'' DAGs underlying unemployment expectation formation for humans (i.e., ``Households-Human'' and ``Experts-Human''), LLM Agents (i.e., ``Households-LLM Agent'' and ``Experts-LLM Agent''), and foundation models with only naive personas (i.e., ``Households-Naive Persona'' and ``Experts-Naive Persona'') in the oil price vignette. The nodes represent categories of intermediate variables, whose definitions and classifications are provided in Supplementary Appendix Table~\ref{tab:a-3}. The aggregated DAGs reveal the most relevant variables (nodes) and causal links in the responses of humans, LLM Agents, and Naive Persona. Node size: The size of the nodes is proportional to the share of responses that refer to the nodes. Node color: Red indicates supply-side variables, blue indicates demand-side variables, green indicates miscellaneous variables, black is used for start and end nodes. Edge thickness: The thickness of the edges is proportional to the share of responses that refer to the causal connections (among humans, LLM Agents and Naive Persona, respectively).
}
\end{figure}

\begin{figure}[!htbp]
\centering
\includegraphics[width=\linewidth,keepaspectratio]{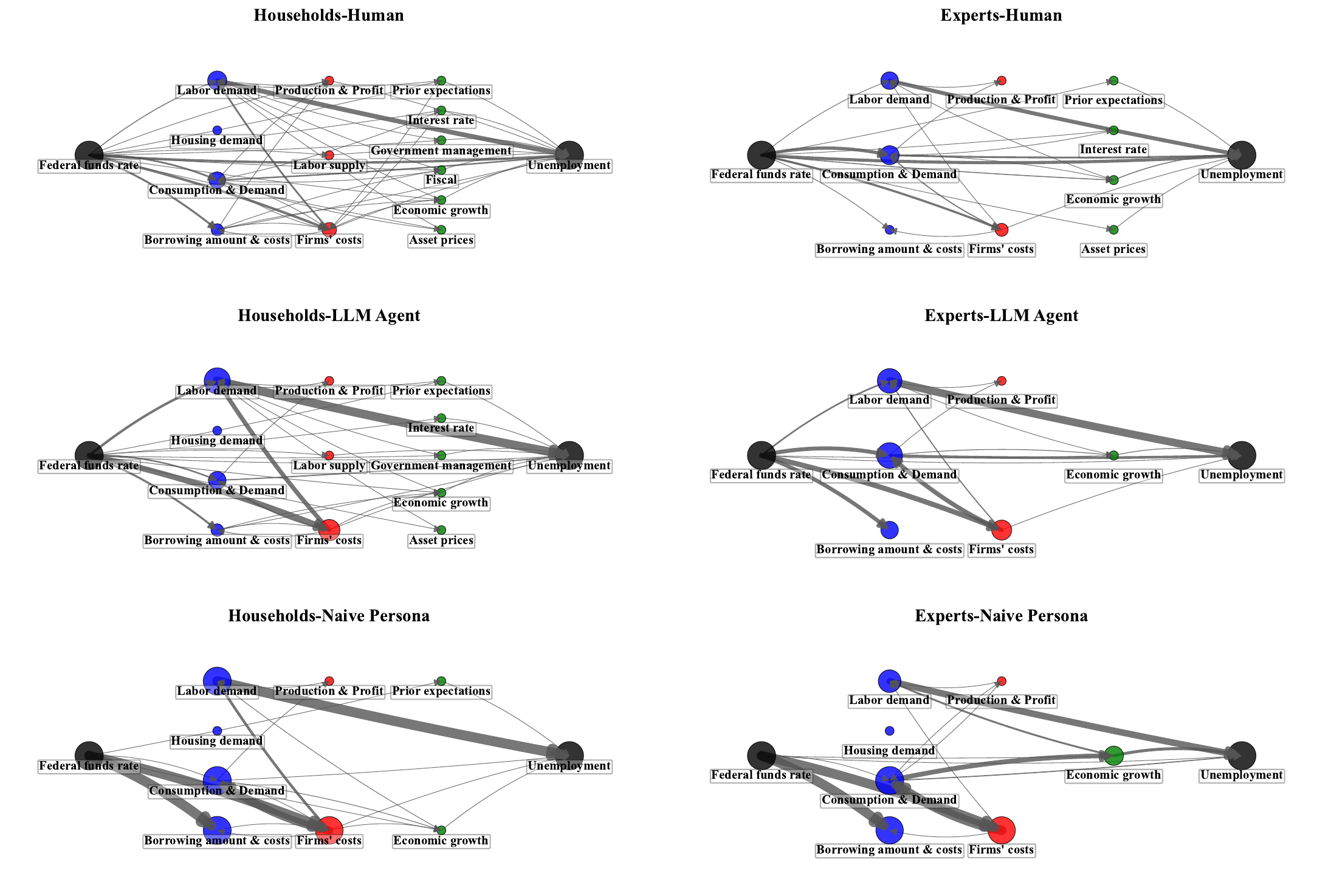}
\caption{The ``average'' DAGs underlying the formation of unemployment expectations in the interest rate vignette}\label{fig:a-14}
\notepar{%
Notes: The figure presents the ``average'' DAGs underlying unemployment expectation formation for humans (i.e., ``Households-Human'' and ``Experts-Human''), LLM Agents (i.e., ``Households-LLM Agent'' and ``Experts-LLM Agent''), and foundation models with only naive personas (i.e., ``Households-Naive Persona'' and ``Experts-Naive Persona'') in the interest rate vignette. The nodes represent categories of intermediate variables, whose definitions and classifications are provided in Supplementary Appendix Table~\ref{tab:a-3}. The aggregated DAGs reveal the most relevant variables (nodes) and causal links in the responses of humans, LLM Agents, and Naive Persona. Node size: The size of the nodes is proportional to the share of responses that refer to the nodes. Node color: Red indicates supply-side variables, blue indicates demand-side variables, green indicates miscellaneous variables, black is used for start and end nodes. Edge thickness: The thickness of the edges is proportional to the share of responses that refer to the causal connections (among humans, LLM Agents and Naive Persona, respectively).
}
\end{figure}

\begin{figure}[!htbp]
\centering
\includegraphics[width=\linewidth,keepaspectratio]{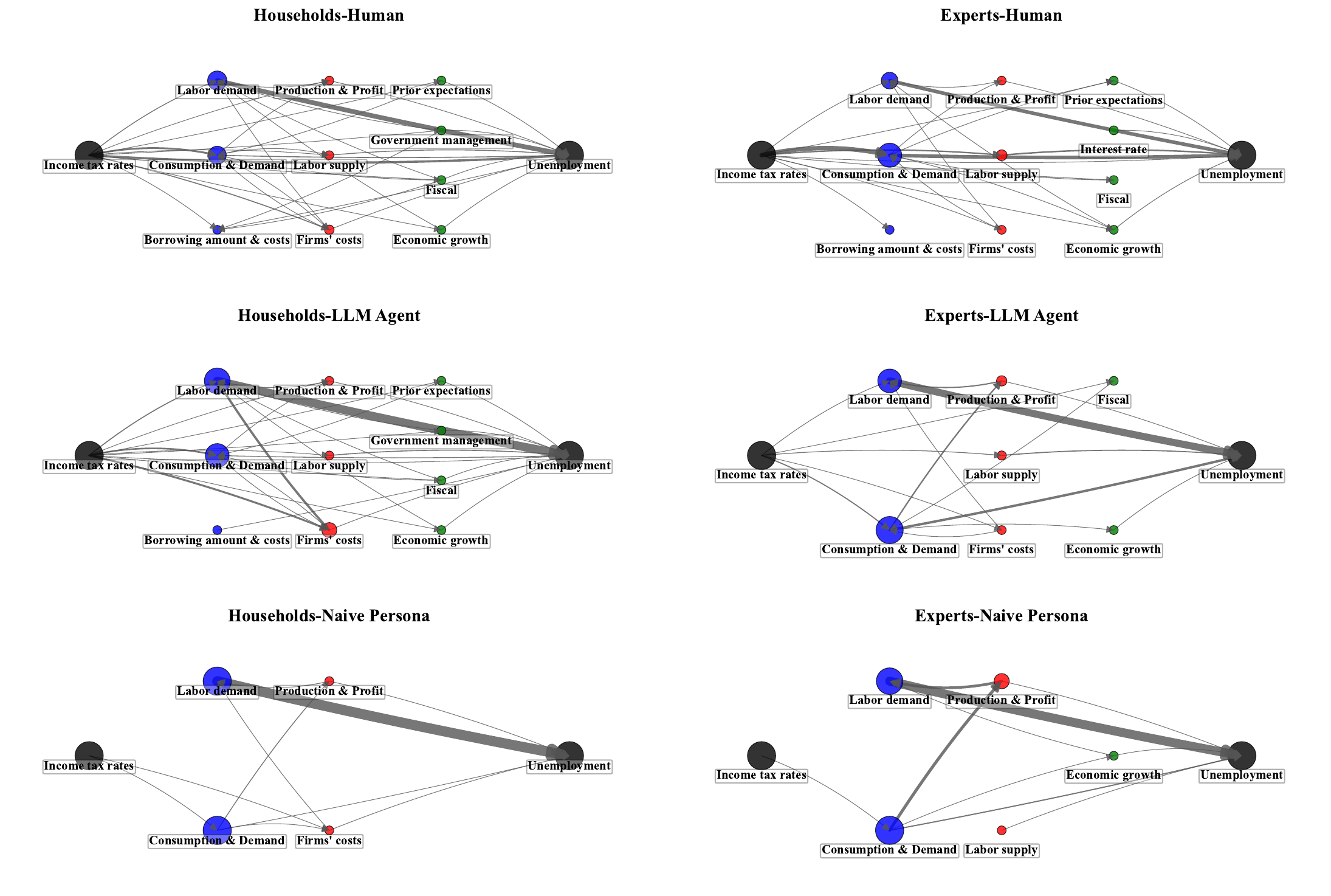}
\caption{The ``average'' DAGs underlying the formation of unemployment expectations in the taxation vignette}\label{fig:a-15}
\notepar{%
Notes: The figure presents the ``average'' DAGs underlying unemployment expectation formation for humans (i.e., ``Households-Human'' and ``Experts-Human''), LLM Agents (i.e., ``Households-LLM Agent'' and ``Experts-LLM Agent''), and foundation models with only naive personas (i.e., ``Households-Naive Persona'' and ``Experts-Naive Persona'') in the taxation vignette. The nodes represent categories of intermediate variables, whose definitions and classifications are provided in Supplementary Appendix Table~\ref{tab:a-3}. The aggregated DAGs reveal the most relevant variables (nodes) and causal links in the responses of humans, LLM Agents, and Naive Persona. Node size: The size of the nodes is proportional to the share of responses that refer to the nodes. Node color: Red indicates supply-side variables, blue indicates demand-side variables, green indicates miscellaneous variables, black is used for start and end nodes. Edge thickness: The thickness of the edges is proportional to the share of responses that refer to the causal connections (among humans, LLM Agents and Naive Persona, respectively).
}
\end{figure}

\begin{figure}[!htbp]
\centering
\includegraphics[width=\linewidth,keepaspectratio]{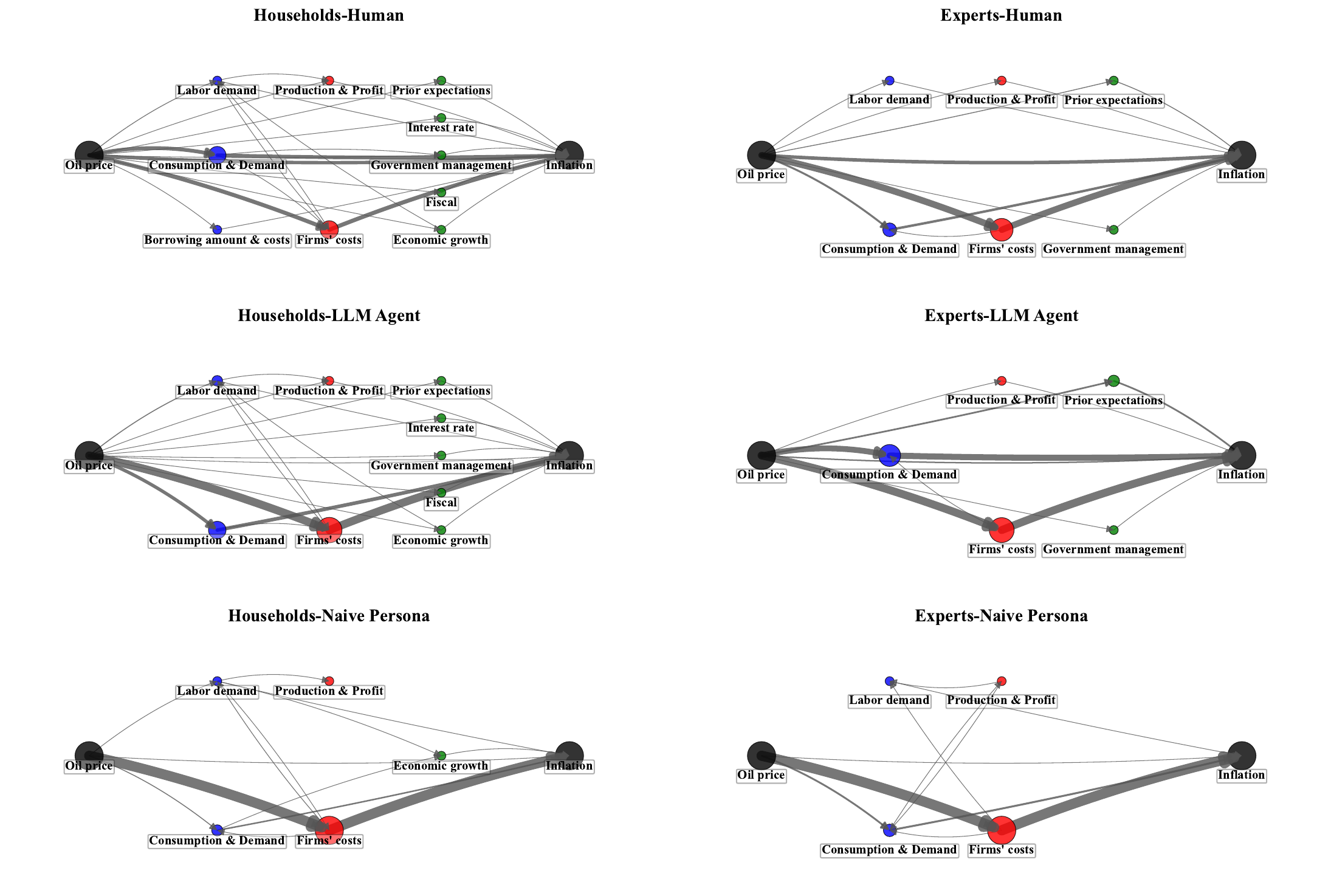}
\caption{The ``average'' DAGs underlying the formation of inflation expectations in the oil price vignette}\label{fig:a-16}
\notepar{%
Notes: The figure presents the ``average'' DAGs underlying inflation expectation formation for humans (i.e., ``Households-Human'' and ``Experts-Human''), LLM Agents (i.e., ``Households-LLM Agent'' and ``Experts-LLM Agent''), and foundation models with only naive personas (i.e., ``Households-Naive Persona'' and ``Experts-Naive Persona'') in the oil price vignette. The nodes represent categories of intermediate variables, whose definitions and classifications are provided in Supplementary Appendix Table~\ref{tab:a-3}. The aggregated DAGs reveal the most relevant variables (nodes) and causal links in the responses of humans, LLM Agents, and Naive Persona. Node size: The size of the nodes is proportional to the share of responses that refer to the nodes. Node color: Red indicates supply-side variables, blue indicates demand-side variables, green indicates miscellaneous variables, black is used for start and end nodes. Edge thickness: The thickness of the edges is proportional to the share of responses that refer to the causal connections (among humans, LLM Agents and Naive Persona, respectively).
}
\end{figure}

\begin{figure}[!htbp]
\centering
\includegraphics[width=\linewidth,keepaspectratio]{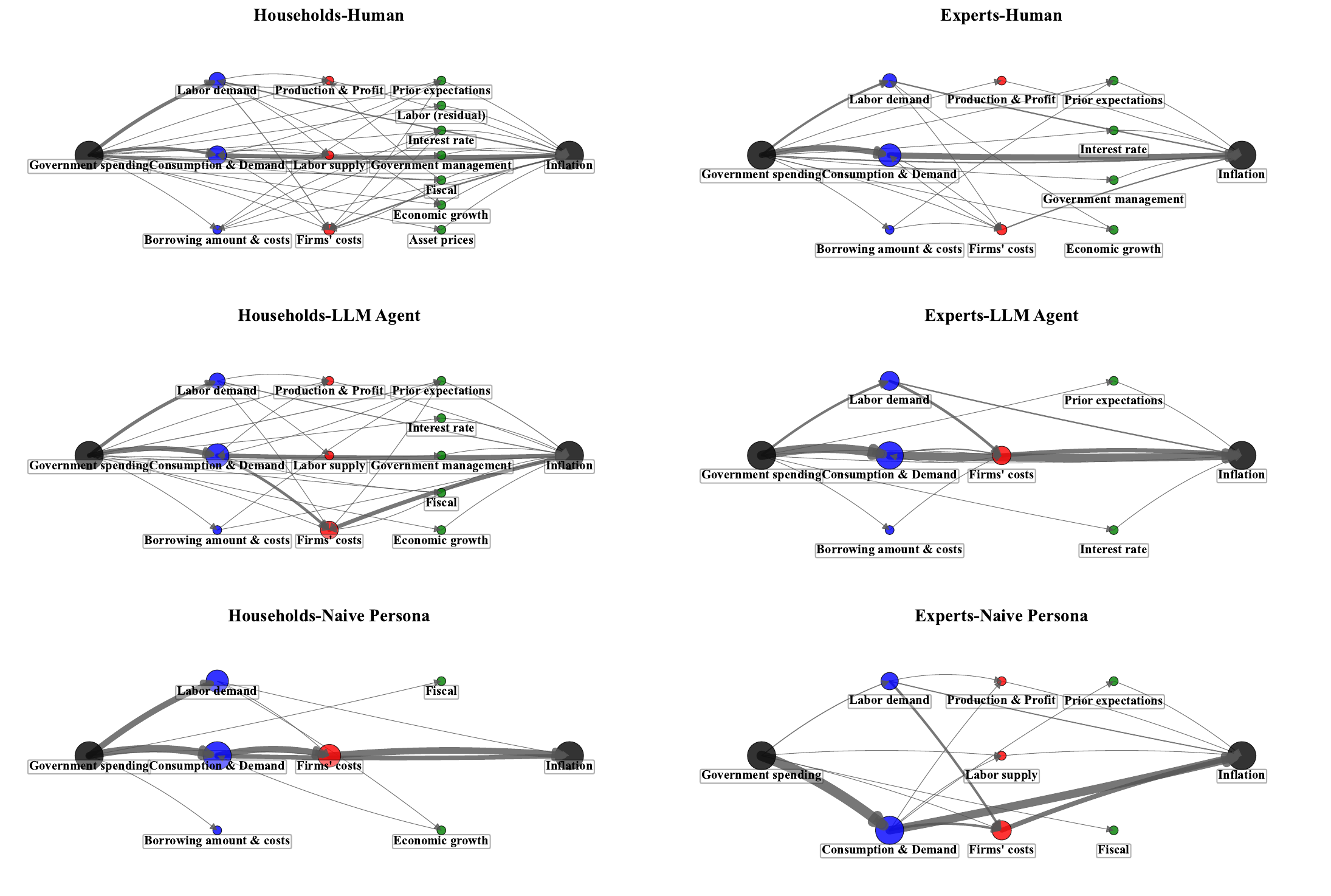}
\caption{The ``average'' DAGs underlying the formation of inflation expectations in the government spending vignette}\label{fig:a-17}
\notepar{%
Notes: The figure presents the ``average'' DAGs underlying inflation expectation formation for humans (i.e., ``Households-Human'' and ``Experts-Human''), LLM Agents (i.e., ``Households-LLM Agent'' and ``Experts-LLM Agent''), and foundation models with only naive personas (i.e., ``Households-Naive Persona'' and ``Experts-Naive Persona'') in the government spending vignette. The nodes represent categories of intermediate variables, whose definitions and classifications are provided in Supplementary Appendix Table~\ref{tab:a-3}. The aggregated DAGs reveal the most relevant variables (nodes) and causal links in the responses of humans, LLM Agents, and Naive Persona. Node size: The size of the nodes is proportional to the share of responses that refer to the nodes. Node color: Red indicates supply-side variables, blue indicates demand-side variables, green indicates miscellaneous variables, black is used for start and end nodes. Edge thickness: The thickness of the edges is proportional to the share of responses that refer to the causal connections (among humans, LLM Agents and Naive Persona, respectively).
}
\end{figure}

\begin{figure}[!htbp]
\centering
\includegraphics[width=\linewidth,keepaspectratio]{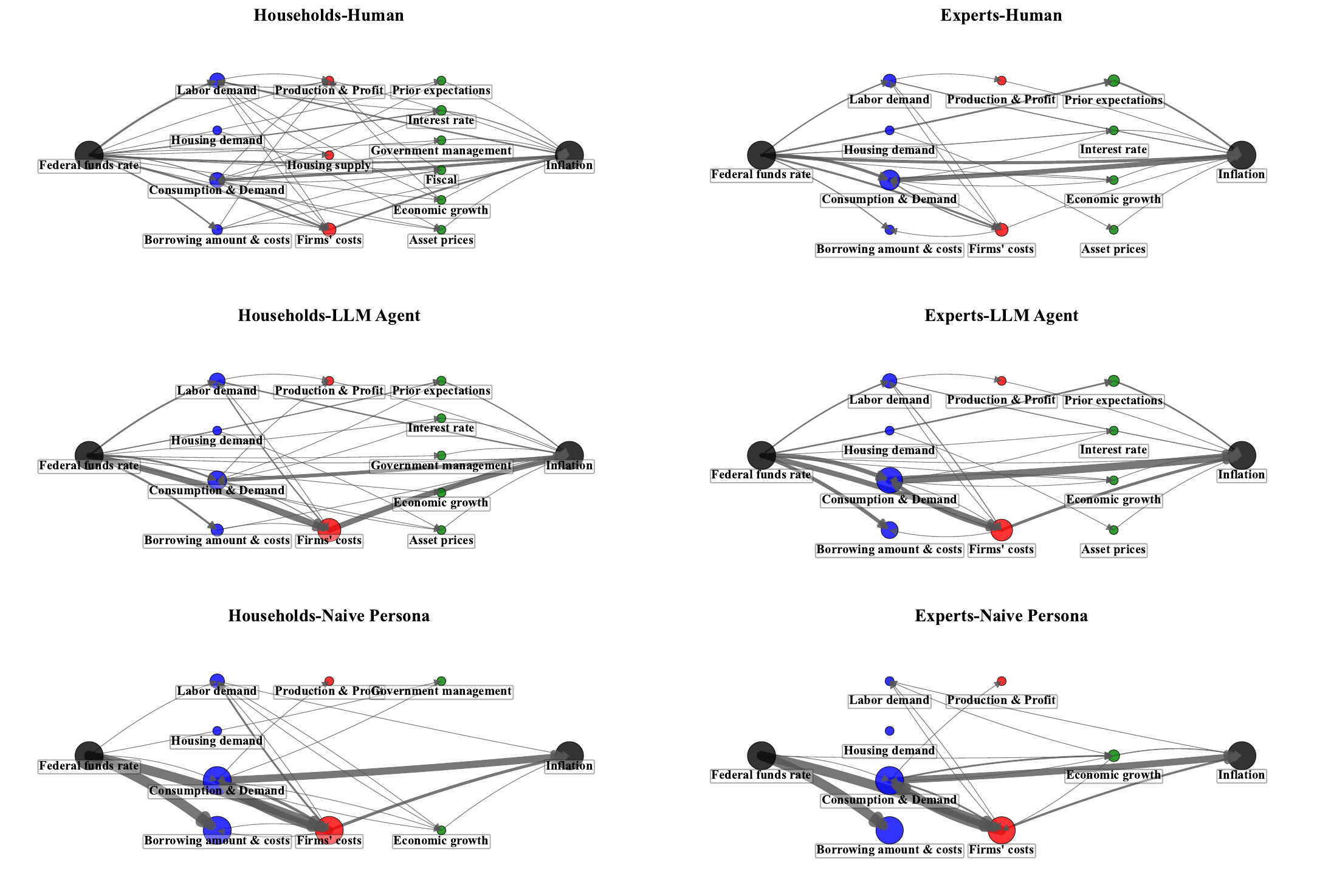}
\caption{The ``average'' DAGs underlying the formation of inflation expectations in the interest rate vignette}\label{fig:a-18}
\notepar{%
Notes: The figure presents the ``average'' DAGs underlying inflation expectation formation for humans (i.e., ``Households-Human'' and ``Experts-Human''), LLM Agents (i.e., ``Households-LLM Agent'' and ``Experts-LLM Agent''), and foundation models with only naive personas (i.e., ``Households-Naive Persona'' and ``Experts-Naive Persona'') in the interest rate vignette. The nodes represent categories of intermediate variables, whose definitions and classifications are provided in Supplementary Appendix Table~\ref{tab:a-3}. The aggregated DAGs reveal the most relevant variables (nodes) and causal links in the responses of humans, LLM Agents, and Naive Persona. Node size: The size of the nodes is proportional to the share of responses that refer to the nodes. Node color: Red indicates supply-side variables, blue indicates demand-side variables, green indicates miscellaneous variables, black is used for start and end nodes. Edge thickness: The thickness of the edges is proportional to the share of responses that refer to the causal connections (among humans, LLM Agents and Naive Persona, respectively).
}
\end{figure}

\begin{figure}[!htbp]
\centering
\includegraphics[width=\linewidth,keepaspectratio]{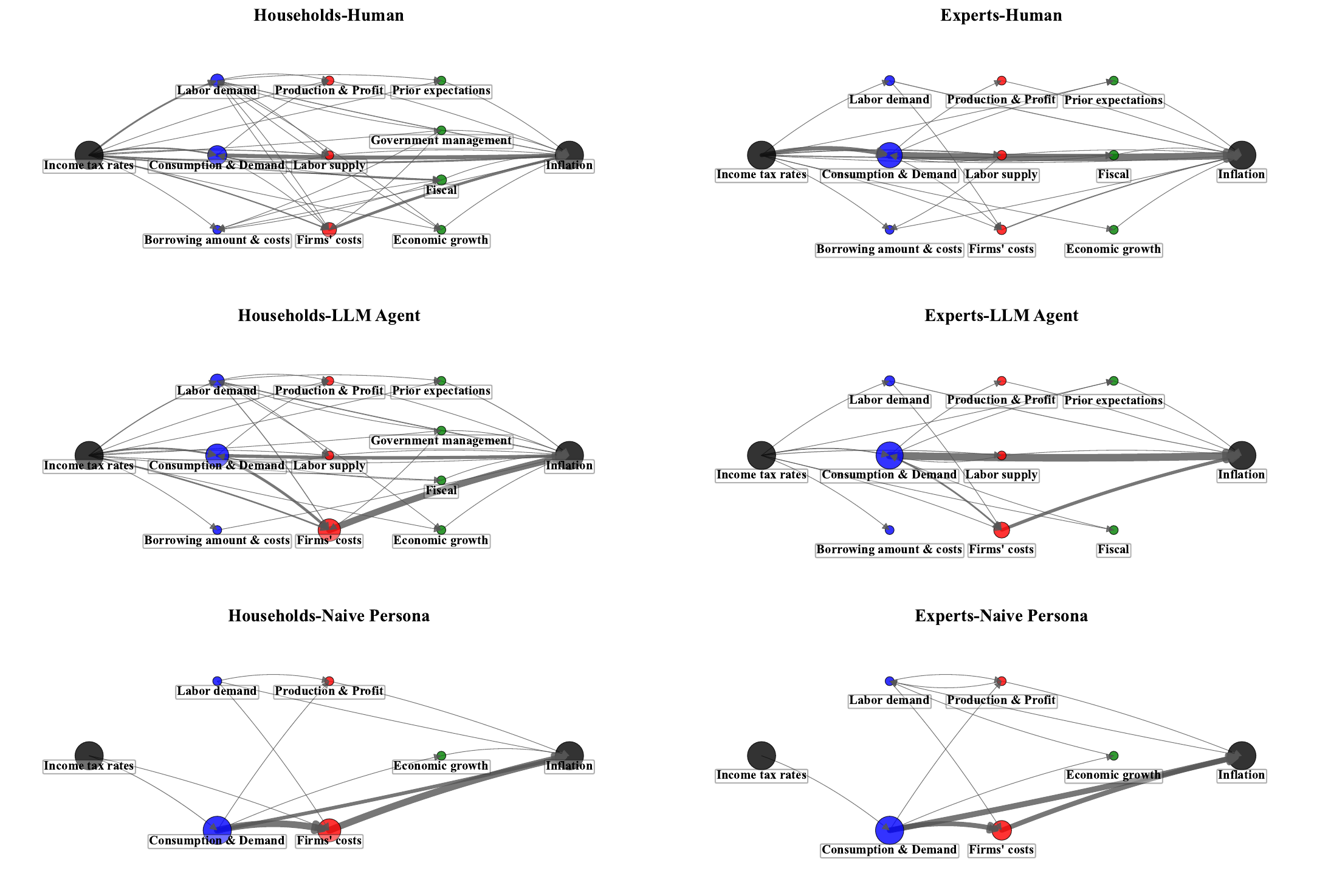}
\caption{The ``average'' DAGs underlying the formation of inflation expectations in the taxation vignette}\label{fig:a-19}
\notepar{%
Notes: The figure presents the ``average'' DAGs underlying inflation expectation formation for humans (i.e., ``Households-Human'' and ``Experts-Human''), LLM Agents (i.e., ``Households-LLM Agent'' and ``Experts-LLM Agent''), and foundation models with only naive personas (i.e., ``Households-Naive Persona'' and ``Experts-Naive Persona'') in the taxation vignette. The nodes represent categories of intermediate variables, whose definitions and classifications are provided in Supplementary Appendix Table~\ref{tab:a-3}. The aggregated DAGs reveal the most relevant variables (nodes) and causal links in the responses of humans, LLM Agents, and Naive Persona. Node size: The size of the nodes is proportional to the share of responses that refer to the nodes. Node color: Red indicates supply-side variables, blue indicates demand-side variables, green indicates miscellaneous variables, black is used for start and end nodes. Edge thickness: The thickness of the edges is proportional to the share of responses that refer to the causal connections (among humans, LLM Agents and Naive Persona, respectively).
}
\end{figure}

\begin{figure}[!htbp]
\centering
\includegraphics[width=\linewidth,keepaspectratio]{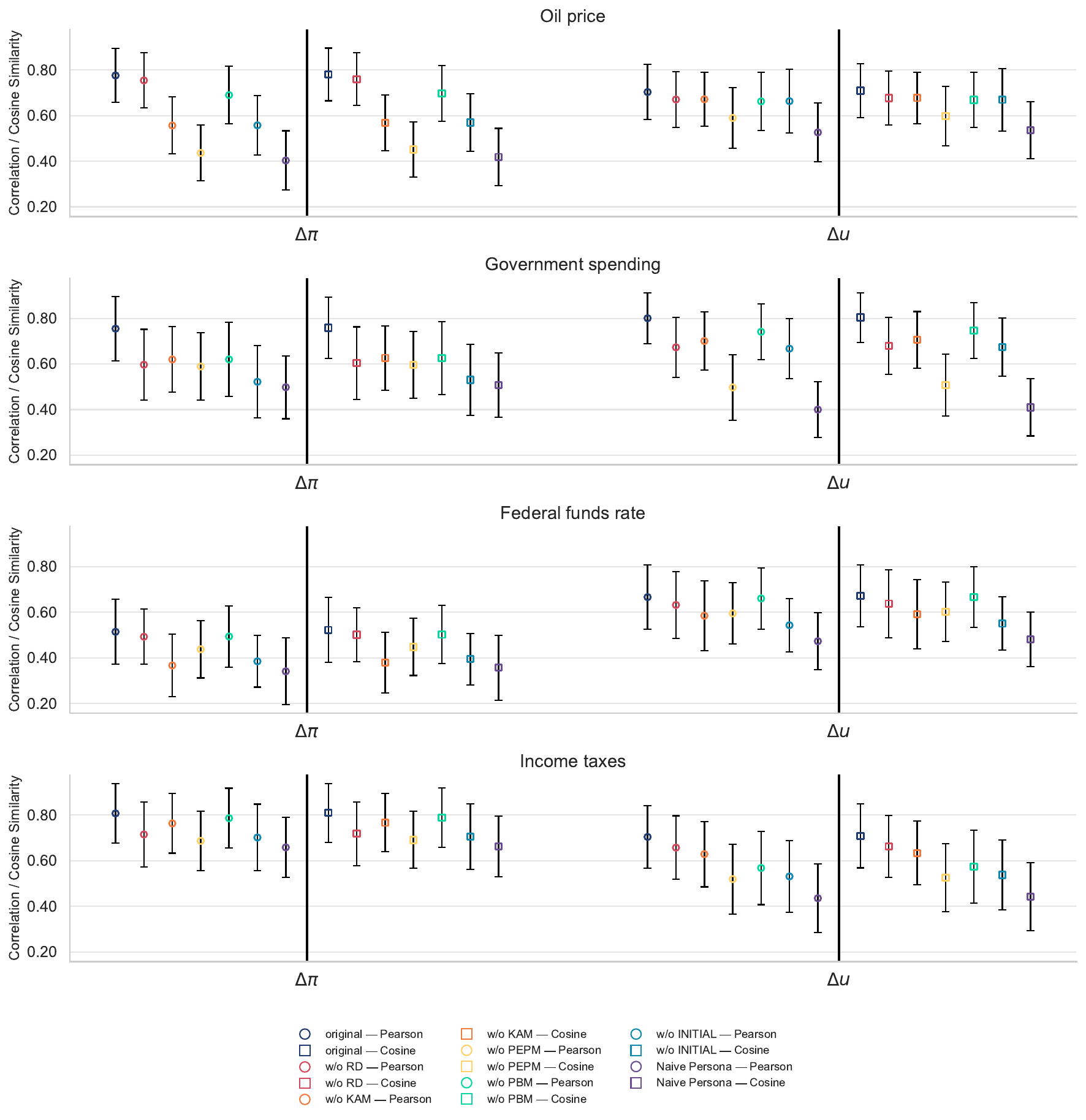}
\caption{Shape similarity between the expectation distributions generated by different Expert Agents and those generated by humans in Experiment 1}\label{fig:a-20}
\notepar{%
Notes: This figure displays the distributional shape similarity, as measured by Pearson correlation (displayed to the left of the bold vertical lines) and cosine similarity (displayed to the right of the bold vertical lines), between the changes in inflation expectations (\(\Delta\ \pi\)) and unemployment expectations (\(\Delta\ u\)) generated by Expert Agents (original and those without different components) and those of experts under four different vignettes. Error bars present two-sided 95\% confidence intervals for the similarity metrics, obtained by bootstrap over histogram-based probability vectors.
}
\end{figure}

\begin{figure}[!htbp]
\centering
\includegraphics[width=\linewidth,keepaspectratio]{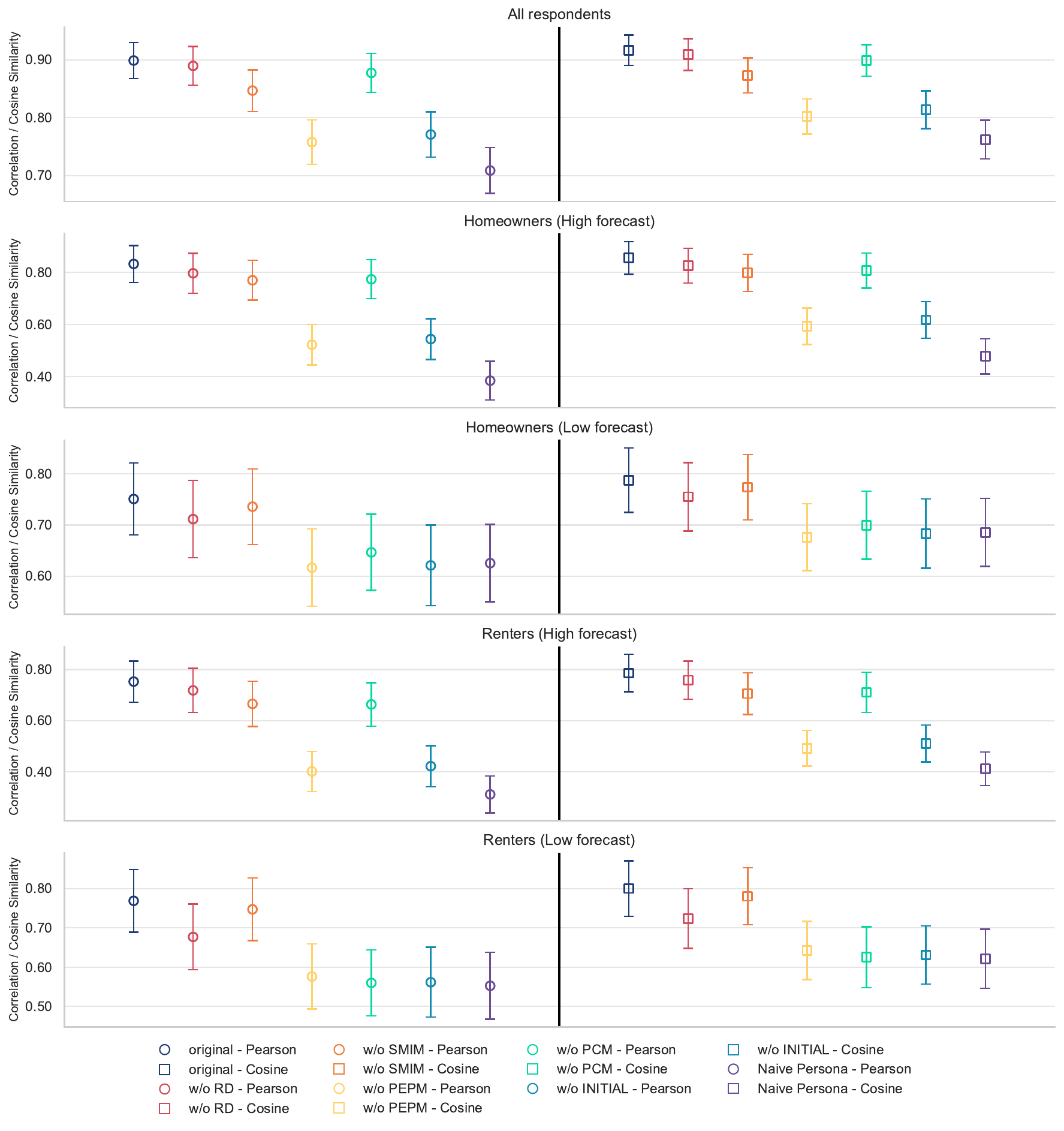}
\caption{Shape similarity between the expectation distributions generated by different LLM Agents and those generated by humans in Sub-Experiment 1 of Experiment 2}\label{fig:a-21}
\notepar{%
Notes: This figure displays the distributional shape similarity, as measured by Pearson correlation (displayed to the left of the bold vertical lines) and cosine similarity (displayed to the right of the bold vertical lines), between the home price expectations generated by LLM Agents (original and those without different components) and those of humans in different treatment groups. Error bars present two-sided 95\% confidence intervals for the similarity metrics, obtained by bootstrap over histogram-based probability vectors.
}
\end{figure}

\begin{figure}[!htbp]
\centering
\includegraphics[width=\linewidth,keepaspectratio]{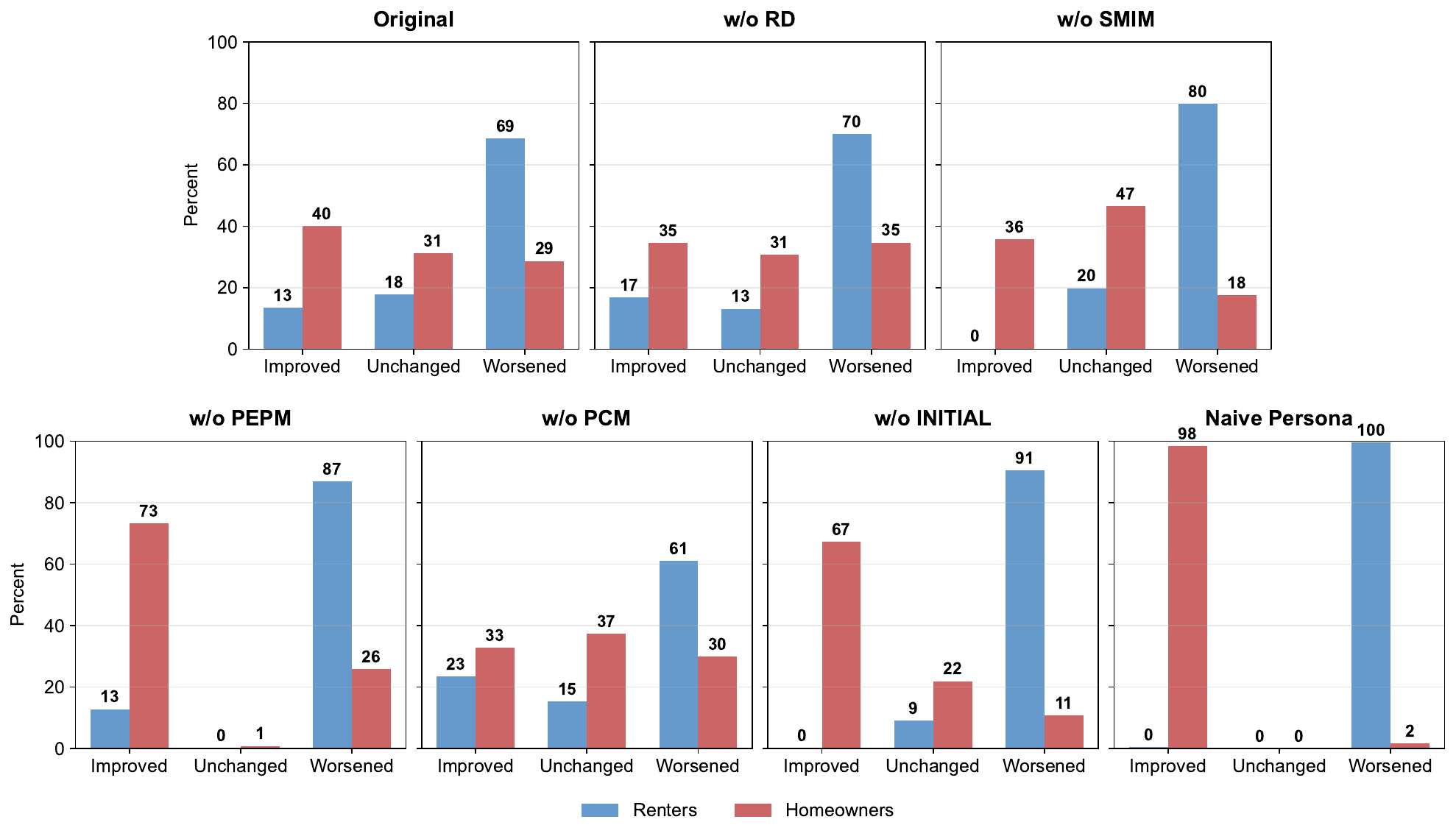}
\caption{Comparison of original LLM Agents' simulated results with those of LLM Agents without different components in Sub-Experiment 2 of Experiment 2}\label{fig:a-22}
\notepar{%
Notes: This figure compares the changes in expectations about their household's future economic situation generated by original LLM Agents with those generated by LLM Agents without different components in Sub-Experiment 2 of the randomized information experiment. The horizontal axis in each subplot represents the three possible directions of changes in expectations (improved, unchanged, worsened), and the vertical axis in each subplot indicates the percentage of respondents selecting each direction.
}
\end{figure}

\begin{figure}[!htbp]
\centering
\includegraphics[width=\linewidth,keepaspectratio]{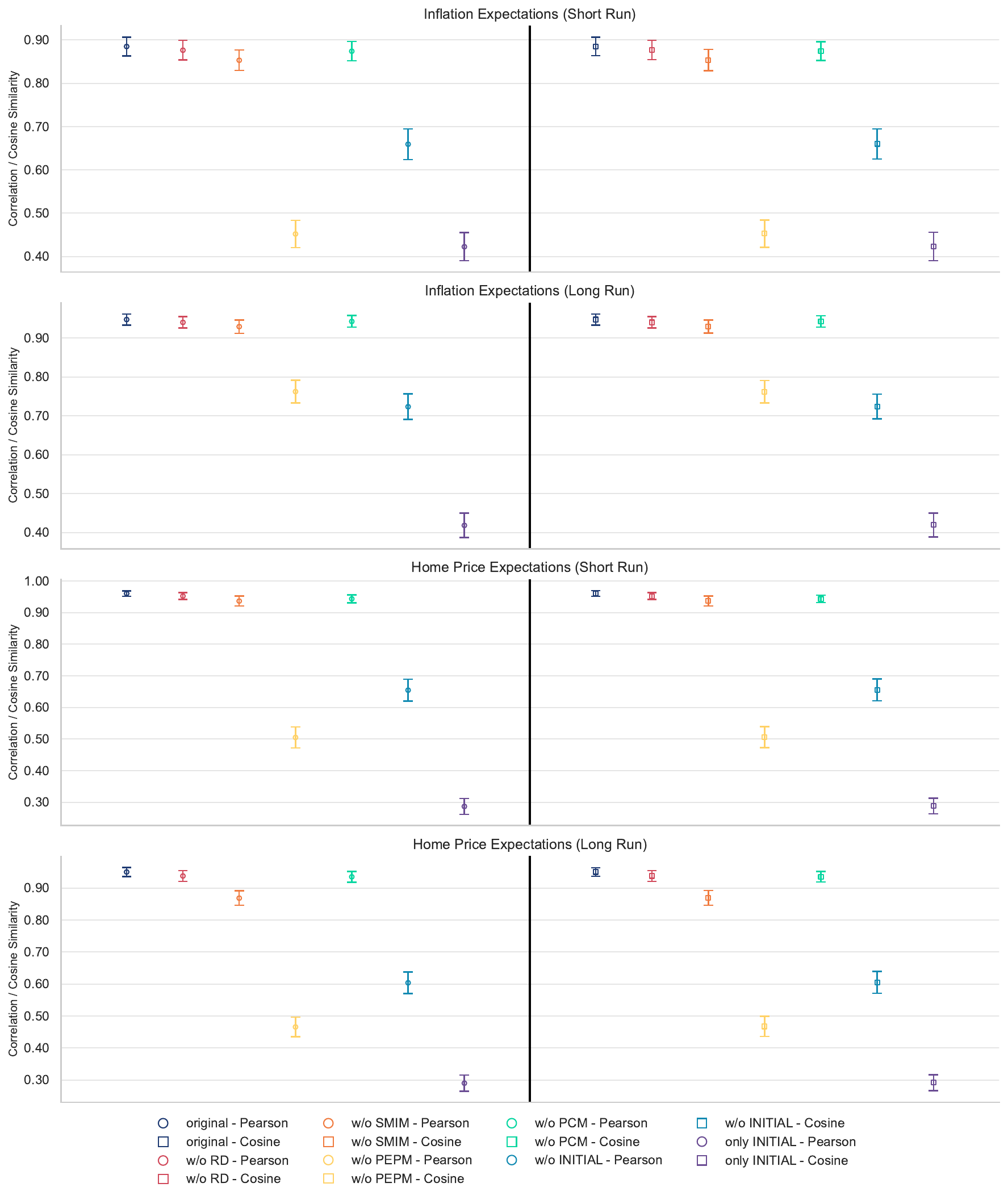}
\caption{Shape similarity between the expectation distributions generated by different LLM Agents and those generated by humans in Experiment 3}\label{fig:a-23}
\notepar{%
Notes: This figure displays the distributional shape similarity, as measured by Pearson correlation (displayed to the left of the bold vertical lines) and cosine similarity (displayed to the right of the bold vertical lines), between long- and short-term inflation expectations and home price expectations generated by LLM Agents (original and those without different components) and those of households in 2025 Michigan Surveys of Consumers. Error bars present two-sided 95\% confidence intervals for the similarity metrics, obtained by bootstrap over histogram-based probability vectors.
}
\end{figure}

\begin{figure}[!htbp]
\centering
\includegraphics[width=\linewidth,keepaspectratio]{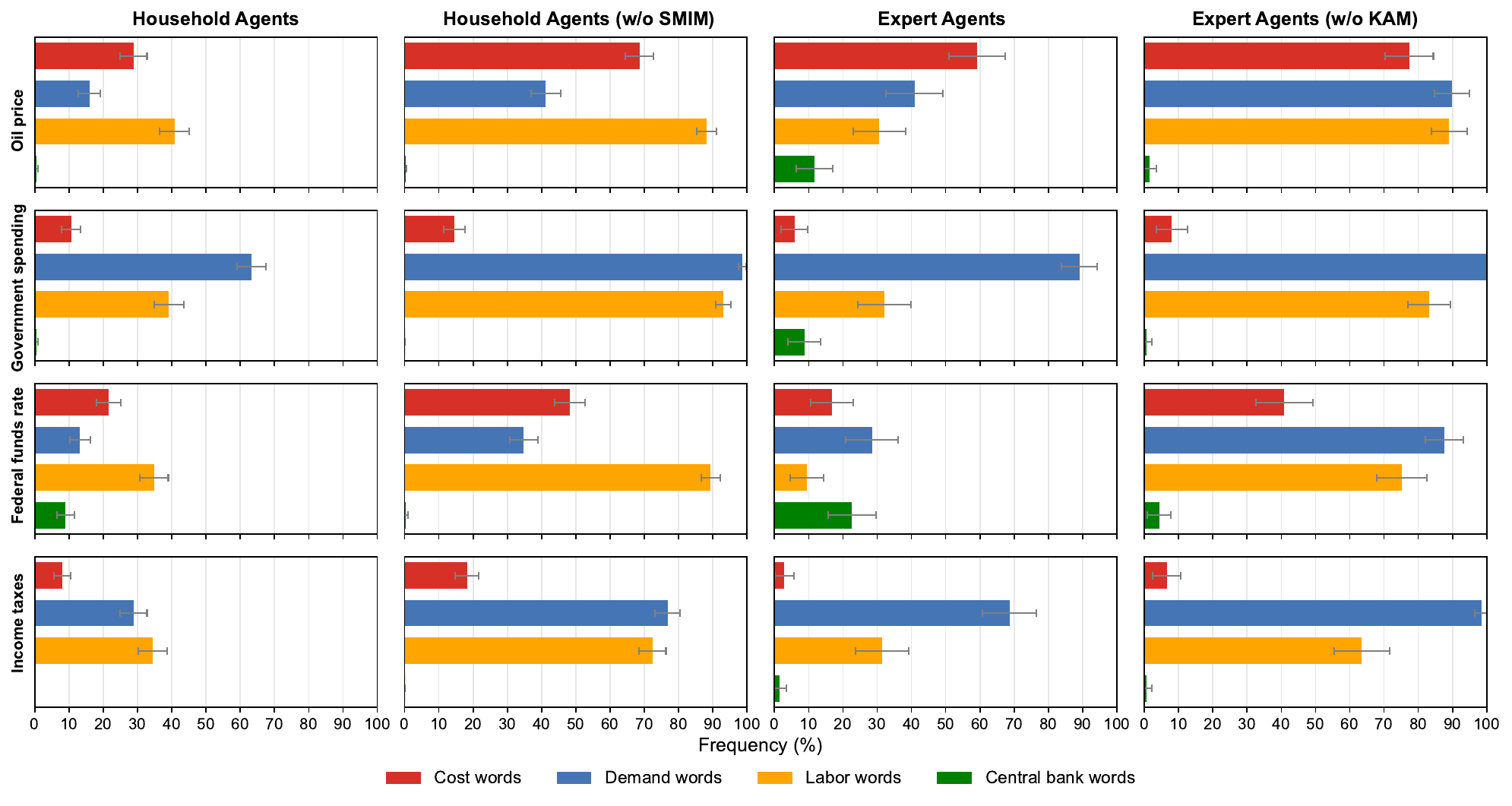}
\caption{Word usage for open-ended responses of different LLM Agents in Experiment 1}\label{fig:a-24}
\notepar{%
Notes: This figure presents the proportions of Household Agents (Column 1), Household Agents without SMIM (Column 2), Expert Agents (Column 3), and Expert Agents without KAM (Column 4) mentioning words from four word groups in their open-ended responses under four different vignettes. The error bars indicate 95\% confidence intervals.
}
\end{figure}

\begin{figure}[!htbp]
\centering
\includegraphics[width=\linewidth,keepaspectratio]{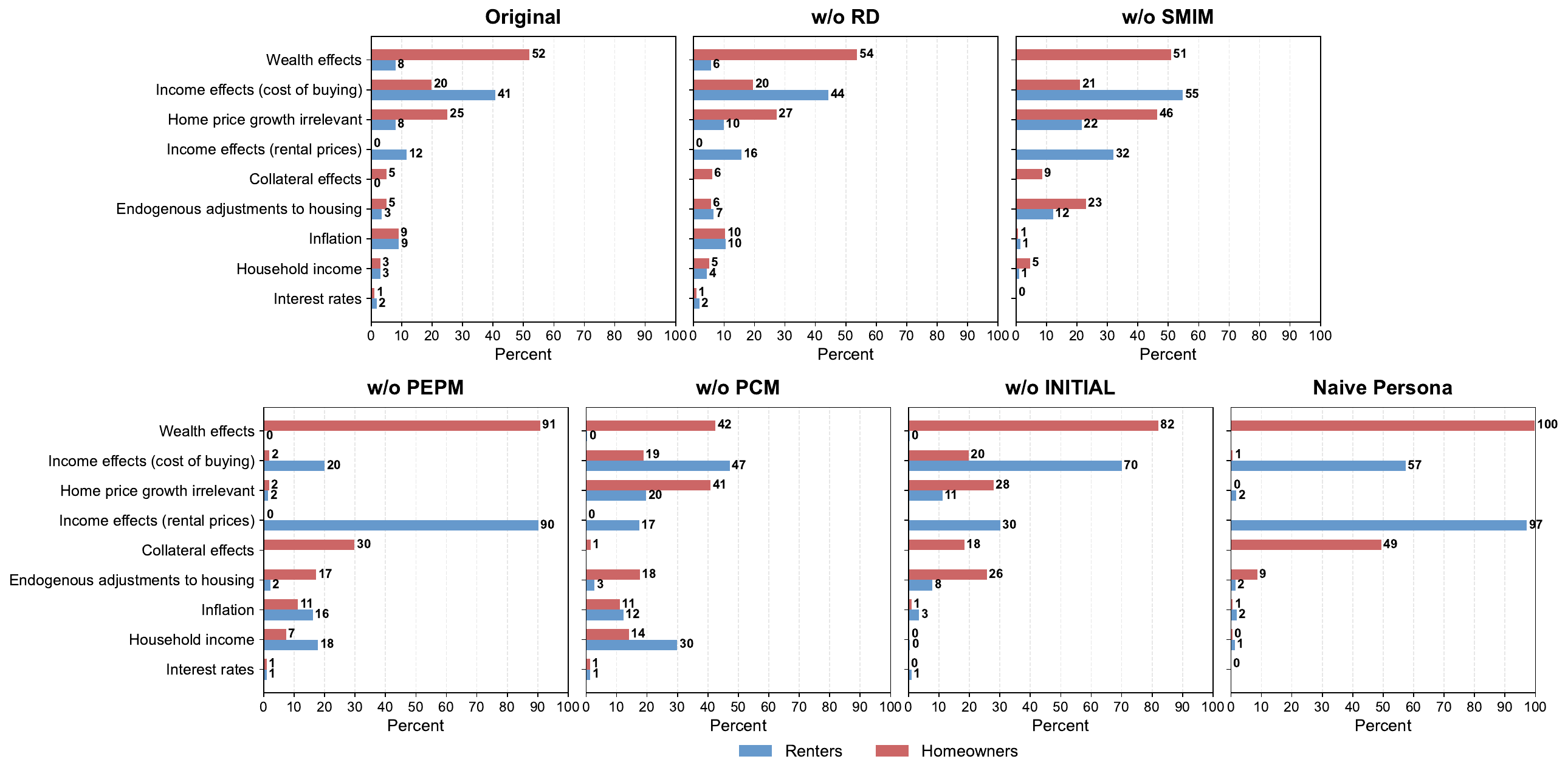}
\caption{Open-ended responses on how higher expected home price growth affects expectations about economic situation generated by different LLM Agents in Sub-Experiment 2 of Experiment 2}\label{fig:a-25}
\notepar{%
Notes: The figure shows the proportion of LLM Agents (original and those without different components) who invoke different arguments to explain why an increase in their expectations about home price growth over the next 10 years would affect their household economic outlook. The open-ended responses from all LLM Agents are automatically classified by an agentic workflow and manually verified.
}
\end{figure}

\begin{figure}[!htbp]
\centering
\subcaptionbox{Inflation expectations\label{fig:sub-a-26-a}}{%
  \includegraphics[width=\linewidth,keepaspectratio]{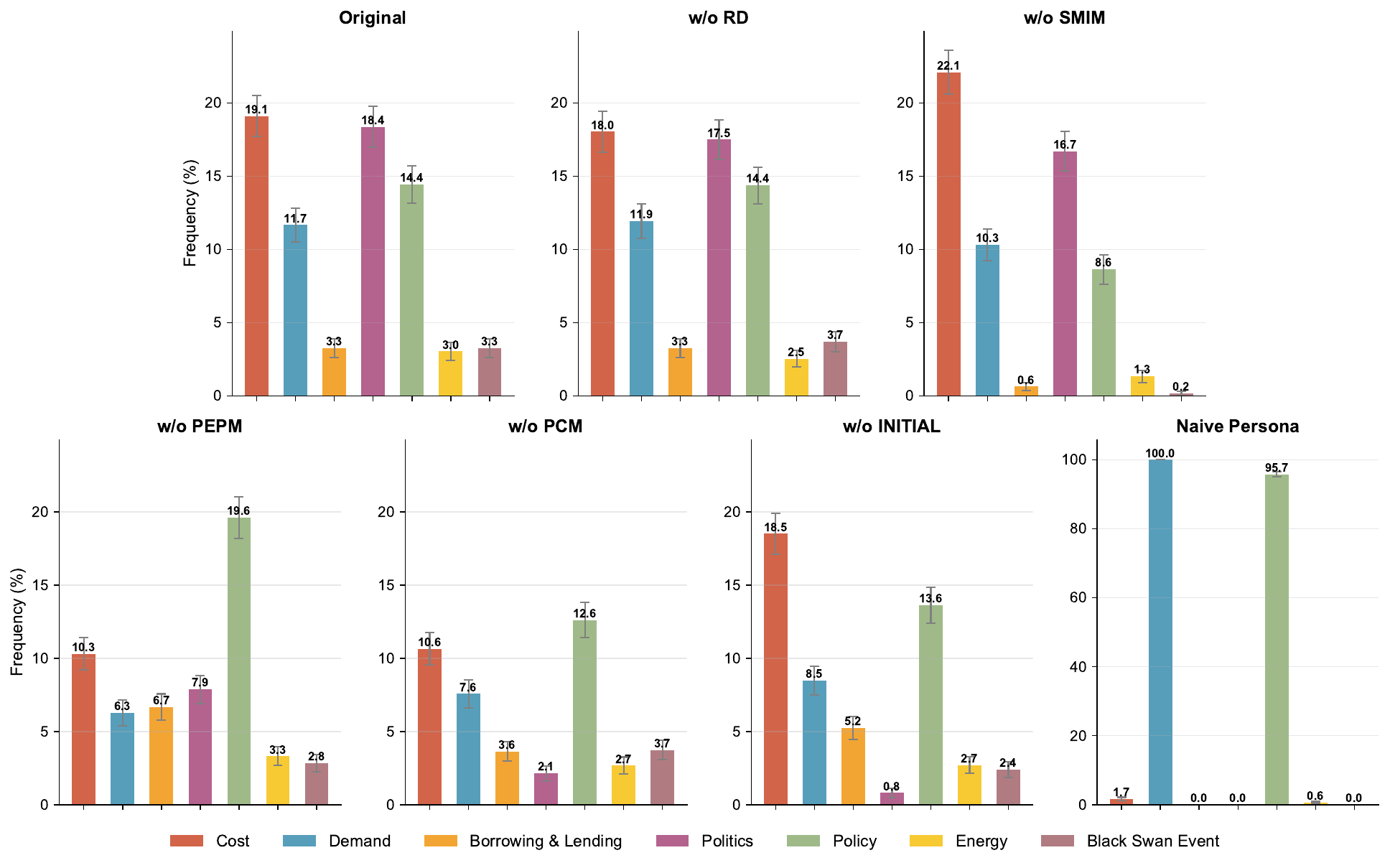}%
}\\[0.6em]
\subcaptionbox{Home price expectations\label{fig:sub-a-26-b}}{%
  \includegraphics[width=\linewidth,keepaspectratio]{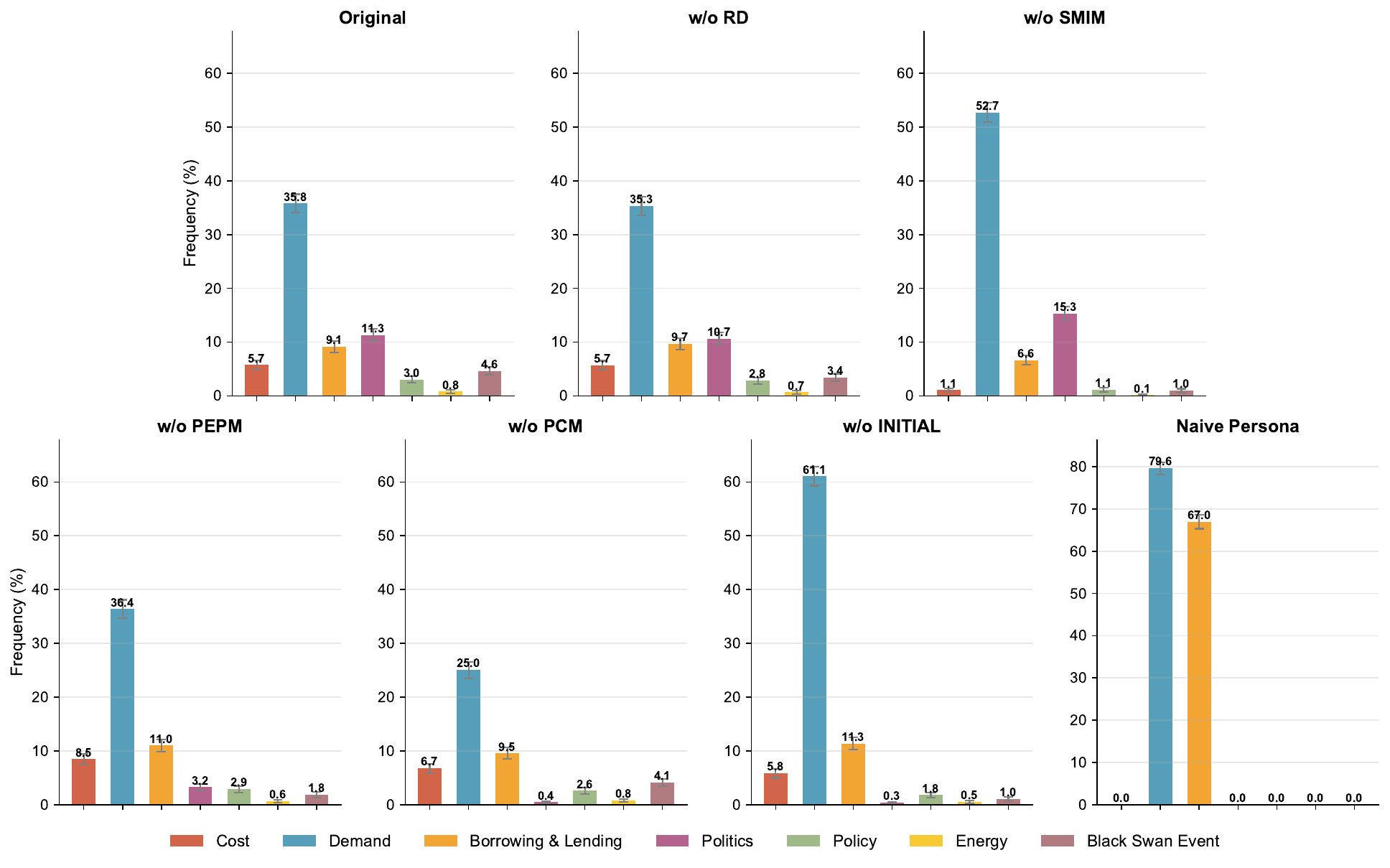}%
}
\caption{Proportion of various channels recalled by different Household Agents in Experiment 3}\label{fig:a-26}
\notepar{%
Notes: This figure displays the proportions of seven channels recalled by Household Agents (original and those without different components) when generating inflation and home price expectations for 2025. Panel (a) presents the results for inflation expectations, while Panel (b) shows the results for home price expectations. Error bars display 95\% confidence intervals.
}
\end{figure}

\newpage

\section{Robustness of Simulation Results}\label{robustness-of-simulation-results}\label{sec:b}

To account for the randomness potentially introduced by the design of certain components in LLM Agents, as well as the inherent randomness in foundation models' outputs\footnote{Across all runs, we hold the `\texttt{seed}' parameter of foundation models fixed. While this substantially reduces output randomness, it cannot eliminate it entirely.}, we run the original experiment multiple times to assess the robustness of the simulation results. Specifically, for Household Agents, we adjust the random seed in each run to alter the process of random matching tweets in the SMIM. For Expert Agents, we adjust the random seed in each run to change the random pairing process in the PBM and the random selection of the most relevant chunks via RAG in the KAM. In addition, for both types of LLM Agents, in each run we vary the random assignment of the two hyperparameters, \texttt{temperature} and \texttt{top-p}.

Given computational costs and runtime, we conduct 20 independent runs using the experiment in the oil price vignette as an example. For Household Agents, we randomly sample 100 cases based on demographic characteristics due to the large original dataset, while for Expert Agents, we utilize all 137 (semi-synthetic) samples but randomly reassign personal information in PBM to priors in PEPM.

\begin{figure}[!htbp]
\centering
\subcaptionbox{Simulation results generated by Household Agents\label{fig:sub-a-27-a}}{%
  \includegraphics[width=\linewidth,keepaspectratio]{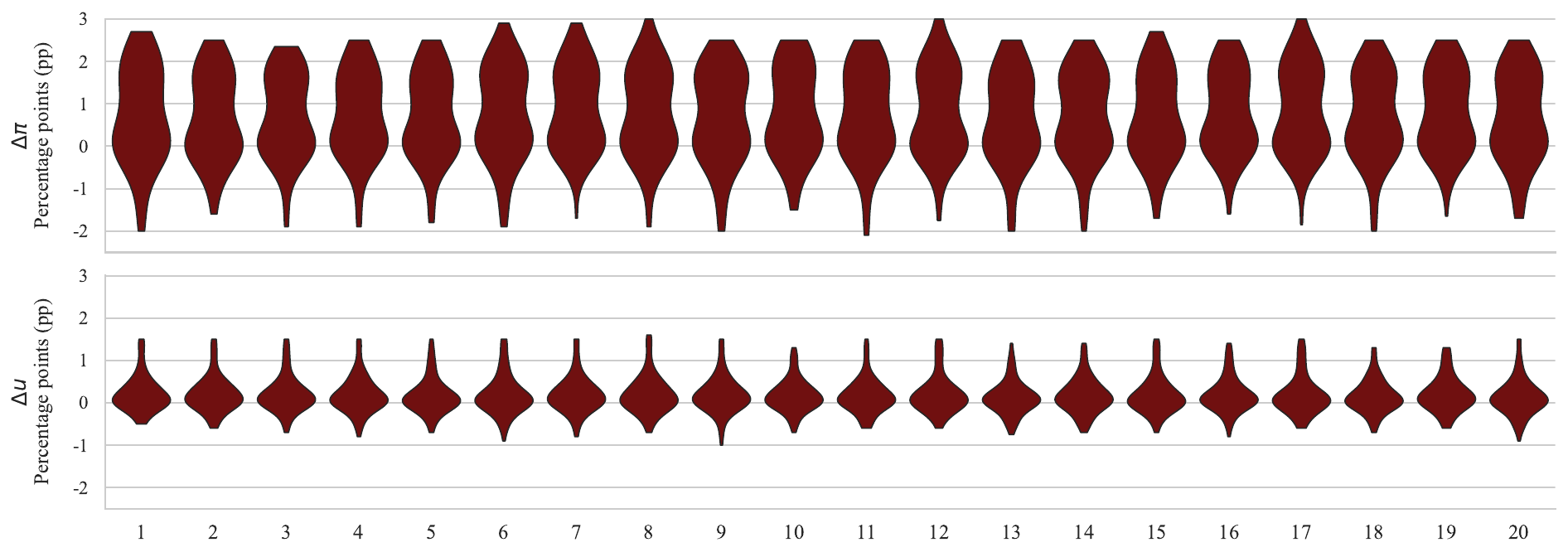}%
}\\[0.6em]
\subcaptionbox{Simulation results generated by Expert Agents\label{fig:sub-a-27-b}}{%
  \includegraphics[width=\linewidth,keepaspectratio]{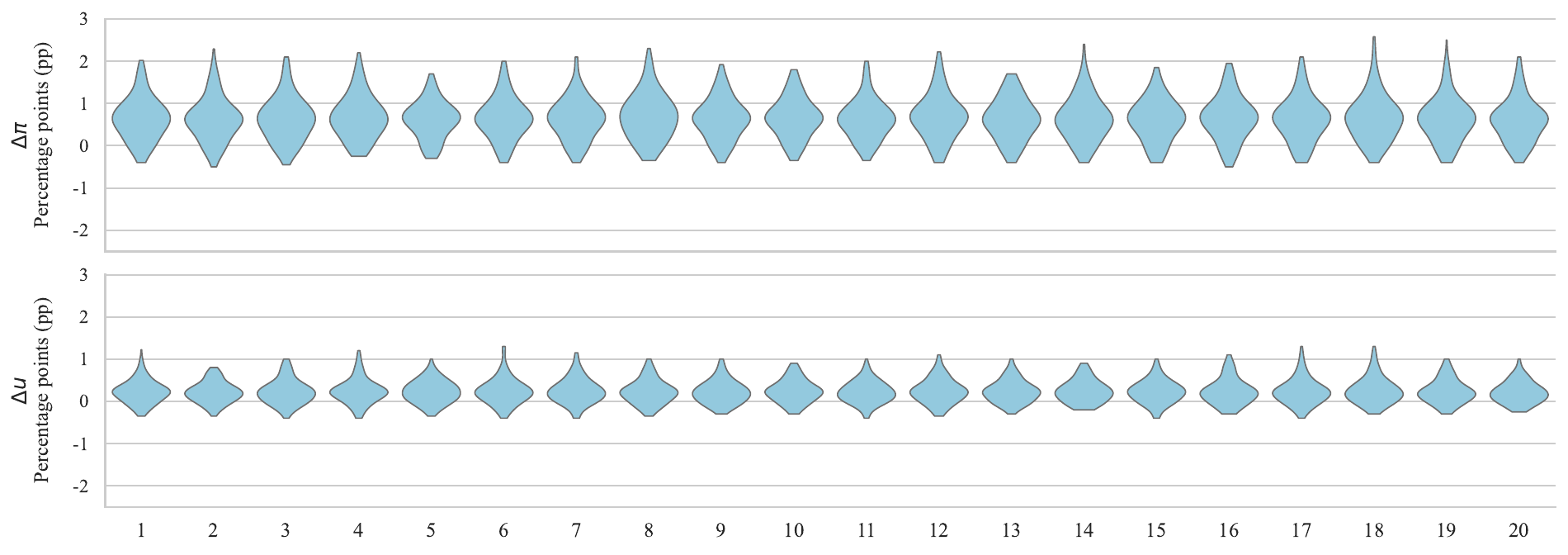}%
}
\caption{Expectation distributions simulated by LLM Agents under oil price shocks in 20 simulations}\label{fig:a-27}
\notepar{%
Notes: Panels (a) and (b) present the results of 20 simulations showing the distributions of changes in expectations (with trimmed 5\% tails) generated by Household Agents and Expert Agents in the oil price vignette, respectively. The x-axis labels the simulations from 1 to 20. The y-axis represents the percentage range of changes in expectations.
}
\end{figure}

Figure~\ref{fig:a-27} shows the distributions of unemployment and inflation expectations from 20 simulation runs. It can be observed that the expectation distributions generated in each run are closely aligned for both Household Agents and Expert Agents. Furthermore, we applied the Kruskal--Wallis test to rigorously examine whether significant differences exist among the distributions from these 20 simulations. As presented in Table~\ref{tab:a-7}, the test results indicate that we cannot reject the null hypothesis ($p$-values are at least 0.85), so the distributions across all 20 simulations originate from the same population and show no statistically significant differences. This suggests that, although the simulation results of LLM Agents exhibit variability due to randomness, the differences are not statistically significant.

Additionally, we examine whether there are significant differences among the 20 thought processes generated by the LLM Agents. Specifically, we first use the Sentence-BERT (SBERT) model all-MiniLM-L6-v2\footnote{{} Detailed information on the model is available at \href{https://huggingface.co/sentence-transformers/all-MiniLM-L6-v2}{{https://huggingface.co/sentence-transformers/all-MiniLM-L6-v2}}.} to obtain embeddings for each open-ended response from every run of the LLM Agents. We then aggregate the embeddings of all open-ended responses in each run into a global embedding---representing the overall semantic information of that run---via four pooling methods: max pooling, mean pooling, min pooling, and weighted mean pooling based on text length. Next, we pair the global embeddings of all 20 runs and identify the lowest cosine similarity among all pairs. As shown in Table~\ref{tab:a-8}, for both Household Agents and Expert Agents, the minimum semantic similarity between any two runs exceeds 0.97, regardless of the pooling method. These results indicate that the randomness introduced by the design of LLM Agents does not compromise the robustness of the simulation results, particularly when the sample size is sufficiently large.

\begin{table}[!htbp]
\centering
\caption{Kruskal--Wallis test on expectation distributions across 20 simulations}\label{tab:a-7}
\small
\begin{tabular}{@{}>{}p{(\linewidth - 10\tabcolsep) * \real{0.2046}}
  >{}p{(\linewidth - 10\tabcolsep) * \real{0.1131}}
  >{}p{(\linewidth - 10\tabcolsep) * \real{0.1372}}
  >{}p{(\linewidth - 10\tabcolsep) * \real{0.2435}}
  >{}p{(\linewidth - 10\tabcolsep) * \real{0.1714}}
  >{}p{(\linewidth - 10\tabcolsep) * \real{0.1289}}@{}}
\toprule
\textbf{Agent} & \textbf{Variable} & \textbf{Total \emph{N}} & \textbf{Mean \emph{N}/Group} & \textbf{H-statistic} & \textbf{\emph{p}-value} \\
\midrule
\multirow{2}{=}{ Household Agents} & \(\Delta\ \pi\) & 3830 & 191.5 & 12.6594 & 0.8555 \\
& \(\Delta\ u\) & 3830 & 191.5 & 11.2398 & 0.9155 \\
\multirow{2}{=}{ Expert Agents} & \(\Delta\ \pi\) & 5214 & 260.7 & 9.0924 & 0.9719 \\
& \(\Delta\ u\) & 5229 & 261.4 & 6.8884 & 0.9948 \\
\bottomrule
\end{tabular}
\notepar{%
Notes: The table presents the Kruskal--Wallis test results on the distributions of changes in inflation (\(\Delta\ \pi\)) and unemployment expectations (\(\Delta\ u\)) generated by Household Agents and Expert Agents over 20 simulations. The null hypothesis states that all independent distributions have the same central tendency and thus originate from the same population; the alternative hypothesis posits that at least one distribution differs in central tendency and therefore originates from a different population.
}
\end{table}

\begin{table}[!htbp]
\centering
\caption{The minimum semantic similarity between the thoughts of LLM Agents across 20 simulations}\label{tab:a-8}
\small
\begin{tabular}{@{}>{}p{(\linewidth - 4\tabcolsep) * \real{0.3286}}
  >{}p{(\linewidth - 4\tabcolsep) * \real{0.3285}}
  >{}p{(\linewidth - 4\tabcolsep) * \real{0.3283}}@{}}
\toprule
\textbf{Agent} & \textbf{Pooling Method} & \textbf{Minimum Similarity} \\
\midrule
\multirow{4}{=}{ Household Agents} & Max & 0.979337 \\
& Mean & 0.997027 \\
& Min & 0.975266 \\
& Weighted Mean & 0.996773 \\
\multirow{4}{=}{ Expert Agents} & Max & 0.977026 \\
& Mean & 0.997873 \\
& Min & 0.976205 \\
& Weighted Mean & 0.997891 \\
\bottomrule
\end{tabular}
\notepar{%
Notes: This table presents the minimum pairwise cosine similarity of LLM Agents' thought processes across 20 simulations, where the overall embeddings of open-ended responses in each simulation are aggregated using max pooling, mean pooling, min pooling, and length-weighted mean pooling, respectively.
}
\end{table}

\clearpage
\section{Analysis of Mental Models}\label{analysis-of-mental-models}\label{sec:c}

\subsection{How to Identify Directed Acyclic Graphs}\label{how-to-identify-directed-acyclic-graphs}\label{sec:c-1}

We develop an agentic workflow to automatically identify and label DAGs in open-ended responses, with its architecture shown in Figure~\ref{fig:a-12}. First, we employ two medium-scale LLMs of different types---deepseek-r1-distill-qwen-32b and qwq-32b---as Causal Analyzers. Each independently identifies causal pathways in open-text responses related to unemployment and inflation expectations. Given the vignette and expectation type, the start and end points of each causal pathway are predetermined; thus, the workflow focuses on identifying and categorizing all intermediate variables (see Table~\ref{tab:a-3}) in these pathways. We illustrate the tasks of the Causal Analyzers in the first round using the example of processing open-text responses about inflation expectations under an oil price vignette, as detailed in the following prompt:

\begin{verbatim}
You are an expert economic analyst specializing in inflation causal structure identification. Your task is to analyze survey responses and identify causal pathways from oil price changes to inflation outcomes.
\end{verbatim}

\begin{verbatim}
Note: The survey is conducted under a scenario of rising oil price and is intended to examine how respondents perceive the causal pathways of the impact of rising oil price on inflation.
\end{verbatim}

\begin{verbatim}
For each text sample (i.e. response), you must identify causal structures in the following format:
\end{verbatim}

\begin{verbatim}
{Oil price}→{Intermediate variable 1 (optional)}→{Intermediate variable 2 (optional)}→...→{Inflation}
\end{verbatim}

\begin{verbatim}
The intermediate variables must come from this predefined list:
\end{verbatim}

[See Table~\ref{tab:a-3}, which will not be elaborated here.]

\begin{verbatim}
It is important to note that if the text is complex and mentions multiple variables and their causal relationships, there may be more than one causal path related to inflation. For example, the causal paths regarding inflation may include:
\end{verbatim}

\begin{verbatim}
{Oil price} → {Inflation};
\end{verbatim}

\begin{verbatim}
{Oil price} → {Intermediate variable 1} → {Inflation};
\end{verbatim}

\begin{verbatim}
{Oil price} → {Intermediate variable 2} → {Intermediate variable 3} → {Inflation}.
\end{verbatim}

\begin{verbatim}
Analyze the following text and identify causal pathways from oil price to inflation:
\end{verbatim}

\begin{verbatim}
<text>{start}</text>
\end{verbatim}

\begin{verbatim}
Identify all causal pathways that lead from oil price increases to inflation and provide corresponding reasons. Focus specifically on inflation-related outcomes.  Please make sure to identify all mentioned intermediate variables; even if these variables are considered unchanged, they should still be included in the causal pathways as long as they are mentioned. Output your response strictly in the following JSON format:
\end{verbatim}

\begin{verbatim}
{
\end{verbatim}

\begin{verbatim}
  “reasons”: “Explain your reasoning for giving this identification. (about 3-4 sentences)”,
\end{verbatim}

\begin{verbatim}
  “pathways”: [
\end{verbatim}

\begin{verbatim}
    “pathway1”, 
\end{verbatim}

\begin{verbatim}
    “pathway2”,
\end{verbatim}

\begin{verbatim}
    “...”
\end{verbatim}

\begin{verbatim}
  ]
\end{verbatim}

\begin{verbatim}
}
\end{verbatim}

\begin{verbatim}
Note: If multiple pathways exist, include the most important ones (no more than three) in the array. If only one pathway exists, include only one string. If no relevant pathway to inflation is found, use an empty array []. Each pathway should be a string in the format: “{Oil price}→{intermediate variables}→{Inflation}”.
\end{verbatim}

Second, upon completion of Round 1, the two Analyzers discuss their initial findings. If their outputs align, they pass the consistency check directly. If not, they engage in iterative discussion until consensus is reached. Should consensus remain elusive after 10 rounds, the entire process is repeated until agreement is achieved.

Third, a Classifier categorizes the intermediate variables within the causal pathways, consolidating diverse variables into generalized nodes. The classification criteria are predefined, as detailed in Table~\ref{tab:a-3}.

Finally, a Results Consolidator integrates the categorized outcomes. The consolidation rules are as follows: (1) If multiple intermediate variables from the same category (node) appear consecutively in a pathway, merge them into a single node. (2) If multiple pathways become identical after the previous step, merge them into a single pathway. (3) For other cases, simply update the pathways with the category information and output directly.

We demonstrate the output of the Classifier and Consolidator with a simple example:

If the agreed-upon pathways after discussion are as follows:

\{Oil price\}→\{cost firms\}→\{firm prices\}→\{Inflation\} and \{Oil price\}→\{costs borrowing firms\}→\{Inflation\}.

After categorization: \{Oil price\}→\{Firms' costs\}→\{Firms' costs\}→\{Inflation\} and \{Oil price\}→\{Firms' costs\}→\{Inflation\}.

After consolidation (final result): \{Oil price\}→\{Firms' costs\}→\{Inflation\} (merged as identical).

\subsection{Similarity of Mental Models}\label{similarity-of-mental-models}\label{sec:c-2}

After completing the identification in Section~\ref{sec:c-1}, each open-ended response is converted into a DAG (i.e., a mental model). First, we represent each DAG as an edge list---a set of unique causal connections.

For example, in a given vignette, suppose there are two human samples: Response 1 has a DAG ``A→B→C'' with an edge list \emph{E}\textsubscript{1} = \{A→B, B→C\}; Response 2 has a DAG ``A→C; B→C'' with \emph{E}\textsubscript{2} = \{A→C, B→C\}. The aggregated edge list for all human responses in this vignette (termed the ``mental model set'') is then \(E_{Human}\) = \{A→B, B→C, A→C\}. Similarly, we assume the mental model set for LLM Agents in this vignette is \(E_{Agent}\) = \{A→B, A→C, C→D\}.

Next, we measure the similarity between the mental models of humans and LLM Agents in this vignette using the Jaccard similarity between their mental model sets

\[Sim(Human,\ Agent) = \frac{\left| E_{Human} \cap E_{Agent} \right|}{\left| E_{Human} \cup E_{Agent} \right|},\]

where \(| \cdot |\) denotes the number of elements in a set.

The Jaccard similarity equals 1 (0) if and only if the two mental models are identical (completely different). This similarity increases as the number of common elements between the two sets grows. For example, in the case above, the similarity between the mental models of humans and LLM Agents is \(Sim(Human,\ Agent) = \ \frac{2}{4}\  = \ \frac{1}{2}\ \).

\clearpage
\section{Measuring the Diversity of the Thoughts Underlying Expectations}\label{measuring-the-diversity-of-the-thoughts-underlying-expectations}\label{sec:d}

In this section, we detail how to measure the diversity of thoughts underlying respondents' expectations, which essentially reflects the semantic diversity in their open-ended responses. Given \emph{n} open-ended responses, we convert them into \emph{n} embeddings using the all-MiniLM-L6-v2 model to capture their semantic information.

Next, we compute the pairwise cosine similarity between all embeddings to obtain a similarity matrix \(\mathbf{S} \in \mathbb{R}^{n \times n}\), where element \(S_{ij} = \frac{\mathbf{e}_{i} \cdot \mathbf{e}_{j}}{\left\| \mathbf{e}_{i} \right\|\left\| \mathbf{e}_{j} \right\|}\) denotes the semantic similarity between response \emph{i} and response \emph{j}, with \(\mathbf{e}_{i}\) and \(\mathbf{e}_{j}\) being their corresponding embeddings.

Then, we compute the eigenvalues \(\lambda_{1},\ \lambda_{2},\ \ldots,\ \lambda_{n}\) of matrix \(\mathbf{S}\). After filtering out the few negative values, these eigenvalues are normalized into a probability distribution \(p_{i} = \frac{\lambda_{i}}{\sum_{j = 1}^{k}\lambda_{j}}\), where \emph{k} is the number of positive eigenvalues. The \emph{Shannon entropy} is then calculated as

\[H = \  - \sum_{i = 1}^{k}{p_{i}\ln}\left( p_{i} \right).\]

A higher entropy value \emph{H} indicates a more uniform eigenvalue distribution, reflecting greater dispersion of texts in the semantic space and thus higher semantic diversity, and vice versa. Finally, to eliminate the influence of text quantity on entropy, \emph{H} is normalized to {[}0, 1{]} by the maximum entropy \(\ln k\)

\[D = \ \frac{H}{\ln k},\]

where \emph{D} represents the metric of semantic diversity used in this paper.

This method is applied to the open-ended responses from three representative experiments, and the resulting diversity of underlying thoughts generated by LLM Agents with different components removed and humans is shown in Table~\ref{tab:a-4} to Table~\ref{tab:a-6}, respectively.

\end{document}